\documentclass[amsmath, amssymb, twocolumn, aps, prx, superscriptaddress, footinbib]{revtex4-1}

\usepackage{mathtools}
\usepackage{mathrsfs, amsmath, wasysym}
\usepackage{bbold}
\usepackage[mathscr]{eucal}
\usepackage{xfrac}
\usepackage{graphicx}
\usepackage{dcolumn}
\usepackage{bm}
\usepackage{color}
\usepackage{tabularx}
\usepackage{natbib}
\usepackage[colorlinks,linkcolor=blue,citecolor=blue,urlcolor=blue]{hyperref}
\usepackage[normalem]{ulem}
\usepackage[all]{hypcap}
\usepackage[dvipsnames,usenames,table]{xcolor}
\usepackage{braket,mleftright}

\newcommand{\nocontentsline}[3]{}
\newcommand{\tocless}[2]{\bgroup\let\addcontentsline=\nocontentsline#1{#2}\egroup}

\newcommand{\bk}{{\bf k}}

\newcommand{\bq}{{\bf q}}

\newcommand{\bw}{{\bf w}}

\newcommand{\bn}{{\bf n}}

\newcommand{\bh}{{\bf h}}

\newcommand{\btau}{\boldsymbol{\tau} }
\newcommand{\bsigma}{\boldsymbol{\sigma} }

\newcommand{\be}{\begin{equation}}
\newcommand{\ee}{\end{equation}}
\newcommand{\beg}{\begin{gather}}
\newcommand{\eeg}{\end{gather}}
\newcommand{\beq}{\begin{eqnarray}}
\newcommand{\eeq}{\end{eqnarray}}
\newcommand{\bea}{\begin{align}}
\newcommand{\eea}{\end{align}}
\newcommand{\beqq}{\begin{eqnarray*}}
\newcommand{\eeqq}{\end{eqnarray*}}

\newcommand{\up}{\uparrow}
\newcommand{\down}{\downarrow}

\newcommand{\ve}{{\varepsilon}}

\begin{document}

\title{Orbital piezomagnetic polarizability of pure insulating altermagnets in two dimensions}

\author{Beryl Bell}
\affiliation{Department of Physics, Drexel University, Philadelphia, PA 19104, USA}%


\author{J. W. F. Venderbos}
\affiliation{Department of Physics, Drexel University, Philadelphia, PA 19104, USA}%
\affiliation{Department of Materials Science \& Engineering, Drexel University, Philadelphia, PA 19104, USA \looseness=-1}%

\begin{abstract}
The distinctive symmetry properties of pure altermagnets make them natural candidates for piezomagnetism. Previous work motivated by the piezomagnetic properties of altermagnets has primarily focused on the spin magnetization response to applied strain. In this paper we study orbital piezomagnetic effects---the orbital magnetization response to applied strain---in minimal lattice models of pure insulating altermagnets in two dimensions. We obtain general microscopic expressions for the orbital magnetization in the presence of strain, as well as the orbital piezomagnetic polarizability, i.e., the defining response characteristic of pure altermagnets. We apply these expressions to three specific tetragonal lattice models, two corresponding to $d$-wave altermagnets and one describing a $g$-wave altermagnet. Whereas the $d$-wave altermagnets are associated with a linear piezomagnetic polarizability, the $g$-wave altermagnet exhibits a nonlinear piezomagnetic effect. Our analysis reveals how the polarizabilities are related to and determined by the Berry curvature of the occupied bands. Connections to materials of current interest are discussed. 
\end{abstract}

\date{\today}

\maketitle

\section{Introduction \label{sec:intro}}

Altermagnets are a category of magnetic materials that have attracted increasing attention in recent years~\cite{Smejkal:2022p031042,Smejkal:2022p040501,Mazin:2022p040002,Turek:2022p094432,Mazin:2021pe2108924118,Song:2025p473,Jungwirth:2025p100162,Jungwirth:2026p837}. Altermagnets have fully compensated collinear magnetic order with no net magnetization, which is similar to antiferromagnets, but nonetheless exhibit a Zeeman spin splitting of the energy bands reminiscent of ferromagnets~\cite{Hayami:2019p123702,Bai:2023p216701,Yuan:2021p014409,Karube:2022p137201,Egorov:2021p2363,Guo:2023p100991,Naka:2019p4305,Yuan:2023pe2211966,Hayami:2020p144441,Gonzalez-Hernandez:2021p127701,Naka:2021p125114,Shao:2021p7061,Ma:2021p2846,Bose:2022p267}. This spin splitting is of non-relativistic origin and is caused by the particular crystallographic arrangement of the magnetic ions. The latter lacks the symmetries that enforce spin-degeneracy. In sharp contrast to ferromagnets, the energy band spin splitting in altermagnets is non-uniform and has a $d$-, $g$-, or $i$-wave nodal structure, which is directly linked to the symmetry and hence response properties of altermagnets~\cite{Smejkal:2022p031042,Smejkal:2022p040501}. These defining characteristics can be viewed as intermediate between ferromagnetism and antiferromagnets, and have fueled renewed interest in the classification of magnets and their fundamental properties. They have further put altermagnets at the forefront of the spintronics frontier~\cite{Song:2025p473}.

While the formal classification of altermagnetism is predicated on the non-relativistic limit~\cite{Smejkal:2022p031042,Liu:2022p021016,Jiang:2024p031039,Xiao:2024p031037,Chen:2024p031038,Schiff:2025p109}, including the effect of spin-orbit coupling gives rise to a variety of interesting phenomena associated with altermagnetic order~\cite{Fernandes:2024p024404,Smejkal:2022p482}. For instance, depending on the orientation of the N\'eel vector the symmetries of the magnetic space group can allow for weak ferromagnetism and an anomalous Hall effect~\cite{Smejkal:2022p482}. Materials of this kind have been referred to as `mixed' altermagnets, since the ordered moments are not strictly but only nearly compensated, giving rise to a net magnetization and a spin polarization of $s$-wave type~\cite{Fernandes:2024p024404}. Instead, in `pure' altermagnets the magnetic space group forbids both a net magnetization and an anomalous Hall effect (since the Hall vector transforms as the magnetization), and moments remain strictly compensated~\cite{Fernandes:2024p024404}. Although an anomalous Hall effect is forbidden by symmetry, pure altermagnets can still show signatures of nontrivial topology, such as mirror Chern bands or topological nodal band crossings~\cite{Antonenko:2025p096703,Fernandes:2024p024404}, and this raises the question whether the Berry curvature associated with these features can be unlocked and exploited by applying external fields. 

One way to achieve this is the application of external strain. The response of pure altermagnets to applied strain is directly tied to the nature of the magnetic order parameter, which, in the case of $d$-wave altermagnets, has the symmetry of a magnetic octupole~\cite{McClarty:2024p176702,Bhowal:2024p011019}. As a result, the reduction of symmetry caused by strain can enable both a net magnetization and an anomalous Hall effect, which makes pure altermagnets natural candidates for harnessing piezomagnetism and for observing elasto-Hall effects~\cite{Ma:2021p2846,McClarty:2024p176702,Aoyama:2024pL041402,Steward:2023p144418,Fernandes:2024p024404,Takahashi:2025p184408}. The piezomagnetic and elasto-Hall responses of pure altermagnets are a particularly appealing feature, and have inspired a number of proposals for manipulating and controlling the properties of altermagnets through strain~\cite{Li:2024p222404,Chakraborty:2024p144421,Karetta:2025p094454,Belashchenko:2025p086701,Naka:2025p083702,Zhang:2025p024415,Chen:2025p102409,Xun:2025p161903,Jiang:2025p053102,Khodas:2025arXiv06,Zhai:arXiv2025,Xun:2025p161903,Smolenski:arXiv2025}.

In the non-relativistic limit, the strain-induced magnetization can only be of spin origin~\cite{McClarty:2024p176702,Khodas:2025arXiv06}, but in general, the induced magnetization has both a spin and an orbital magnetization component~\cite{Sorn:2025p245115,Ye:2026p014413}. In this paper we focus on the latter and study the strain-induced \emph{orbital} magnetization of pure altermagnets in two dimensions (2D). Previous work has addressed the spin magnetization of a strained 2D altermagnet (i.e., the Lieb lattice model \cite{Brekke:2023p224421,Antonenko:2025p096703}) and found that it is only nonzero in a metal, when the spin-polarized bands are filled unequally~\cite{Takahashi:2025p184408,Khodas:2025arXiv06}. In contrast, the orbital magnetization can be nonzero the insulating regime, and in this paper we exclusively consider altermagnetic insulators for which the spin magnetization response vanishes. 

We specifically consider three microscopic models for tetragonal altermagnets in 2D. These models are minimal models for altermagnets with distinct symmetry~\cite{Roig:2024p144412}, in particular for altermagnets with $d_{x^2-y^2}$-wave~\cite{Brekke:2023p224421,Antonenko:2025p096703}, $d_{xy}$-wave~\cite{Maier:2023pL100402}, and $g$-wave symmetry. As such, these representative minimal models serve as an appropriate foundation for a general survey of orbital piezomagnetism in 2D. In particular, whereas the two $d$-wave altermagnets exhibit a linear piezomagnetic effect, i.e., a linear dependence of the orbital magnetization on applied strain, here we show that the $g$-wave altermagnet exhibits a non-linear orbital piezomagnetic effect~\cite{Khodas:2025arXiv06}. 

To examine the orbital piezomagnetic response, we rely on the modern theory of orbital magnetization, which provides the basis for computing the orbital magnetization of periodic systems in terms of band energies and eigenstates~\cite{Xiao:2005p137204,Thonhauser:2005p137205,Ceresoli:2006p024408}. For the general class of pure altermagnets considered here, we obtain a simple formula for the orbital magnetization, which is directly expressed in terms of the parametrization of the Hamiltonian~\cite{Mitscherling:2025arXiv12}. We furthermore derive microscopic expressions for orbital piezomagnetic polarizability, i.e., the piezomagnetic response tensor. Based on these expressions we compute the polarizability for each of the three minimal models and analyze the result in terms of relevant band structure properties, specifically the Berry curvature distribution of the occupied bands. Together, this constitutes a general theory of orbital piezomagnetism in two-dimensional altermagnets.

\section{Model altermagnets in two dimensions}
\label{sec:problem}

In this section we introduce and examine lattice models of pure altermagnets in 2D, with N\'eel vector oriented along the perpendicular $\hat z$ direction. We first present a general formulation of the models of interest, and then discuss specific lattice realizations with tetragonal symmetry. 

\subsection{General Hamiltonian of pure 2D altermagnets}
\label{ssec:general-H}

The altermagnets we consider have two magnetic sites in the unit cell, labeled $A$ and $B$, and these sites define the sublattice degree of freedom. To construct the tight-binding Hamiltonian, we introduce the electron destruction operators $c_{\bk \alpha \sigma}$, where $\alpha=A,B$ and $\sigma=\up,\down$ corresponds to spin. The Hamiltonian is then expressed as
\be
H = \sum_{\bk} c^\dagger_\bk H_\bk c_\bk, \;\; c_{\bk} = ( c_{\bk A\up},c_{\bk A\down},c_{\bk B\up},c_{\bk B\down})^T, \label{eq:H-def}
\ee
where $H_\bk$, the momentum-dependent Hamiltonian matrix, is the sum of two terms: 
\be
H_\bk = H_{0,\bk} + \phi W_\bk. \label{eq:H_k}
\ee
The first term is the Hamiltonian of the altermagnet without strain and the second term describes the effect of strain. Strain is parametrized by a strain field $\phi$. Following notation introduced in Ref.~\onlinecite{Roig:2024p144412}, the unstrained Hamiltonian $H_{0,\bk} $ takes the general form
\be
H_{0,\bk} = \varepsilon_\bk + t_{x,\bk} \tau^x+t_{z,\bk} \tau^z+ \lambda_{z,\bk } \tau^y \sigma^z + N_z \tau^z \sigma^z, \label{eq:H_0k}
\ee
where we have made use of two sets of Pauli matrices, $\btau = (\tau^x,\tau^y,\tau^z)$ and $\bsigma = (\sigma^x,\sigma^y,\sigma^z)$, corresponding to the sublattice and spin degree of freedom, respectively. The functions $t_{x,\bk}$ and $t_{z,\bk}$ describe intersublattice and intrasublattice hopping processes, respectively, and $\lambda_{z,\bk }$ describes an allowed spin-orbit coupling term~\cite{Roig:2024p144412}. As mentioned above, the collinear N\'eel order is always chosen along the $\hat z$ direction, such that, in the absence of strain, a pure altermagnetic state without net magnetization is realized. Magnetic order enters the Hamiltonian via the N\'eel order parameter $N_z$. 

A key symmetry constraint is imposed by the mirror reflection $\mathcal M_z$, which remains a symmetry of the magnetic state when the N\'eel vector points along the $\hat z$ direction. Since $\mathcal M_z$ does not affect $\bk = (k_x,k_y)$, it must commute with $H_{0,\bk} $ and therefore imposes the form of Eq.~\eqref{eq:H_0k}. Specifically, the mirror symmetry $\mathcal M_z$ implies that $H_{0,\bk} $ is diagonal in spin space (i.e., spin-orbit terms featuring $\sigma^x$ and $\sigma^y$ are forbidden) and can be considered in each spin sector separately~\footnote{Technically, the Hamiltonian decomposes into two diagonal mirror sectors. The spin label $\sigma$ may be used here to label these mirror sectors.}. A convenient way to write the Hamiltonian in each spin sector is
\be
H^\sigma_{0,\bk}  = \varepsilon_\bk +\bn^\sigma_\bk \cdot \btau, \label{eq:H_0^sigma}
\ee
where $\bn^\sigma_\bk$ is given by
\be
\bn^\sigma_\bk = (t_{x }, \sigma \lambda_z, t_z + \sigma N_z ) ,\label{eq:n_k}
\ee
and $\sigma $ should be read as $\pm1$ for $\up$/$\down$.

Next, consider strain. The effect of strain is described by a term $W_\bk$ in the Hamiltonian, which we further parametrize as
\be
W_\bk = \chi_\bk +  \bw_\bk \cdot \btau . \label{eq:W_k}
\ee
Since strain preserves the mirror symmetry $\mathcal M_z$, the full Hamiltonian can still be decomposed into separate spin sectors. We express the full Hamiltonian as
\be
H^\sigma_{\bk}  =\xi_\bk+ \bh^\sigma_\bk \cdot \btau ,  \label{eq:H^sigma}
\ee
where $\xi_\bk $ and $\bh^\sigma_\bk$ are given by
\be
\xi_\bk =  \varepsilon_\bk + \phi \chi_\bk , \quad \bh^\sigma_\bk=\bn^\sigma_\bk + \phi \bw_\bk. \label{eq:xi_k-h_k}
\ee
This general formulation of the Hamiltonian provides a convenient framework for our analysis in the remainder of this paper. Since the Hamiltonian in each spin sector is a two-band Hamiltonian, it is straightforward to obtain the energy bands. We denote the energy bands $E^\sigma_{\bk n} $, where $n=1,2$ is the band label. Explicitly, the band energies are given by 
\be
E^\sigma_{\bk1,2} = \xi_\bk \mp | \bh^\sigma_\bk|. \label{eq:E_n}
\ee
It is similarly straightforward to determine the projectors onto the eigenspaces associated with the energy bands $E^\sigma_{\bk n} $. These projectors, denoted $P^\sigma_{\bk n} $, are given by
\be
P^\sigma_{\bk1,2} = \frac12 (\mathbb{1} \mp \bh^\sigma_\bk \cdot \btau /  | \bh^\sigma_\bk|),  \label{eq:P_n}
\ee
and will be used to obtain a general formula for the orbital magnetization [see Eqs.~\eqref{eq:Mz-2} and \eqref{eq:Mz-altermagnet}].

We conclude this subsection with a general remark regarding crystal symmetries and their implications for the energy spectrum. A key characteristic of pure altermagnets is that certain point group symmetries of the crystal lattice exchange the magnetic sublattices but do not flip the spin. In the case of a tetragonal $d$-wave altermagnet an example is fourfold rotation symmetry $\mathcal C_{4z}$. When combined with time-reversal ($\mathcal T$) such point group symmetries leave the magnetic state invariant. Let $g$ be such a point group symmetry. Then $\mathcal T g$ is a symmetry of the pure (unstrained) altermagnet described by Hamiltonian \eqref{eq:H_0^sigma}, and the functions $\varepsilon_{\bk}$ and $ \bn^\sigma_{\bk}$ satisfy
\be
\varepsilon_{-\bk} = \varepsilon_{g\bk}, \quad  \bn^\up_{-\bk} = R\bn^\down_{g\bk}, \label{eq:symmetries}
\ee
where $R = \text{diag}(1,1,-1)$ is a diagonal matrix. Since the Hamiltonian of the pure altermagnet is also manifestly even under inversion, we further have $\varepsilon_{-\bk} =\varepsilon_{\bk}$ and $\bn^\sigma_{-\bk}=\bn^\sigma_{\bk} $. These relations can be used to show that the orbital magnetization must vanish in the absence of strain (see Sec.~\ref{ssec:Mz}). 


\subsection{Model realizations with tetragonal symmetry}
\label{ssec:models}

We now consider three different realizations of the general model introduced above, all with tetragonal symmetry and point group $D_{4h}$. In each of the three models the site symmetry group of the magnetic sublattices is different, however, such that they give rise to altermagnets with $d_{x^2-y^2}$- $d_{xy}$-, and $g_{yx^3-xy^3}$-wave symmetry. 

\begin{figure}
	\includegraphics[width=\columnwidth]{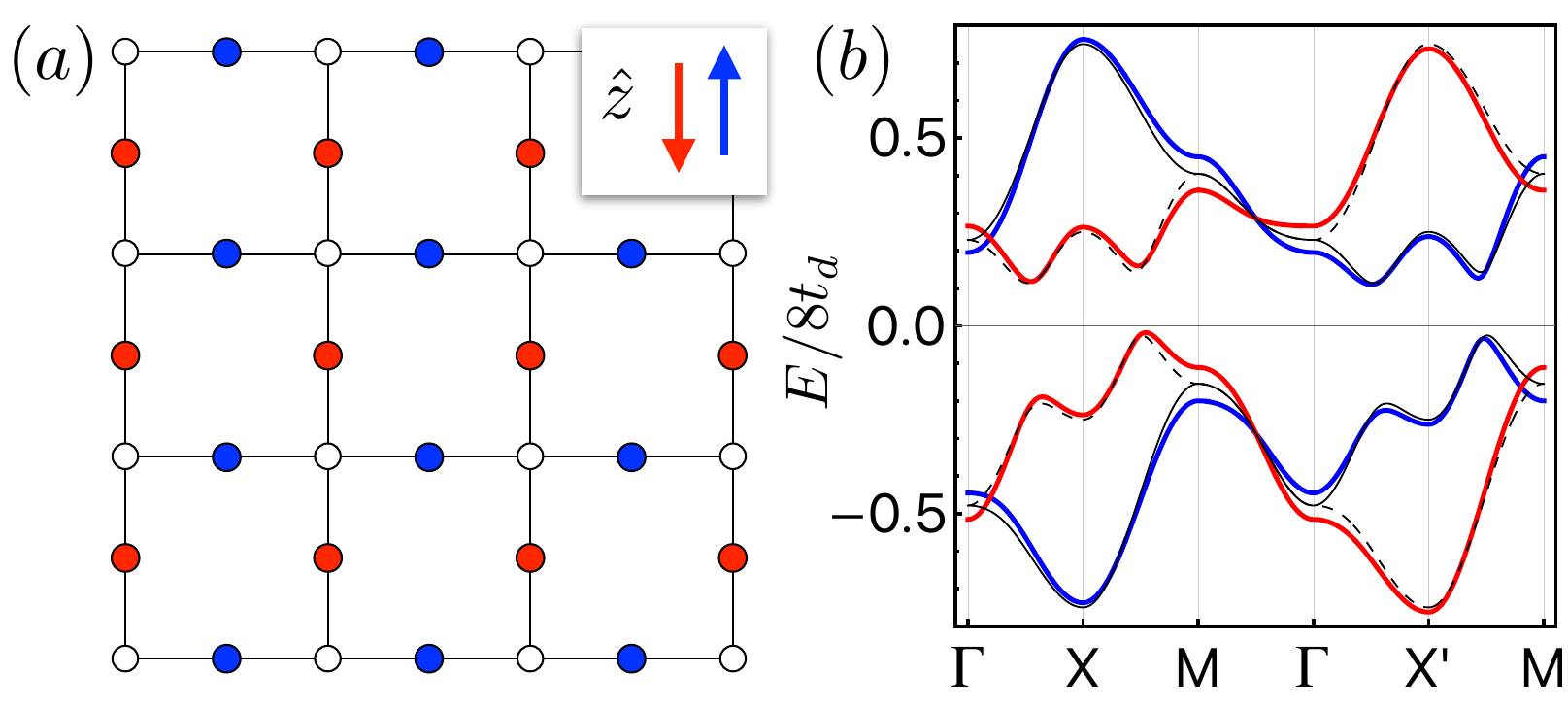}
	\caption{{\bf Lieb lattice model.} (a) Schematic of the two-dimensional Lieb lattice. The magnetic $A$ and $B$ sites are shown in blue and red, and the nonmagnetic site is shown in white. The ordered moments of the collinear altermagnetic N\'eel state point out of the plane, as indicated. (b) Energy spectrum with and without strain (red/blue and solid/dashed black, respectively) of the Lieb lattice model specified by Eqs.~\eqref{eq:eps_k-t_xk}--\eqref{eq:Lieb-strain} with parameters $(t_0,t_1,\lambda,N_z) = (0.25t_d,0.5t_d,0.25t_d,2.0t_d)$ and $\phi=0.1$ for the case with strain. Blue and red bands correspond to $\sigma=\up$ and $\sigma=\down$, respectively.  }
	\label{fig:Lieb}
\end{figure}

{\it Lieb lattice model.} The first model is the Lieb lattice altermagnet shown in Fig.~\ref{fig:Lieb}(a), which is currently one of the primary minimal models for the study of altermagnetism~\cite{Brekke:2023p224421,Antonenko:2025p096703,Yershov:2024p144421,Kaushal:2025p156502,Durrnagel:2025p036502,Fu:2025arXiv07}. The Lieb lattice has two magnetically active sites in the unit cell (i.e., the $A$ and $B$ sites), shown in blue and red in Fig.~\ref{fig:Lieb}(a). The two colors indicate the anti-alignment of spins in the N\'eel state. The additional nonmagnetic site, shown in white, gives rise to the anisotropic local crystal environment seen by the $A$ and $B$ sites. The Lieb lattice has (symmorphic) space group $P4/mmm$, and with the origin chosen at the nonmagnetic site the magnetic sites sit at Wyckoff positions $(\tfrac12,0)$ and $(0,\tfrac12)$. Importantly, the two magnetic sublattices are not exchanged by the inversion $\mathcal I$, but are exchanged by the fourfold rotation $\mathcal C_{4z}$ and the twofold rotation about the $[11]$ direction. This makes the Lieb lattice an altermagnet with $d_{x^2-y^2}$ symmetry. 

The uniform dispersion $\varepsilon_\bk$ and the intersublattice hopping $t_{x,\bk} $, both defined in Eq.~\eqref{eq:H_0k}, are given by
\be
\varepsilon_\bk = -2t_0 (c_x +c_y), \quad  t_{x,\bk} = -4t_1 c_{x/2}c_{y/2}, \label{eq:eps_k-t_xk}
\ee
where we have defined the abbreviations $c_i \equiv \cos k_i$ and $c_{i/2} \equiv \cos( k_i/2)$, with $i=x,y$. Importantly, since the form of $\varepsilon_\bk$ and $t_{x,\bk} $ is not dependent on the site symmetry group of the $A$ and $B$ sublattices, it is the same for all three lattice models. The form of $t_{z,\bk}$ and $\lambda_{z,\bk}$ does depend on the sublattice site symmetry groups, and in the case of the Lieb lattice these functions are given by 
\be
t_{z,\bk}  = -2t_d (c_x - c_y), \quad   \lambda_{z,\bk }=4\lambda s_{x/2}s_{y/2}. \label{eq:Lieb}
\ee
Here we have used the abbreviation $s_{i/2} \equiv \sin( k_i/2)$. To introduce strain effects in the Lieb lattice model, we follow Ref.~\onlinecite{Takahashi:2025p184408} and choose
\be
\chi_\bk = -2t_0 (c_x - c_y), \quad w_{z,\bk} = -2t_d(c_x + c_y ). \label{eq:Lieb-strain}
\ee
This form of the strain field coupling is mandated by symmetry, since the applied strain must be of a particular symmetry type in order to give rise to a nonzero magnetization. In the case of $d$-wave altermagnets, strain must have the same symmetry as the altermagnet~\cite{McClarty:2024p176702,Bhowal:2024p011019,Takahashi:2025p184408,Khodas:2025arXiv06}. Here, the applied strain must have $d_{x^2-y^2}$ symmetry, which means the nonzero strain tensor component is $\varepsilon_{xx}-\varepsilon_{yy} $.

The energy bands $E^\sigma_{\bk n}$ of the Lieb lattice model, as given by Eq.~\eqref{eq:E_n}, are shown in Fig.~\ref{fig:Lieb}(b) for a set of parameters that give rise to a fully gapped spectrum. The blue ($\up$) and red ($\down$) bands include the effect of strain, whereas the black solid ($\up$) and dashed ($\down$) bands correspond to the unstrained system. Note that we have chosen the unrealistically large strain value of $\phi=0.1$ ($10\%$) to make the effect of strain visible. As is clear, in the presence of strain the energy bands are no longer related by $\mathcal T \mathcal C_{4z}$ symmetry [see Eq.~\eqref{eq:symmetries}] or the mirror symmetry $\mathcal M_{[11]}$ (i.e., the mirror reflection in a plane normal to the $[11]$ direction). In particular, the manifest degeneracy of $\sigma=\up$ and $\sigma=\down$ bands on the $\Gamma M$ line, which is characteristic of $d_{x^2-y^2}$-wave altermagnets, is lifted. 

An essential feature of the Lieb altermagnet model is the nontrivial topology of the energy bands. Previous work pointed out that the two valence bands with energies $E^\up_{\bk 1}$ and $E^\down_{\bk 1}$ have an equal but opposite Chern number, giving rise to mirror Chern bands~\cite{Antonenko:2025p096703}. The nontrivial topology of the Lieb lattice bands can be intuitively understood from the Dirac points that exist on the Brillouin zone (BZ) boundary in the limit of vanishing spin-orbit coupling (i.e., $\lambda = 0$)~\cite{Sun:2009p046811,Weeks:2010p085310,Antonenko:2025p096703}. Spin-orbit coupling acts as a mass term for the two Dirac points in each spin sector and the sign structure of the mass terms is such that the spin sectors have nonzero but opposite Chern numbers.

The location of these Dirac points on the $k_x=\pi$ and $k_y = \pi$ lines is determined from the condition that the $\hat z$-component of $\bn^\sigma_\bk$ in \eqref{eq:n_k} vanishes: $t_{z,\bk}+\sigma N_z =0 $~\cite{Antonenko:2025p096703}. This condition defines a critical value of $N_z$ (denoted $N_{z,c}$) beyond which no solutions exist and the bands cannot be understood from a gapped Dirac point picture. In the Lieb lattice model this critical value is given by
\be
N_{z,c} = 4t_d, \label{eq:Nc-Lieb}
\ee
and marks a topological gap closing transition at which the mirror Chern number $C_M$ changes from $\pm 1$ to zero~\cite{Antonenko:2025p096703}. Below, in Sec.~\ref{sec:M-orb}, we will show how the topology of the Lieb lattice band structure is reflected in the piezomagnetic response. 

\begin{figure}
	\includegraphics[width=\columnwidth]{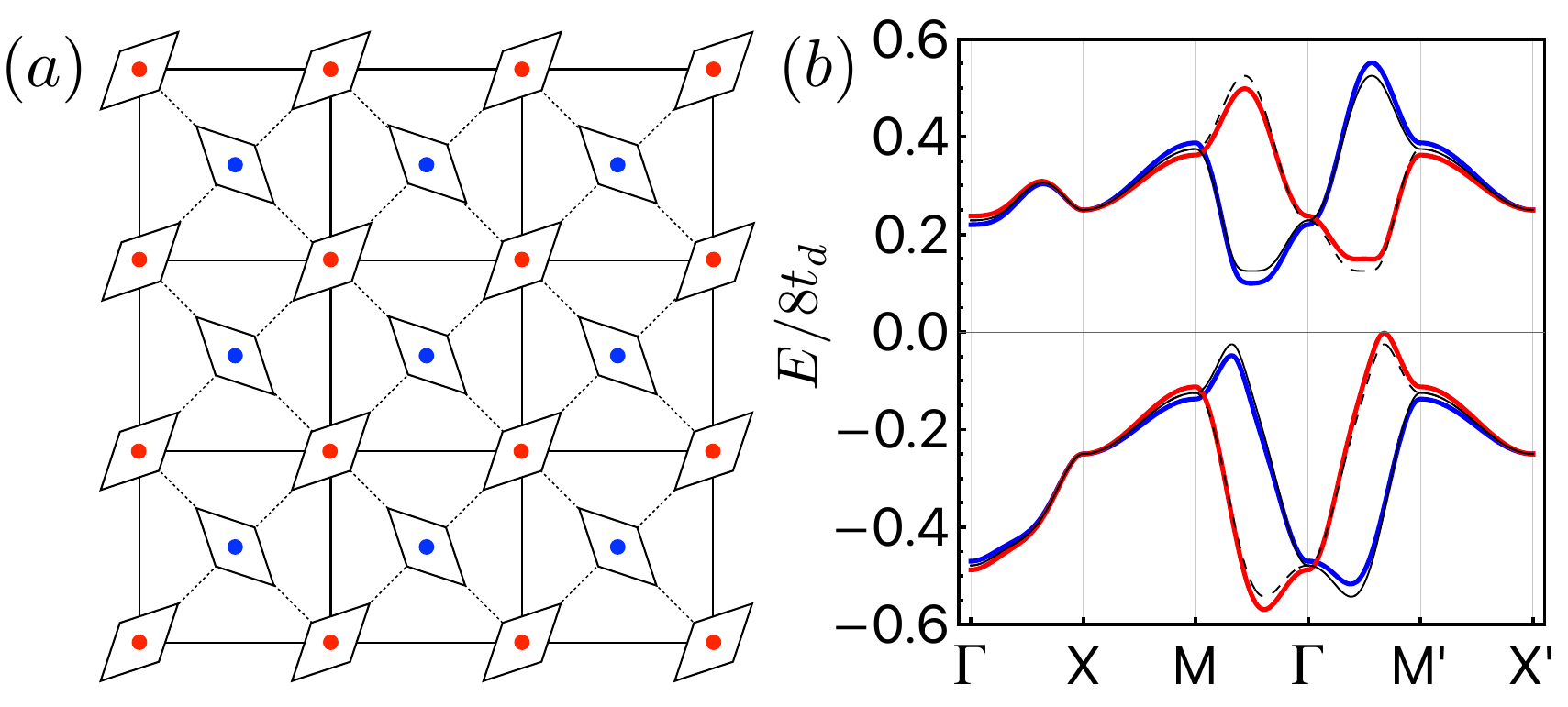}
	\caption{{\bf 2D rutile lattice model.} (a) Schematic of the 2D rutile lattice with layer group $P4/mbm$. The magnetic sites (red and blue) have an anisotropic crystallographic environment. (b) Energy spectrum with and without strain (red/blue and solid/dashed black, respectively) of the model defined by Eq.~\eqref{eq:eps_k-t_xk} and Eqs.~\eqref{eq:rutile}--\eqref{eq:rutile-strain} with parameters $(t_0,t_1,\lambda,N_z) = (0.25t_d,0.5t_d,0.25t_d,2.0t_d)$ and $\phi=0.1$ for the strained case. As in Fig.~\ref{fig:Lieb}(b), blue and red bands correspond to $\sigma=\up$ and $\sigma=\down$, respectively. }
	\label{fig:rutile}
\end{figure}

{\it 2D rutile lattice model.} The second model we consider was introduced in Ref.~\onlinecite{Maier:2023pL100402} and is shown in Fig.~\ref{fig:rutile}(a). As in the Lieb lattice model, the magnetic sites experience different local crystal environments as a result of the arrangement of nonmagnetic sites, which is represented by alternatingly oriented rhombuses in Fig.~\ref{fig:rutile}(a). A crystal lattice of this type can be viewed as the 2D variant of the 3D rutile lattice, the lattice of candidate altermagnets RuO$_2$~\cite{Sun:2017p235104,Berlijn:2017p077201,Ahn:2019p184432,Smejkal:2020peaaz8809,Feng:2022p735,Smejkal:2022p031042,Smejkal:2022p040501} and MnF$_2$~\cite{Yuan:2020p014422,Bhowal:2024p011019}. For this reason, we refer to this model as the ``2D rutile'' lattice model.

The layer group of the 2D rutile lattice is $P4/mbm$. With the origin chosen at one of the magnetic sites, the space group generators are $\{\mathcal C_{4z}|\tfrac12 \tfrac12\}$, $\{\mathcal C_{2x}|\tfrac12 \tfrac12\}$, and $\{\mathcal I| 00 \}$. The fourfold and the twofold rotations exchange the magnetic sublattices and as a result, this model realizes a $d_{xy}$ altermagnet. 

As mentioned above, $\varepsilon_\bk$ and $ t_{x,\bk} $ are the same as in Lieb lattice model and are given by Eq.~\eqref{eq:eps_k-t_xk}. The functions $t_{z,\bk} $ and $ \lambda_{z,\bk }$ reflect the specific symmetries of the crystal lattice, in particular the site symmetry group of the $A$ and $B$ sites, and are therefore different for a $d_{xy}$ altermagnet. In the 2D rutile lattice model these functions are given by 
\be
t_{z,\bk}  = -2t_d s_x s_y, \quad   \lambda_{z,\bk }=4\lambda c_{x/2}c_{y/2}(c_x-c_y), \label{eq:rutile}
\ee
where the same abbreviation convention as above are used. This specifies the microscopic model without strain. To obtain a piezomagnetic response in a $d_{xy}$-wave altermagnet, the applied strain must be of the corresponding symmetry, and therefore we include an $\ve_{xy}$ strain component. The functions $\chi_\bk$ and $w_{z,\bk} $, which parametrize the strain microscopically via Eq.~\eqref{eq:W_k}, have the form
\be
\chi_\bk = -2t_ds_x s_y, \quad w_{z,\bk} = -2t_0(c_x + c_y ).  \label{eq:rutile-strain}
\ee

The energy bands of the 2D rutile lattice model are shown in Fig.~\ref{fig:rutile}(b), both with and without strain, using the same convention as in Fig.~\ref{fig:Lieb}(b). The symmetry breaking effects of strain on the energy bands are clearly visible. In the absence of strain the valence bands of opposite spin are degenerate both on the $\Gamma X$ and on the $XM$ lines (the same is true for the conduction bands). The degeneracy on the $\Gamma X$ line is expected for an altermagnet with $d_{xy}$-wave spin splitting, as is the large spin splitting on the diagonal $\Gamma M$ line. The degeneracy on the BZ boundary $XM$ line is a consequence of non-symmorphic space group symmetries. In fact, in the nonmagnetic $\mathcal T$-symmetric state, the symmetries of the space group $P4/mbm$ require that all energy levels on the BZ boundary are fourfold degenerate (see Appendix \ref{app:P4/mbm}). This leads to the important observation that the 2D rutile lattice model cannot maintain an energy gap, and thus cannot describe an insulator, in the $N_z \rightarrow 0 $ limit. When examining the dependence of the orbital piezomagnetic polarizability on the magnetic order parameter $N_z$ one must account for this, since the computation of $M_z$ tacitly assumes an insulating ground state. 

In sharp contrast to the Lieb lattice model, the two valence bands of the rutile model are topologically trivial and do not have a Chern number. As in the Lieb lattice model it is possible to understand the topology of the bands---in this case the trivial topology of the bands---from a continuum Dirac model. A detailed construction of such a Dirac model is presented in Appendix~\ref{app:2D-rutile}. Here we summarize the features necessary for a discussion of the strain-induced orbital magnetization. In the case of the 2D rutile lattice, the (gapped) Dirac points are located on the body diagonals of the BZ, i.e., the $k_x = \pm k_y$ lines, and their location on the diagonals is again determined from the condition $t_{z,\bk}+\sigma N_z =0 $. As shown in Appendix~\ref{app:2D-rutile}, as long as $0< |N_z| < 2t_d $ this equation has four solutions in each spin sector (i.e., on each body diagonal). These solutions define two pairs of Dirac points with opposite topological charge, which implies that the Chern number in each spin sector vanishes. We again find a critical value of $N_z$ beyond which a Dirac model construction is no longer possible and this value is given by
\be
N_{z,c} = 2t_d. \label{eq:Nc-rutile}
\ee
In the 2D rutile model no gap closing transition occurs at this value of $N_z$. 

Two important differences with the Dirac model of the Lieb lattice altermagnet should be emphasized. First, in the case of the rutile lattice, the Dirac model is only meaningful when spin-orbit coupling is included. Recall that in the case of the Lieb lattice model the Dirac points already exist in the absence of spin-orbit coupling and are gapless, and acquire a gap when spin-orbit coupling is included. Instead, here the spin-orbit coupling provides the linear dispersion in the direction perpendicular to the body diagonal, rather than the Dirac mass (see Appendix~\ref{app:2D-rutile}). The second key difference is the number of (gapped) Dirac points in each spin sector. Whereas the Lieb lattice model has one pair (per spin sector), the rutile lattice model has two such pairs. The members of each pair have the same topological charge, but the overall topological charge vanishes. 

Fleshing out these differences provides a basis for the discussion of the orbital magnetization response to strain in the next section. 

\begin{figure}
	\includegraphics[width=\columnwidth]{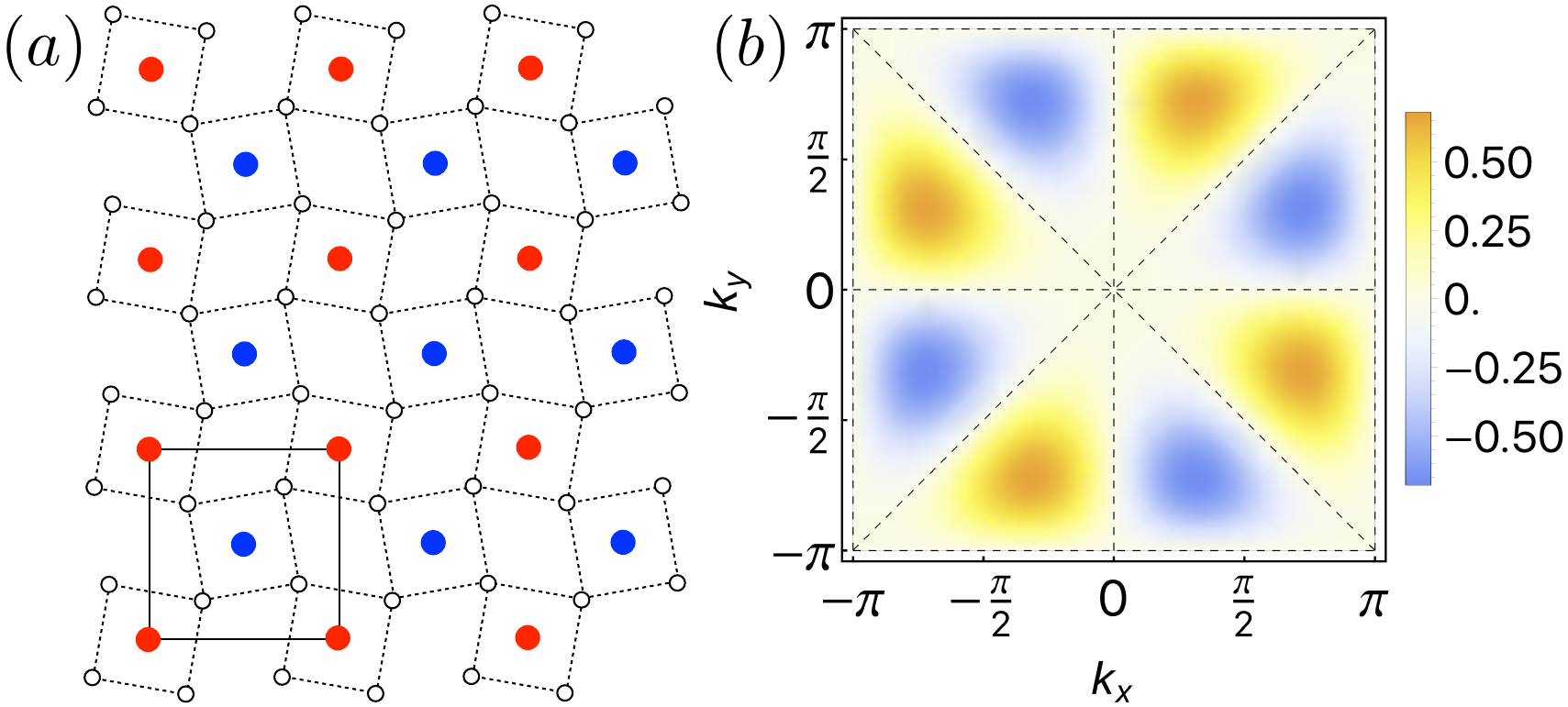}
	\caption{{\bf Octahedral rotation lattice.} (a) A lattice of magnetic sites (shown in blue and red) surrounded by rotated square cages of nonmagnetic sites (shown in white). (b) Momentum-space resolved plot of the energy difference $(E^\up_{\bk1} -E^\down_{\bk1})/8t_g $ between the valence bands of opposite spin in the absence of strain. The $g$-wave structure of the spin splitting is evident. Energies are calculated for the model defined in Eqs.~\eqref{eq:eps_k-t_xk} and~\eqref{eq:octahedral} with parameters $(t_0,t_1,\lambda_2,N_z) = (0.25t_g,0.5t_g,0.5t_g,2.0t_g)$. }
	\label{fig:octahedral}
\end{figure}

{\it Octahedral rotation model.} The third and final model we consider is shown in Fig.~\ref{fig:octahedral}(a). The unit cell has two magnetic sites, again shown in red and blue, as well as nonmagnetic sites, shown in white. The magnetic sites are surrounded by square plaquettes of nonmagnetic sites, which are rotated in opposite directions for the two magnetic sublattices. It is this rotation of the nonmagnetic squares that is responsible for the different crystal environment seen by the magnetic sites. Rotations of surrounding ligand cages are common in magnetic oxides, where they are referred to as octahedral rotations (of the octahedral oxygen cages)~\cite{Glazer:1972p3384}, and the square lattice model considered here can therefore be viewed as a 2D variant of altermagnetism induced by octahedral rotations~\cite{Schmitz:2005p144412,Bousquet:2011p197603,Wang:2022p144404,Fernandes:2024p024404,Rooj:2025p014434}. For this reason, we refer to this third model as the octahedral rotation model.

The layer group of this crystal structure is the same as that of the 2D rutile model, $P4/mbm$. Crucially, the site symmetry group is different, however. Choosing the origin at one of the magnetic sites, the space group generators are $\{\mathcal C_{4z}|00\}$, $\{\mathcal C_{2x}|\tfrac12 \tfrac12\}$, and $\{\mathcal I| 00 \}$. It follows that the fourfold rotation does not exchange the magnetic sublattices, but all twofold rotations with in-plane rotation axes do. The octahedral rotation model therefore realizes a $g_{yx^3-xy^3}$-wave altermagnet. 

The functions $\varepsilon_\bk$ and $ t_{x,\bk} $ are still given by Eq.~\eqref{eq:eps_k-t_xk}, while the form of $t_{z,\bk} $ and $ \lambda_{z,\bk }$ is determined the site symmetry group of the magnetic sites. For the octahedral rotation model we find
\be
\begin{split}
t_{z,\bk} & = -2t_g s_{x}s_{y}(c_x-c_y), \\
   \lambda_{z,\bk }& =4\lambda_1 c_{x/2}c_{y/2}+4\lambda_2 c_{x/2}c_{y/2}(c_x+c_y).
\end{split}\label{eq:octahedral}
\ee
Here $t_g$ describes an intrasublattice fourth-nearest neighbor hopping. Note further that here we have introduced two spin-orbit terms, with parameters $\lambda_1$ and $\lambda_2$, respectively. This is necessary to ensure that $\lambda_{z,\bk }$ and $t_{x,\bk }$ are not simply linearly proportional to each other; in what follows we will set $\lambda_1=0$, unless otherwise specified. 

Introducing the effect of strain in this $g$-wave model requires generalizing Eq.~\eqref{eq:W_k} to include two symmetry-distinct strain components. The full Hamiltonian of the octahedral rotation lattice model therefore takes the form
\be
H_\bk = H_{0,\bk}+ \phi_1 W_{1,\bk}+\phi_2 W_{2,\bk}, \label{eq:W_k-nonlinear}
\ee
where $\phi_1$ and $\phi_2$ are strain fields with $\varepsilon_{xx}-\varepsilon_{yy} $ and $\varepsilon_{xy}  $ symmetry, respectively. The strain terms $W_{1,\bk}$ and $ W_{2,\bk}$ are expanded as in Eq.~\eqref{eq:W_k}, and the nonzero expansion coefficients are chosen as
\be
\begin{split}
\chi_{1,\bk}&= w_{2,z,\bk}= -2t_0(c_x - c_y ), \\
 w_{1,z,\bk} &=\chi_{2,\bk}= -2t_ds_xs_y.
\end{split}  \label{eq:octa-strain}
\ee
Introducing two symmetry-distinct strain components $\phi_1$ and $\phi_2$ is necessary to obtain a magnetization response in a $g$-wave altermagnet. We will show in Sec.~\ref{sec:nonlinear} that the nonlinear piezomagnetic effect is proportional to $\phi_1\phi_2$.

In Fig.~\ref{fig:octahedral}(b) we show a density plot of the energy difference between the two valence bands $E^\up_{\bk1} $ and $ E^\down_{\bk1}  $. As expected for a square lattice $g$-wave altermagnet, the energy difference vanishes both on the principal axes and on the body diagonals. It is furthermore clear from Fig.~\ref{fig:octahedral}(b) that the energy difference respects fourfold rotation symmetry and breaks all vertical mirror symmetries. 

The energy difference also vanishes on the BZ boundary, which, as in the case of the 2D rutile lattice, is a consequence of space group symmetry (as detailed in Appendix \ref{app:P4/mbm}). As for the 2D rutile lattice, these space group symmetries also imply that the energy bands must be fourfold degenerate on the BZ boundary in the nonmagnetic state with $N_z=0$. As a result, the model cannot maintain an insulating gap as $N_z\rightarrow0$ either.

\section{Orbital magnetization and linear piezomagnetic polarizability}
 \label{sec:M-orb}

We now turn to the main objective of this paper: the study of orbital piezomagnetism in altermagnets. As mentioned in the introduction, $d$-wave altermagnets are naturally piezomagnetic (i.e., exhibit a linear piezomagnetic effect) and we therefore focus first on the two $d$-wave models introduced in the previous section. We begin by computing the orbital magnetization in the presence of strain based on the modern theory~\cite{Thonhauser:2005p137205,Xiao:2005p137204,Ceresoli:2006p024408}, and then compute the corresponding linear response coefficient, the linear orbital piezomagnetic polarizability. The piezomagnetic polarizability describes how strong the orbital magnetization response to strain is and is a property of pure altermagnets. 

\subsection{Strain-induced orbital magnetization}
\label{ssec:Mz}

According to the modern theory, the orbital magnetization $M_z$ of a general 2D system is given by~\cite{Thonhauser:2005p137205,Xiao:2005p137204,Ceresoli:2006p024408}
\be
M_z = \frac{e}{\hbar}  \int\frac{d^2\bk}{(2\pi)^2} \sum_{n} \text{Im} \langle \partial_x u_{\bk n} | (H_\bk + E_{\bk n}) | \partial_y u_{\bk n} \rangle. \label{eq:Mz}
\ee
Here $H_\bk$ is the Bloch Hamiltonian and $| u_{\bk n}\rangle$ and $E_{\bk n}$ are the Bloch eigenstates and band energies of $H_\bk$, respectively. The derivatives $\partial_j \equiv \partial / \partial k_j$ are momentum derivatives; $n$ is the band label, and the sum is over all occupied bands. This expression for the orbital magnetization tacitly assumes it is calculated for an insulator with vanishing Chern number~\cite{Ceresoli:2006p024408}, which is applicable to the case at hand. It is useful to rewrite the expression for the orbital magnetization in terms of derivatives of the Hamiltonian $H_\bk$, rather than of eigenstates $| u_{\bk n}\rangle$. The orbital magnetization can be recast as
\begin{multline}
M_z = \frac{e}{2\hbar} \int\frac{d^2\bk}{(2\pi)^2}\sum_{n,m} \frac{E_{\bk n}+E_{\bk m}}{(E_{\bk n}-E_{\bk m})^2} \\
\times \text{Im} \text{Tr}[P_{\bk n}(\partial_x H_\bk) P_{\bk m}(\partial_y H_\bk)],   \label{eq:Mz-2}
\end{multline}
where $P_{\bk n}$ are the projectors onto the eigenspace of energy band $E_{\bk n}$, as given by Eq.~\eqref{eq:P_n}. In Eq.~\eqref{eq:Mz-2} the sum on $n$ is over all occupied bands and the sum on $m$ is over \emph{unoccupied} bands. 

\begin{figure}
	\includegraphics[width=\columnwidth]{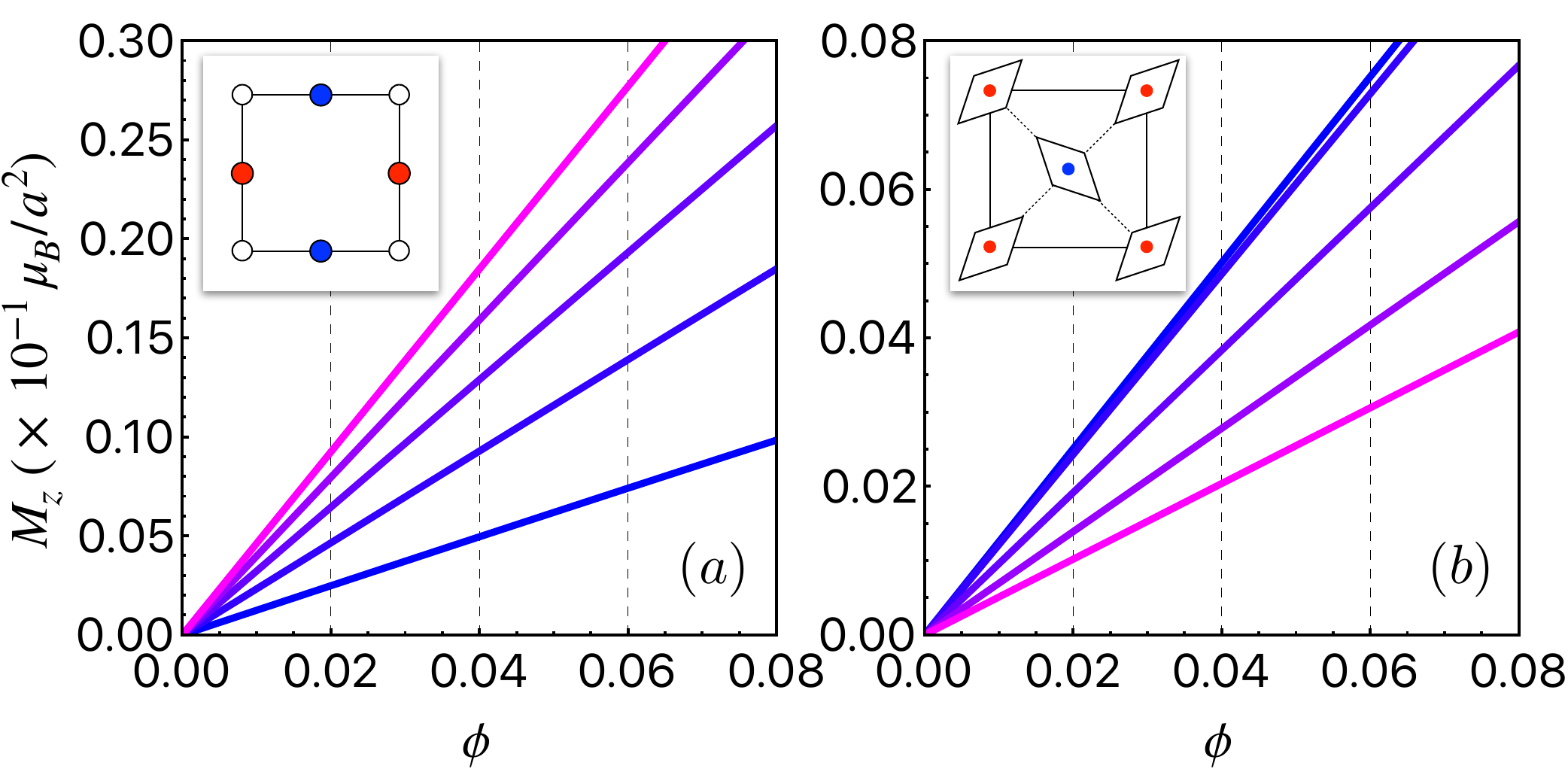}
	\caption{{\bf Orbital magnetization.} (a) Orbital magnetization $M_z$ as a function of the dimensionless strain field $\phi$, calculated for the Lieb lattice model (see inset) using Eq.~\eqref{eq:Mz-altermagnet}. We have used the parameters $(t_0,t_d,\lambda) = (0.5t_1,2t_1,0.5t_1)$ and different curves correspond to $N_z/t_1= 1.0,2.0,3.0,4.0,5.0$ (from blue to magenta). Here $a$ is a length scale on the order of the lattice spacing defined via $\hbar^2/2m_e t_1 \equiv  a^2$. Note further that the value of the $M_z$ has been multiplied by a factor $10$. (b) Same as in (a) but for the 2D rutile lattice model (see inset). For the rutile lattice model we have used the same model parameters $(t_0,t_d,\lambda)$ but different curves correspond to $N_z/t_1= 2.0,\ldots,6.0$ (from blue to magenta). }
	\label{fig:Mz-dwave}
\end{figure}

The expression for the magnetization given by Eq.~\eqref{eq:Mz-2} is valid for a general 2D Hamiltonian. Here we apply it to the 2D altermagnets defined by Eq.~\eqref{eq:H^sigma}. The contributions from the two spin sectors can be considered separately and then added. Each spin sector has an occupied band, given by $E^\sigma_{\bk 1}$, and an unoccupied band, given by $E^\sigma_{\bk 2}$ [see Eq.~\eqref{eq:E_n}], with associated projectors given by Eq.~\eqref{eq:P_n}. Substituting these expressions into Eq.~\eqref{eq:Mz-2} yields the appealing result
\be
M_z = -\frac{e}{\hbar} \sum_\sigma \int\frac{d^2\bk}{(2\pi)^2} \xi_\bk \frac{\bh^\sigma_\bk \cdot \partial_x \bh^\sigma_\bk \times \partial_y \bh^\sigma_\bk}{2|\bh^\sigma_\bk|^3} , \label{eq:Mz-altermagnet}
\ee
which shows that $M_z$ can be expressed directly in terms of the parametrization of the Hamiltonian (i.e., $\xi_\bk$ and $\bh^\sigma_\bk$). We further observe that the integrand is the product of $\xi_\bk$ with the Berry curvature of the occupied valence bands. (See also Ref.~\onlinecite{Mitscherling:2025arXiv12}.)

It is straightforward to deduce from Eq.~\eqref{eq:Mz-altermagnet} that the orbital magnetization must be zero in the absence of strain. Without strain one has $\xi_\bk \rightarrow \varepsilon_\bk$ and $\bh^\sigma_\bk \rightarrow \bn^\sigma_\bk$, as is clear from Eq.~\eqref{eq:xi_k-h_k}. We can then exploit the fact that $\bn^\up_\bk$ and $\bn^\down_\bk$ are related via Eq.~\eqref{eq:symmetries} to show that
\be
\frac{\bn^\up_\bk \cdot \partial_x \bn^\up_\bk \times \partial_y \bn^\up_\bk}{2|\bn^\up_\bk|^3}=- \frac{\bn^\down_{g\bk} \cdot \partial_x \bn^\down_{g\bk} \times \partial_y \bn^\down_{g\bk}}{2|\bn^\down_{g\bk}|^3},
\ee
where we have used that $\text{Det}(R)=-1$. In the case of $d$-wave altermagnets it is natural to take $g$ as the fourfold rotation $\mathcal C_{4z}$. Performing a change of variables in the integral from $\bk$ to $\bk'=g\bk$ shows that the two terms in \eqref{eq:Mz-altermagnet} cancel and that $M_z$ must vanish in the absence of strain.

\begin{figure}
	\includegraphics[width=\columnwidth]{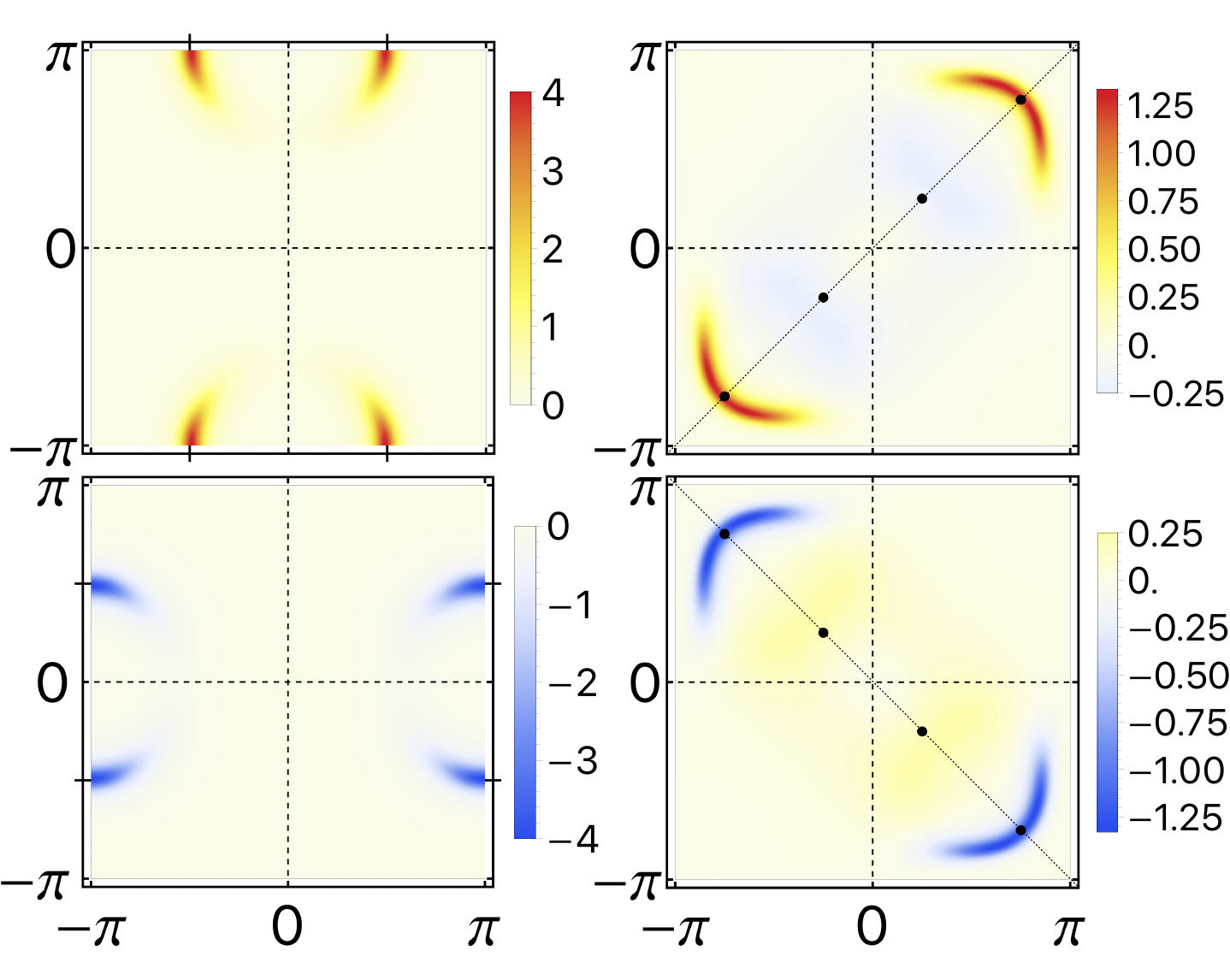}
	\caption{{\bf Berry curvature distribution.} (Left) Plot of the Berry curvature of the Lieb lattice model without strain. The upper and lower panels correspond to the $\sigma=\up$ and $\sigma=\down$ valence bands, respectively. The Berry curvature is concentrated around the gapped Dirac crossings, which exist on the $k_y=\pi$ and $k_x=\pi$ lines (respectively) and are indicated by black markers. Here we have used $( t_d,\lambda,N_z) = (2t_1,0.3t_1,4.0t_1)$. (Right) Plot of the Berry curvature of the (unstrained) 2D rutile lattice model, with upper and lower panels corresponding to $\sigma=\up$ and $\sigma=\down$, respectively. For the rutile model we have used $(t_d,\lambda,N_z) = (2t_1,0.5t_1,2.0t_1)$. The location of the Dirac points on the body diagonal, obtained as described in Appendix~\ref{app:2D-rutile}, is indicated by black dots.}
	\label{fig:berry}
\end{figure}

In Fig.~\ref{fig:Mz-dwave} we show the orbital magnetization $M_z$ as a function of strain, calculated using Eq.~\eqref{eq:Mz-altermagnet}, for the Lieb lattice model in panel (a) and the 2D rutile lattice model in panel (b). Different curves (color coded) correspond to different values of $N_z$, and thus shed light on the dependence on the strength of (alter)magnetic order. As expected, $M_z$ vanishes in the absence of strain. We further find that, for both models and the chosen parameter sets, the dependence of $M_z$ on strain is essentially linear for reasonably realistic values of strain. We have verified that deviations from linearity occur at larger (and thus more unrealistic) values of strain. 

\begin{figure}
	\includegraphics[width=\columnwidth]{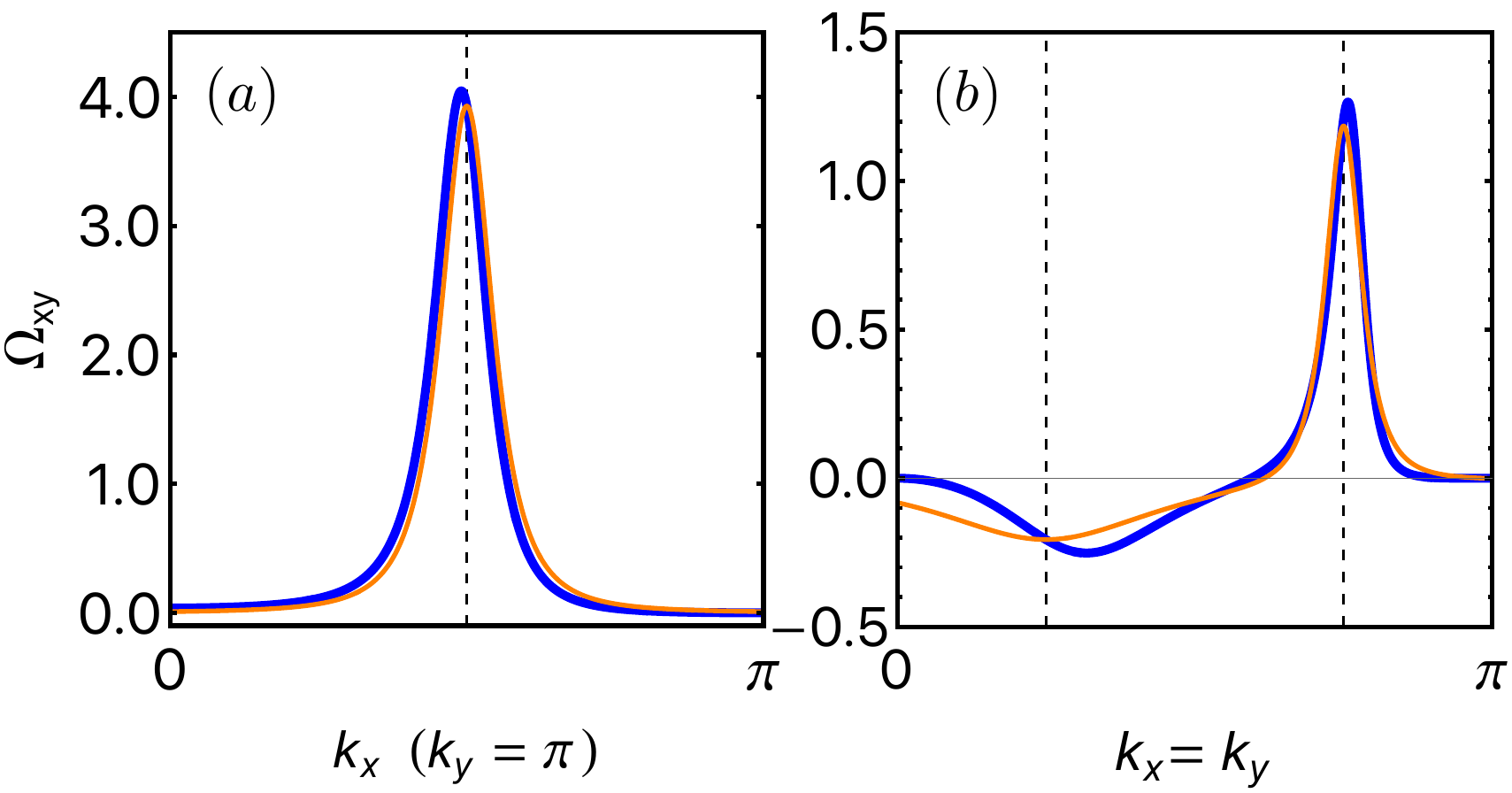}
	\caption{{\bf Berry curvature approximation.} (a) Berry curvature $\Omega_{xy}(\bk)$ of the Lieb lattice model without strain ($\sigma=\up$ valence band) on the $k_y=\pi$ line as a function of $k_x$. The blue curve corresponds to the full lattice model Berry curvature and the orange curve shows the approximation obtained from a linearized Dirac model at $k_{x,D} = 2\arccos(\sqrt{N_z/4t_d})$. Parameters are chosen as in Fig.~\ref{fig:berry}. (b) Same as in (a) but for the 2D rutile model. The orange curve is an approximation based on two Dirac points, at $k_{x,D} = \arcsin(\sqrt{N_z/2t_d})$ and $\pi-k_{x,D}$ (see Appendix~\ref{app:2D-rutile}), with opposite topological charge and different mass.}
	\label{fig:berry-approx}
\end{figure}

Two differences between the two $d$-wave models can be observed. The first is that the slope of the magnetization curves increases with $N_z$ for the Lieb lattice model, whereas it decreases for the rutile lattice beyond $N_z\approx 3.0t_1$. This will be discussed in more detail below, when we compute the linear polarizability (see also Fig.~\ref{fig:linear}) and examine its dependence on $N_z$. The second is that the magnitude of $M_z$ at the same strain is significantly larger for the Lieb lattice model. This can be traced back to the specific band topology of the Lieb lattice model in the unstrained limit. As mentioned above (see also Ref.~\onlinecite{Antonenko:2025p096703}), the energy bands of the Lieb lattice model have a nonzero mirror Chern number in the regime $|N_z| < N_{z,c}$, and this nontrivial topology can be understood from a model of gapped Dirac points on the BZ boundary. To visualize this, we show the Berry curvature distribution of the two spin-projected valence bands (without strain) in the left two panels of Fig.~\ref{fig:berry}, where the top and bottom panels correspond to the $\sigma=\up$ and $\sigma=\down$ bands, respectively. The Berry curvature is clearly seen to be concentrated close to where the Dirac crossings occur (see Sec.~\ref{ssec:models} and Ref.~\cite{Antonenko:2025p096703}). To further demonstrate that the Berry curvature shown in Fig.~\ref{fig:berry} can be understood from a Dirac model, we show the Berry curvature of the $\sigma=\up$ band as a function of $k_x$ on the $k_y=\pi$ line in Fig.~\ref{fig:berry-approx}(a). The blue curve corresponds to the Berry curvature of the full lattice model (as also shown in the upper left panel of Fig.~\ref{fig:berry}) and the orange curve is an approximation of the Berry curvature obtained from a linear expansion around the Dirac point. This comparison shows that the linearized Dirac model provides an excellent approximation of the Berry curvature. 

The Berry curvature distribution of the 2D rutile model (without strain) is shown in the right two panels of Fig.~\ref{fig:berry}. As mentioned in Sec.~\ref{ssec:models}, and further explained in Appendix~\ref{app:2D-rutile}, the band topology of the 2D rutile model can also be understood from a continuum Dirac model, but with a crucial difference. Each spin-projected band can be described by two pairs of Dirac points, rather than one pair, and the two pairs have opposite topological charge. This can be seen in the top right panel of Fig.~\ref{fig:berry}, where the Dirac points on the $k_x=k_y$ diagonal are indicated by black dots. The Berry curvature in the vicinity of the Dirac points is seen to have opposite sign for the two pairs. It also has different magnitude, which is a result of different Dirac masses for the two pairs. To show this more clearly, we plot the Berry curvature on the $k_x=k_y$ line in Fig.~\ref{fig:berry-approx}(b), where, as in panel (a), the blue curve corresponds to the Berry curvature of the full lattice model. The orange curve is an approximation based on the continuum Dirac model developed in Appendix~\ref{app:2D-rutile}, showing good agreement in particular in the vicinity of one of the Dirac points. The agreement is not as strong for the other Dirac point, which is a result of the larger Dirac mass. A larger Dirac mass makes the continuum approximation less adequate, since the continuum approximation assumes that the mass is much smaller than $\sim v(\pi/a)$. 

The difference observed in Fig.~\ref{fig:Mz-dwave}, i.e., the stronger orbital magnetization response of the Lieb lattice model, can be understood from the difference in topology exposed by the structure of the Berry curvature. As Eq.~\eqref{eq:Mz-altermagnet} suggests, the Berry curvature plays a key role in the orbital magnetization response. We now examine this further by calculating the linear piezomagnetic polarizability.

\subsection{Linear piezomagnetic polarizability}
 \label{ssec:Mz-polarizability}

In the case of pure $d$-wave altermagnets the occurrence of an orbital magnetization in the presence of strain is an example of a linear piezomagnetic effect~\cite{Ma:2021p2846,McClarty:2024p176702,Aoyama:2024pL041402,Steward:2023p144418,Fernandes:2024p024404,Naka:2025p083702,Takahashi:2025p184408,Khodas:2025arXiv06}. The linear magnetization response to applied strain is indeed evident from Fig.~\ref{fig:Mz-dwave}. A linear piezomagnetic effect can be expressed in terms of a general response equation as
\be
M_i = \Lambda_{ijk} \varepsilon_{jk}, \label{eq:linear-piezo}
\ee
where $M_i$ is magnetization, $\varepsilon_{jk}$ is the strain tensor, and $\Lambda_{ijk}$ is the piezomagnetic tensor. The piezomagnetic tensor relates the strain tensor to the magnetization and is therefore a quantity that characterizes the unstrained system: it quantifies the strength of the magnetization response once strain is applied, and is therefore governed by the symmetries of the unstrained material. Clearly, time-reversal symmetry must be broken, since $M_i $ and $ \varepsilon_{ij}$ are odd and even under time-reversal, respectively. Since the symmetries of $d$-wave altermagnets naturally allow for nonzero components of the piezomagnetic tensor, they have been recognized as natural candidates for piezomagnetism.

When applied to the two specific models for 2D altermagnets, the general response equation \eqref{eq:linear-piezo} simplifies and can be written more directly as $M_z = \Lambda \phi$. Here $\phi$ is the strain field introduced in Sec.~\ref{sec:problem} and corresponds to the relevant strain tensor component. In both $d$-wave models only one independent component of the piezomagnetic tensor is nonzero and we denote it $\Lambda$. In the Lieb lattice model, which realizes a $d_{x^2-y^2}$ altermagnet, the piezomagnetic polarizability $\Lambda$ corresponds to the piezomagnetic tensor components $\Lambda \simeq \Lambda_{zxx}=-\Lambda_{zyy}$, whereas in the case of the 2D rutile lattice (a $d_{xy}$ altermagnet) it corresponds to the tensor components $\Lambda \simeq \Lambda_{zxy}=\Lambda_{zyx}$. It is, of course, these fundamental symmetry properties that have determined which strain terms were included in these two models, specifically Eqs.~\eqref{eq:Lieb-strain} and \eqref{eq:rutile-strain}.

The general form of the microscopic expression for $\Lambda $ is given by~\cite{Venderbos:arXiv2025-2}
\begin{multline}
\Lambda = 
\frac{e}{\hbar} \int \frac{d^2\bk}{(2\pi)^2}  \bigg( \text{Im}  \text{Tr} [W_\bk(P_\bk-Q_\bk)\partial_x  P_\bk \partial_y  P_\bk    ]   \\
 +\sum_{n,m} \frac{2 \text{Im}  \text{Tr} [W_\bk P_{\bk n} \{ \partial_x  H_{0,\bk},\partial_y  P_\bk\}P_{\bk m}    ] }{E_{\bk n}-E_{\bk m}} \bigg), \label{eq:Lambda-gen}
\end{multline}
where $n$ denotes the occupied bands and $m$ denotes the unoccupied bands. Here it is important to stress that all quantities appearing in the integrand refer to the \emph{unperturbed} (i.e., unstrained) Hamiltonian $ H_{0,\bk}$. In particular, the energies $E_{\bk n}$ and projectors $P_{\bk n}$ in Eq.~\eqref{eq:Lambda-gen} are determined from Eq.~\eqref{eq:H_0^sigma}, such that the projectors are given by $P^\sigma_{\bk1,2} = (\mathbb{1} \mp \bn^\sigma_\bk \cdot \btau /  | \bn^\sigma_\bk|)/2$. Furthermore, $P_\bk  = \sum_n P_{\bk n}$ is the sum over all occupied bands and thus the projector onto the occupied subspace; $Q_\bk = \mathbb{1}- P_\bk$. After substituting the expressions for the energies, projectors, and the perturbation $W_\bk$, the expression for $\Lambda $ reduces to 
\begin{multline}
\Lambda = 
- \frac{e}{\hbar} \sum_\sigma \int \frac{d^2\bk}{(2\pi)^2}\bigg\{ \chi_\bk \frac{\bn^\sigma_\bk \cdot \partial_x \bn^\sigma_\bk \times \partial_y \bn^\sigma_\bk}{2|\bn^\sigma_\bk|^3}  \\
 +\bigg[ \partial_x \varepsilon_\bk   \frac{\bw_\bk \cdot \bn^\sigma_\bk \times \partial_y \bn^\sigma_\bk}{2|\bn^\sigma_\bk|^3} -  (x\leftrightarrow y) \bigg] \bigg\}. \label{eq:Lambda-AM}
\end{multline}
In this way, the orbital piezomagnetic polarizability is directly expressed in terms of the parametrization of the Hamiltonian ($\varepsilon_\bk$ and $ \bn^\sigma_\bk$) and the perturbation ($\chi_\bk$ and $\bw_\bk $). Note that $\Lambda$ is a sum over contributions from the two spin sectors, as was $M_z$.

\begin{figure}
	\includegraphics[width=\columnwidth]{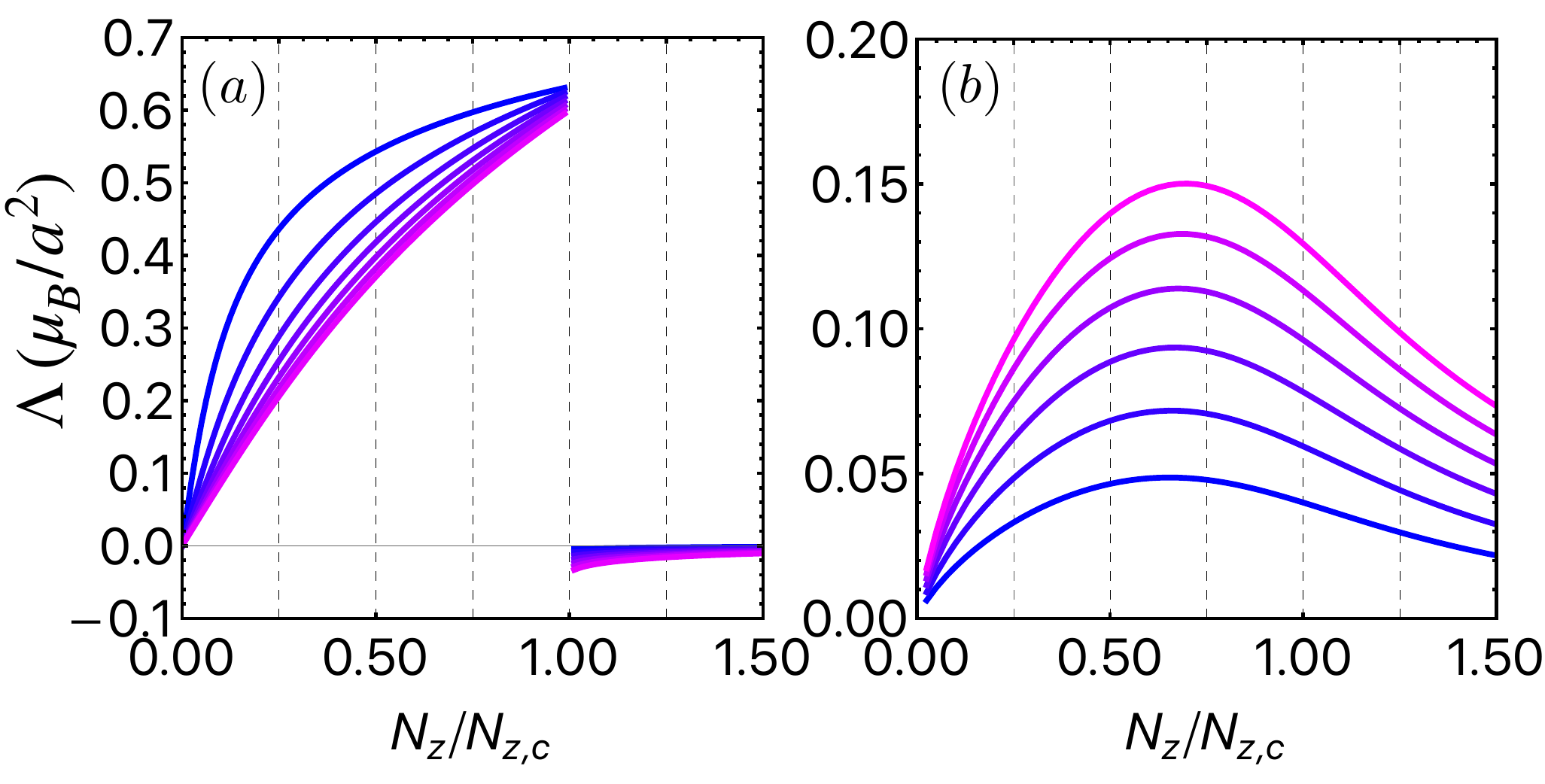}
	\caption{{\bf Linear piezomagnetic polarizability.} (a) Linear orbital piezomagnetic polarizability $\Lambda$ [given by Eq.~\eqref{eq:Lambda-AM}] as a function of $N_z/N_{z,c}$ of the Lieb lattice model, with $N_{z,c}$ given by Eq.~\eqref{eq:Nc-Lieb}. We have used the parameters $(t_0,t_d) = (0.5t_1,2t_1)$ and different curves correspond to $\lambda/t_1= 0.1,\ldots,0.7$ (from blue to magenta). (b) Same as in (a) but for the 2D rutile lattice model, with $N_{z,c}$ given by Eq.~\eqref{eq:Nc-rutile}. For the rutile lattice model we have used the parameters $(t_0,t_d) = (0.2t_1,2t_1)$ and the curves correspond to $\lambda /t_1= 0.2,\ldots,0.7$ (from blue to magenta). }
	\label{fig:linear}
\end{figure}

In Fig.~\ref{fig:linear} we show the linear piezomagnetic polarizability for both the Lieb lattice model (panel a) and the 2D rutile model (panel b). The polarizability, computed using Eq.~\eqref{eq:Lambda-AM}, is shown as a function of $N_z$ in units of $N_{z,c}$. Recall that $N_{z,c}$ is the critical value of $N_z$ at which a continuum Dirac model ceases to be meaningful [see Eqs.~\eqref{eq:Nc-Lieb} and \eqref{eq:Nc-rutile}]. Different curves correspond to different values of spin-orbit coupling strength $\lambda$. 

Consider first the polarizability of the Lieb lattice model, as shown in Fig.~\ref{fig:linear}(a). Three features are worth noting. The first and most striking feature is the large discontinuous jump at $N_z = N_{z,c}$. This discontinuity is a consequence of the topological transition which occurs at $N_{z,c}$ and is caused by a Dirac fermion mass inversion in each spin sector. (The significance of a topological transition for the orbital piezomagnetic polarizability is discussed elsewhere~\cite{Radhakrishnan:2016arXiv02}.) The second feature is the dependence of $\Lambda$ on spin-orbit coupling, as evidenced by the different curves in Fig.~\ref{fig:linear}(a). It can be seen that the polarizability decreases as $\lambda$ is increased, which at first sight is counterintuitive. This behavior can be understood, however, from the Berry curvature distribution discussed above in Sec.~\ref{ssec:Mz} and shown in Fig.~\ref{fig:berry}. Since the spin-orbit coupling strength determines the mass of the Dirac fermions in the continuum model, decreasing $\lambda$ leads to a more concentrated Berry curvature, which in turn gives rise to a larger piezomagnetic polarizability. The latter follows directly from the observation that the first term in Eq.~\eqref{eq:Lambda-AM} is simply the Berry curvature of the unstrained model multiplied by $\chi_\bk$, and highlights the importance of band topology for the piezomagnetic polarizability. The third feature is the magnitude of $\Lambda$, specifically in comparison with the rutile lattice model shown in panel (b). The comparatively large polarizability of the Lieb lattice model can be similarly understood from the Berry curvature distribution associated with the gapped Dirac points. 

Now consider the polarizability of the rutile lattice model shown in Fig.~\ref{fig:linear}(b). This model exhibits the expected increase of $\Lambda$ with increasing spin-orbit coupling strength. The absence of any discontinuities is due to the absence of a topological transition as a function of $N_z$. We furthermore see that the polarizability reaches a maximum at a value of $N_z$ on the order of (but smaller than) $N_{z,c}$, and decreases monotonically beyond $N_{z,c}$. Given that the Berry curvature also decreases as $N_z$ increases, this is expected. In the case of the rutile lattice model it is important to note that the energy band structure becomes metallic as $N_z \rightarrow 0$. One may still calculate $\Lambda$, since the two valence bands remain disconnected from the conduction bands, but the indirect band gap closes as $N_z$ decreases. 

\mbox{}

\section{Nonlinear piezomagnetic effect}
\label{sec:nonlinear}

Our discussion of the orbital piezomagnetic effect has so far exclusively focused on the $d$-wave altermagnets. The reason is that piezomagnetism in $g$-wave altermagnets is qualitatively different from their $d$-wave counterparts, and therefore warrants a separate discussion. This section is devoted to such a discussion.

\begin{figure}
	\includegraphics[width=\columnwidth]{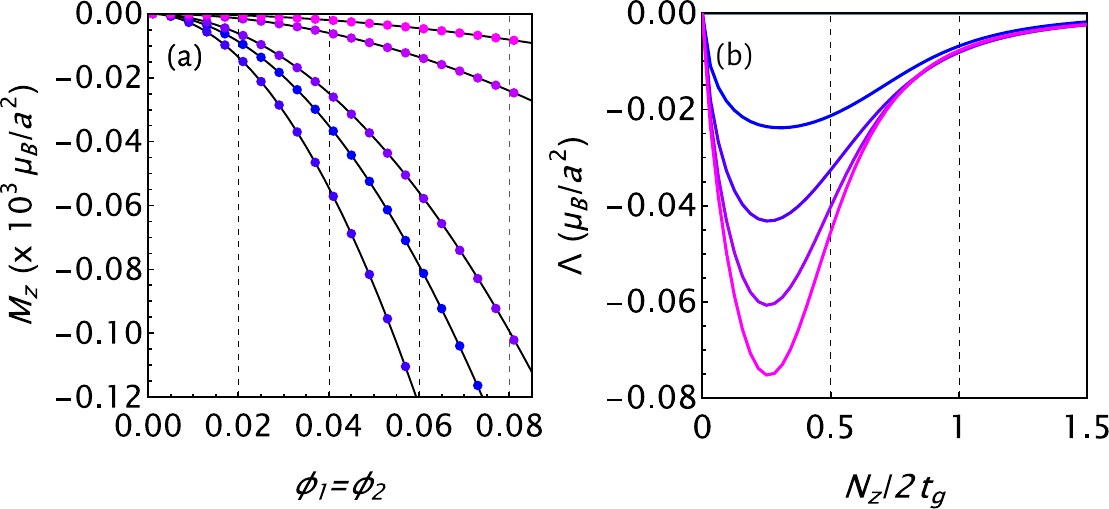}
	\caption{{\bf Non-linear piezomagnetic polarizability.} (a) Orbital magnetization $M_z$ as a function of the dimensionless strain fields $\phi_1=\phi_2$, calculated for the octahedral rotation model using Eq.~\eqref{eq:Mz-altermagnet}. We have used the parameters $(t_0,t_d, t_g, \lambda_2) = (0.5 t_1,t_1,2 t_1,t_1)$ and different curves (dotted lines) correspond to $N_z/t_1= 1.0,2.0,4.0,6.0,8.0$ (from blue to magenta). The solid black lines are computed using the nonlinear response equation $M_z = \Lambda^{(2)} \phi_1 \phi_2$ (see text). (b) Calculation of the nonlinear polarizability $\Lambda^{(2)}$ as a function $N_z$ for different values of $\lambda_2$. We have used the same model parameters as in (a); different curves correspond to $\lambda_2/t_1= 0.5, 1.0 ,1.5, 2.0$ (from blue to magenta).}
	\label{fig:Mz-gwave}
\end{figure}

The distinction between $d$-wave and $g$-wave altermagnets follows from the key observation that a linear piezomagnetic effect is forbidden in $g$-wave altermagnets. The response equation of Eq.~\eqref{eq:linear-piezo} clearly reflects this. It relates an electric quadrupole (the strain tensor $ \varepsilon_{jk}$) to a magnetic dipole (the magnetization $M_i$) and as a result, the piezomagnetic tensor must vanish for a $g$-wave altermagnet~\cite{McClarty:2024p176702,Khodas:2025arXiv06}. While a linear piezomagnetic effect is therefore not allowed, a \emph{nonlinear} piezomagnetic effect is in general allowed for $g$-wave altermagnets. A nonlinear piezomagnetic effect can be expressed via the nonlinear response equation
\be
M_i = \Lambda^{(2)}_{ijklp} \varepsilon_{jk}\varepsilon_{lp},
\ee
where the magnetization depends quadratically on the strain tensor components. Here we have introduced $\Lambda^{(2)}_{ijklp}$ as the second order nonlinear piezomagnetic polarizability tensor. 

When applied to the specific $g$-wave model introduced in Sec.~\ref{ssec:models} (i.e., the octahedral rotation model), the nonlinear response equation simplifies and becomes $M_z = \Lambda^{(2)} \phi_1 \phi_2$, where $\phi_1 $ and $ \phi_2$ are the two symmetry-distinct strain components defined in Eq.~\eqref{eq:W_k-nonlinear}, and $\Lambda^{(2)}$ is the symmetry-allowed component of the nonlinear polarizability tensor. Our goal in this section is to study this nonlinear response equation by computing the orbital magnetization of the octahedral rotation model in the presence of both strain components, and to obtain an expression for the nonlinear response coefficient $\Lambda^{(2)}$.

We note in passing that the nonlinear response equation may also be understood from the perspective of a strain-controlled transition from $g$-wave to $d$-wave altermagnetism~\cite{Karetta:2025p094454}. Indeed, for fixed nonzero $\phi_2$ (or $\phi_1$) the response equation describes a linear response to $\phi_1$ (or $\phi_2$), which reflects a transition from $g$-wave to $d$-wave. 

In Fig.~\ref{fig:Mz-gwave}(a) we show the orbital magnetization as a function of strain for a number of different values of $N_z$, calculated for the $g$-wave octahedral rotation model. Here we have set $\phi_1 = \phi_2$, such that a quadratic dependence on strain is expected based on the nonlinear response equation. This is indeed what is shown in Fig.~\ref{fig:Mz-gwave}(a) and confirms that the octahedral rotation model is an appropriate minimal model for nonlinear orbital piezomagnetism in two dimensions. In Fig.~\ref{fig:Mz-gwave}(a) the colored dotted curves are computed directly using Eq.~\eqref{eq:Mz-altermagnet}, whereas the black solid lines are computed using the response equation $M_z = \Lambda^{(2)} \phi_1 \phi_2$, with $ \Lambda^{(2)}$ determined as described in Appendix~\ref{app:nonlinear}. 

A noteworthy observation is that the orbital magnetization decreases in magnitude as $N_z$ is increased, at least for the chosen values of $N_z$. To understand this further, we compute and show the nonlinear polarizability $ \Lambda^{(2)}$ as a function of $N_z$ in  Fig.~\ref{fig:Mz-gwave}(b). Note that here we measure $N_z$ in terms of $2t_g$, the energy scale associated with the $g$-wave spin splitting of the octahedral rotation model. This energy scale plays a similar role as $4t_d$ and $2t_d$ in the Lieb lattice and 2D rutile models, respectively. We find that $ \Lambda^{(2)}$ reaches a maximum at around $N_z \approx 0.5t_g$ and then monotonically decreases. This explains why the orbital magnetization decreases with $N_z$ in Fig.~\ref{fig:Mz-gwave}(a) for the chosen values of $N_z$. Consistent with the results of Sec.~\ref{ssec:Mz-polarizability}, in particular of Fig.~\ref{fig:linear}, it further emphasizes that the manifestation of band topology, as reflected in the orbital magnetization, is strongest when $N_z$ is smaller than the characteristic altermagnetic spin splitting energy scale.

\section{Discussion and Conclusion \label{sec:discuss}}

The distinctive symmetry properties of altermagnets---in particular of $d$-wave altermagnets---have led to the recognition that altermagnets are natural candidates for piezomagnetism. Piezomagnetism broadly refers to the magnetization response to applied strain, which in general has both spin magnetization and orbital magnetization contributions. The spin magnetization response can occur in the non-relativistic limit, when spin-orbit coupling is ignored, which may be understood on an phenomenological level from Landau theory~\cite{McClarty:2024p176702}, or more microscopically from the non-relativistic spin splitting allowed in altermagnets~\cite{Smejkal:2022p031042,Khodas:2025arXiv06}. The microscopic viewpoint also suggests an important distinction between metallic and insulating altermagnets (in the non-relativistic limit). Whereas in metallic altermagnets strain has the effect of imbalancing the population of energy bands of opposite spin, hence producing net spin magnetization, such an imbalance does not occur in insulators, since the energy bands of opposite spin both remain fully occupied. 

When spin-orbit coupling is included, the magnetization response may also have an orbital contribution and in general is a sum of spin and orbital contributions for both metallic and insulating altermagnets. Since the orbital contribution has received much less attention~\cite{Sorn:2025p245115,Ye:2026p014413}, in this paper we developed a theory for the orbital magnetization response in two dimensional altermagnets. We have specifically focused on a family of altermagnetic models for which the spin magnetization response vanishes, such that the response originates only from the orbital motion of electrons. Two key assumptions underlie this family of models. First, an easy-axis magnetocrystalline anisotropy is assumed, such that the N\'eel vector is perpendicular to the plane of the 2D crystal lattice. This preserves two important symmetries, $\mathcal T g$, where $g$ is a crystal symmetry which exchanges the magnetic sublattices, and $\mathcal M_z$, a horizontal mirror symmetry. The former forbids a net magnetization and thus protects the ``pure'' altermagnet, whereas the latter ensures that energy bands of opposite spin cannot mix. The second assumption is that energy spectrum has a gap at half filling and that the system is thus insulating. Under these assumptions strain cannot cause an imbalance in the population of bands of opposite spin. The family of models considered in this paper is therefore ideally suited to study the orbital magnetization response. 

Three specific models with tetragonal symmetry have been examined in detail, all defined on a square lattice. These models are realizations of altermagnets with $d_{x^2-y^2}$-wave, $d_{xy}$-wave, and $g_{yx^3-xy^3}$-wave symmetry, and therefore represent all possible tetragonal altermagnets in 2D. Each model is a generic minimal model for an altermagnet of the given symmetry type, in the sense that all couplings allowed by the constraining symmetries are included, in particular also the effect of spin-orbit coupling. Generic models do not have any non-essential features or properties which would disappear by simply including additional couplings. The three models considered in this paper therefore provide a complete set of microscopic minimal models for the description and study of (tetragonal) altermagnetism in 2D. In the particular case of (orbital) piezomagnetism, this generality is exemplified by the ability to describe both linear and nonlinear piezomagnetic responses. More specifically, in this paper we have microscopically demonstrated that the $d$-wave altermagnet models exhibit a linear piezomagnetic response, while the $g$-wave model is shown to have a nonlinear piezomagnetic response, as expected from symmetry arguments. 

The Lieb lattice model has been introduced in previous work and has been used to investigate a variety of different aspects of altermagnetism~\cite{Brekke:2023p224421,Antonenko:2025p096703,Yershov:2024p144421,Kaushal:2025p156502}. Here we have focused on orbital piezomagnetism, which is of particular interest in the context of the Lieb lattice model given its topological properties. The model we refer to as the 2D rutile lattice was considered previously as a heuristic toy model for altermagnetism, but only in the limit of vanishing spin-orbit coupling~\cite{Maier:2023pL100402}. Here we have reconsidered this model as a 2D variant of the 3D rutile lattice structure, which makes more explicit reference to the underlying crystal lattice and further establishes a direct connection with experimentally accessible materials in the rutile class. In fact, the 2D model examined here is equal to the 3D rutile model (considered in its generic form in Refs.~\onlinecite{Antonenko:2025p096703} and \onlinecite{Roig:2024p144412}) when restricted to the $k_z=0$ plane. The octahedral rotation model for the $g$-wave altermagnet has, to the best of our knowledge, not been studied previously. Its construction is motivated by the recognition that a number of magnetic perovskite oxides realize altermagnetism when the oxygen octahedra are rotated. The octahedral rotations give rise to a multiplicity of magnetic sublattices, and therefore to a crystal structure which admits altermagetic order. The octahedral rotation model considered here is a 2D variant of this notion of ligand rotations. In this 2D version the ``octahedral'' rotations produce a $g$-wave altermagnet.

All three models can therefore be viewed as individual layers, or more generally 2D versions, of 3D bulk materials currently considered altermagnets or candidate altermagnets. In the case of the Lieb lattice, the essential structural Lieb lattice motif has been identified in a number of different systems~\cite{Lin:2018p075132,Ma:2021p2846,Li:2024p222404,Wei:2025p024402,Jiang:2025p754,Zhang:2025p760,Chang:2025arXiv08}, and currently counts as one of the primary venues for altermagnetism. The insight gained by studying orbital piezomagnetism in the 2D models is therefore expected to form an important basis for future studies addressing 3D bulk systems.

A key result of this paper is the microscopic expression for the orbital piezomagnetic polarizability of the class of models considered. In the case of the linear piezomagnetic effect, the polarizability expression is an application of a more general formula for the orbital magnetization response to an arbitrary perturbation obtained by one of the authors~\cite{Venderbos:arXiv2025-2}. The expression for the nonlinear polarizability has been obtained by expanding the orbital magnetization, given by Eq.~\eqref{eq:Mz-altermagnet}, to second order in the strain fields $\phi_1$ and $\phi_2$. It is therefore specific to the class of models considered here. Generalizations of microscopic nonlinear polarizability formulas will be an interesting direction for future work.

The expressions for the orbital magnetization and the orbital piezomagnetic polarizability by Eqs.~\eqref{eq:Mz-altermagnet} and \eqref{eq:Lambda-AM}, respectively, suggest a direct link between the orbital magnetization response and the electronic band topology, specifically the Berry curvature distribution of the occupied valence bands. The analysis of the three lattice models confirms this. In the case of the Lieb lattice model, the orbital magnetization response can be understood from its known topological properties, in particular the mirror Chern band structure. The latter can in turn be understood from a continuum Dirac model, which explains how the piezomagnetic polarizability is to the the Berry curvature distribution associated with gapped Dirac points. In the case of the 2D rutile lattice model we have exposed and examined a similar Dirac structure, albeit with the crucial difference that each spin-projected valence band can be described by two pair of Dirac points with opposite topological charge. The analysis of the band topology of the 2D rutile lattice, in particular the construction of the continuum Dirac model, can be seen as further evidence that there is a deeper connection between altermagnetism and topology. In a broader sense, a comparison of the obtained piezomagnetic polarizabilities reveals that the manifestation of topology in responses is tied to an important relative energy scale: the strength of magnetic order (here quantified by $N_z$) relative to the energy scale associated with anisotropic hopping (here represented by $t_d$ or $t_g$). The latter encodes the anisotropic crystal environment seen by the magnetic sublattices and is therefore the origin of the altermagnetic spin splitting. Our analysis shows that two regimes can be distinguished, depending on which of these two energy scales is larger. The observable consequences of topology are most pronounced when $N_z$ is smaller than the spin splitting.

\mbox{}

\section*{Acknowledgements}

We gratefully acknowledge insightful conversations with Rafael Fernandes and Daniel Agterberg. We are particularly indebted to Harini Radhakrishnan and Carmine Ortix for collaboration on a related project. Both B.B. and J.W.F.V. were supported by the U.S. Department of Energy under Award No. DE-SC0025632.

\appendix

\section{The space group $P4/mbm$ and band degeneracies}
\label{app:P4/mbm}

In this Appendix we demonstrate that the algebraic relations between the symmetry elements of the space group $P4/mbm$ mandate either fourfold (nonmagnetic case) or twofold (magnetic case) degeneracies on the BZ boundary. In establishing this, we follow a line of argument first laid out in Ref.~\onlinecite{Sun:2017p235104} for the space group $P4_2/mnm$ (i.e., the space group of the 3D rutile lattice) and subsequently reformulated in Ref.~\onlinecite{Antonenko:2025p096703}.
 
Consider first the case without magnetism, i.e., when time-reversal ($\mathcal T$) is present. We examine the algebraic relations between the mirror reflection $ \{ \mathcal M_{z} | 00 \} $ and space group symmetry $\{ \mathcal X | \tfrac12 \tfrac12  \}$, where $\mathcal X$ is either $\mathcal M_x$ or $\mathcal C_{2y}$. All these symmetries map the $k_x=\pi$ line of the BZ boundary to itself. Here we will focus only on the $k_x=\pi$ line; the argument is similar for the $k_y=\pi$ line. One may furthermore establish the relation
\begin{align}
\{ \mathcal  X | \tfrac12 \tfrac12   \}  \{\mathcal  M_{z} | 00  \} &= \{ \mathcal X \mathcal M_z | \tfrac12 \tfrac12  \} =\mathcal R  \{ \mathcal M_z \mathcal X | \tfrac12 \tfrac12 \} ,\\
&= \mathcal R  \{ \mathbb{1} | 00 \}  \{ \mathcal M_{z} | 00 \}  \{ \mathcal X | \tfrac12 \tfrac12  \} ,\label{X-M_z}
\end{align}
where $\mathcal R$ denotes a rotation by $2\pi$, which is equal to $-1$ ($+1$) for spinful (spinless) electrons. This relation  implies that representations of these symmetries in the space of Bloch states anti-commute at each $\bk$. We denote these representations as $U_{\mathcal  M_z}$ and $U_{\mathcal  X}$. The eigenvalues of $U_{\mathcal  M_z}$ must be $\pm i$, since $\{ \mathcal M_{z} | 00 \}^2 = \mathcal R  \{ \mathbb{1} | 00 \} $. Now let $\ket{\psi}$ be an eigenstate of the Hamiltonian, but also of $U_{\mathcal  M_z}$ with eigenvalue $\pm i$. Then it follows that $U_{\mathcal  X}\ket{\psi}$ is also an eigenstate of $U_{\mathcal  M_z}$ with eigenvalue $\mp i$, and is thus orthogonal to $\ket{\psi}$. This establishes a manifest twofold degeneracy of all energy levels on the $k_x=\pi$ line. 

Next, consider the product of time-reversal and inversion, which is given by $ \mathcal T\{ \mathcal I | 00 \}$. This product symmetry leaves the $k_x=\pi$ line invariant and furthermore commutes with the mirror reflection $\{ M_{z} | 00 \}$. Therefore, since $\mathcal T$ is an anti-unitary operation, the state $\mathcal T U_{\mathcal I} \ket{\psi}$ is an eigenstate of $U_{\mathcal  M_z}$ with eigenvalue $\mp i$. It then follows that the state  $\mathcal T U_{\mathcal I} U_{\mathcal X} \ket{\psi}$ is an eigenstate of $U_{\mathcal  M_z}$ with eigenvalue $\pm i$. Hence, $\ket{\psi}$ and $\mathcal T U_{\mathcal I} U_{\mathcal X} \ket{\psi}$ are both eigenstates of $U_{\mathcal  M_z}$ with eigenvalue $\pm i$. To prove that these states must be orthogonal, it is sufficient to show that $ \mathcal T\{ \mathcal I | 00 \}\{ \mathcal X | \tfrac12 \tfrac12  \} $ is a anti-unitary symmetry which squares to $-1$ on the $k_x=\pi$ line. Let us take $\mathcal X=\mathcal  M_x$ as an example and evaluate the square of $ \mathcal T\{ \mathcal I | 00 \}\{ \mathcal M_x | \tfrac12 \tfrac12  \} $:
\begin{align}
(\mathcal T\{ I | 00 \}\{ \mathcal M_x | \tfrac12 \tfrac12  \} )^2&=(\mathcal T\{ \mathcal C_{2x} | \tfrac{\bar 1}{2} \tfrac{\bar 1}{2}  \} )^2 ,\\
& =\mathcal T^2 \mathcal R \{ \mathbb{1} | \bar 10 \} .\label{T.I.M_x_squared}
\end{align}
From this it follows that (the representation of) $ \mathcal T\{ \mathcal I | 00 \}\{ \mathcal M_x | \tfrac12 \tfrac12  \} $ indeed squares to $-1$ on the $k_x=\pi$ line. For spin-$1/2$ fermions one has $\mathcal T^2=-1$ and $ \mathcal R=-1$, and the representation of the translation $ \{ \mathbb{1} | \bar 10 \}$ in the space of Bloch states is $e^{i k_x}$, which equals $-1$ when $k_x=\pi$. 

We have now shown that, in the absence of magnetic order, all energy levels must be fourfold degenerate on the BZ boundary, since we have constructed four states which have the same energy eigenvalue and must be orthogonal. 

When the system becomes magnetic time-reversal symmetry is broken (while inversion is preserved) and the argument therefore no longer holds. The fourfold degeneracy is generally lifted. Twofold degeneracies of all energy levels may still persist, depending on the spatial symmetries that remain preserved. For instance, in the case of the 2D rutile model and the octahedral rotation model, when the N\'eel vector points along the $\hat z$ axis, as is assumed in this work, the $ \{ \mathcal M_{z} | 00 \} $, $\mathcal M_x$, and $\mathcal C_{2y}$ are preserved. The twofold degeneracy then follows from Eq.~\eqref{X-M_z}.

\section{Continuum model expansion of the 2D rutile model}
 \label{app:2D-rutile}

In this Appendix we collect further details on the continuum model expansion of the 2D rutile model. As explained in the main text, the Berry curvature of the two spin-projected valence bands can be qualitatively understood in terms of a continuum Dirac model. Here we present a derivation of this conclusion.

The starting point for the continuum model expansion is the $\hat z$-component of $\bn^\sigma_\bk$ given by $n^{z,\sigma}_{\bk}  = -2t_d \sin k_x \sin k_y + \sigma N_z $, from which we determine the location of the Dirac nodal points by setting $n^{z,\sigma}_{\bk} =0 $ on the lines $k_x=k_y$ and $k_x=-k_y$. 

\begin{figure}
	\includegraphics[width=0.5\columnwidth]{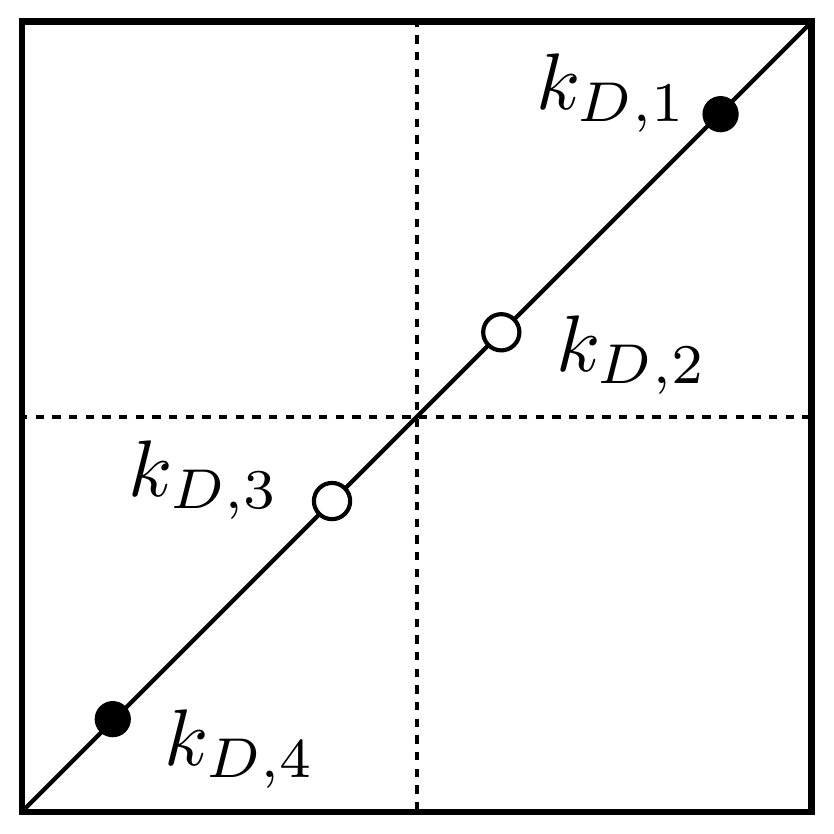}
	\caption{Sketch of the four Dirac points which are solutions of \eqref{app:k_D}. The Dirac point momenta are labeled $k_{D,j}$ ($j=1,2,3,4$), and are related as $k_{D,4}=-k_{D,1}$ and $k_{D,3}=-k_{D,2}$.}
	\label{fig:dirac_rutile}
\end{figure}

Let us consider the former and define $k_x=k_y\equiv k$.  We then obtain 
\be
-2t_d \sin^2 k + \sigma N_z = 0, \quad \Rightarrow \quad  \sin^2 k = \frac{\sigma N_z }{2t_d}, \label{app:k_D}
\ee
from which it follows that a solution exists whenever $\sigma N_z  > 0$. Hence, when $N_z > 0$ solutions only exist in the $\sigma=\up$ sector. Denoting the momenta which solve \eqref{app:k_D} as $k_D$, we then find that $\sin k_D = \pm \sqrt{\sigma N_z/ 2t_d}$. From this, in turn, it follows that there are four solutions on the $k_x=k_y$ line, as long as $0< |N_z|/2t_d < 1$, which we denote $k_{D,j}$ ($j=1,2,3,4$). These four solutions give rise to four Dirac points, shown schematically in Fig.~\ref{fig:dirac_rutile}, which form two pairs. These pairs are $(k_{D,1},k_{D,4})$ and $(k_{D,2},k_{D,3})$, with Dirac point momenta related as $k_{D,4}=-k_{D,1}$ and $k_{D,3}=-k_{D,2}$. 

The equation $\sin k_D = \pm (\sigma N_z/ 2t_d)^{1/2}$ also yields the critical value of $N_z$, which has been defined in main text as $N_{z,c}=2t_d$, since no solutions are obtained for $N_z > N_{z,c}$. 

The next step is to expand the lattice Hamiltonian around each of the four Dirac points. Expanding the $\hat z$-component of $\bn^\sigma_\bk$ amounts to an expansion of the form factor $\sin k_x \sin k_y$, for which we find to linear order in small momentum $\bq = (q_x,q_y)$
\begin{align}
\sin k_x \sin k_y &= \sin(k_D + q_x) \sin (k_D + q_y) \\
& \approx \sin^2 k_D + \sin k_D\cos k_D (q_x + q_y). 
\end{align}
Note that the sign of $\sin k_D\cos k_D$ depends on the Dirac point in question. Given the expansion of the form factor, we find for the $\hat z$-component of $\bn^\sigma_\bk$:
\be
n^{z,\sigma}_{\bq} \approx  -2t_d \sin k_D\cos k_D (q_x + q_y). \label{app:n^z-expand}
\ee
In case of the $\hat x$-component, we find that it does not vanish when evaluated at the Dirac points. We therefore only need to expand to zeroth order in $\bq$. Specifically, we obtain
\be
n^{x,\sigma}_{\bq} \approx  -4 t_1\cos^{2}(k_D/2) = -2t_1 (1+ \cos k_D). \label{app:n^z-expand}
\ee
Since $\cos k_D$ has a different sign for the two pairs of Dirac points, their Dirac mass is of different magnitude [see Eq.~\eqref{app:Dirac-m} below]. Finally, we obtain for the $\hat y$-component: 
\be
n^{y,\sigma}_{\bq} \approx  2\sigma\lambda (1+ \cos k_D)\sin k_D (-q_x + q_y). \label{app:n^y-expand}
\ee 
It follows from this expression that the Dirac velocity perpendicular to the body diagonal $k_x=k_y$ is different in magnitude for the two pairs of Dirac points [see Eq.~\eqref{app:Dirac-v2} below].

Putting these results together we obtain the following for the Dirac points on the $k_x=k_y$ line. For the Dirac point pair $(k_{D,1},k_{D,4}) = (k_{D,1},-k_{D,1})$ we find the Hamiltonian
\be
\mathcal H^\pm _{\bq} =\pm v_1 (q_x + q_y) \tau^z \pm  v_2 (q_y - q_x)\tau^y - m\tau^x , \label{app:H-Dirac-1}
\ee
and for the Dirac point pair $(k_{D,2},k_{D,3}) = (k_{D,2},-k_{D,2})$ we find 
\be
\tilde{\mathcal H}^\pm _{\bq} =\mp  v_1 (q_x + q_y) \tau^z \pm  \tilde v_2 (q_y - q_x)\tau^y - \tilde m \tau^x. \label{app:H-Dirac-2}
\ee
Here $v_1$, $v_2$, and $\tilde v_2$ are Dirac velocities defined as
\be
v_1 = 2t_d\sqrt{\frac{\sigma N_z }{2t_d} \left(1-\frac{\sigma N_z }{2t_d}\right) }, \label{app:Dirac-v1}
\ee
and 
\be
v_2 , \tilde v_2= 2\sigma \lambda \left(1\mp \sqrt{1-\frac{\sigma N_z }{2t_d} }\right)\sqrt{\frac{\sigma N_z }{2t_d} },  \label{app:Dirac-v2}
\ee
whereas $m$ and $\tilde m$ are Dirac masses given by 
\be
m , \tilde m= 2t_1 \left(1\mp \sqrt{1-\frac{\sigma N_z }{2t_d} }\right). \label{app:Dirac-m}
\ee
Note that throughout this derivation we have set $\hbar=1$ and also set the lattice constant to unity ($a=1$). 

Given the Dirac Hamiltonians of Eqs.~\eqref{app:H-Dirac-1} and \eqref{app:H-Dirac-2}, it is straightforward to compute the Berry curvatures of the valence bands for each Dirac point. We find
\be
\Omega^\pm_{xy} = \frac{mv_1 v_2}{\sqrt{m^2 + v_1^2 (q_x+q_y)^2 + v_2^2 (q_x-q_y)^2}} \label{app:Dirac-Berry-1},
\ee
and 
\be
\tilde \Omega^\pm_{xy} = - \frac{\tilde mv_1 \tilde v_2}{\sqrt{\tilde m^2 + v_1^2 (q_x+q_y)^2 + \tilde v_2^2 (q_x-q_y)^2}},  \label{app:Dirac-Berry-2}
\ee
respectively, and it is these expressions that have been used to construct the approximate Berry curvature shown in Fig.~\ref{fig:berry-approx} in orange.

\section{Non-linear orbital piezomagnetic polarizability}
 \label{app:nonlinear}
 
The nonlinear piezomagnetic polarizability defined by the response equation $M_z = \Lambda^{(2)} \phi_1 \phi_2$ can be obtained from Eq.~\eqref{eq:Mz-altermagnet} of the main text as follows. One simply makes the substitutions
\be
\xi_\bk = \ve_\bk + \phi_1 \chi_{1,\bk}+ \phi_2 \chi_{2,\bk},
\ee
 and
 \be
 \bh^\sigma = \bn^\sigma_\bk +  \phi_1 \bw_{1,\bk}+ \phi_2 \bw_{2,\bk},
 \ee
 and then expands the integrand in $\phi_1$ and $\phi_2$, keeping only terms of order $\phi_1 \phi_2$. This is essentially trivial for the numerator and is also straightforward in case of the denominator. Collecting all terms yields
\begin{widetext}
 \begin{multline}
   \Lambda^{(2)} = -\frac{e}{\hbar} \sum_\sigma \int\frac{d^2\bk}{(2\pi)^2} \Bigg\{
          \frac{\varepsilon}{|\mathbf{n}^{\sigma}|^3}\big[ \mathbf{n}^{\sigma}\cdot(\partial_x\mathbf{w}_{1} \times \partial_y \mathbf{w}_{2})+\mathbf{w}_{1}\cdot(\partial_x\mathbf{w}_{2}\times \partial_y \mathbf{n}^{\sigma} )+\mathbf{w}_{1}\cdot(\partial_x\mathbf{n}^{\sigma}\times \partial_y \mathbf{w}_{2}) +(1\leftrightarrow 2)  \big]  \\
     + 3\frac{ \mathbf{n}^{\sigma} \cdot (\partial_x\mathbf{n}^{ \sigma} \times \partial_y \mathbf{n}^{\sigma}) }{|\mathbf{n}^{\sigma}|^5}\Big[  5 \ve(\hat{\mathbf{n}}^{\sigma}\cdot \mathbf{w}_{1})(\hat{\mathbf{n}}^{\sigma}\cdot \mathbf{w}_{2}) - \ve\, \mathbf{w}_{1}\cdot\mathbf{w}_{2} -\chi_{1} \, \mathbf{n}^{\sigma}\cdot \mathbf{w}_{2}-\chi_{2}\, \mathbf{n}^{\sigma}\cdot \mathbf{w}_{1} \Big] \\
     -\varepsilon\frac{3 }{|\mathbf{n}^{\sigma}|^5}\ \Big[( \mathbf{n}^{\sigma}\cdot \mathbf{w}_{1})[\mathbf{n}^{\sigma}\cdot(\partial_x\mathbf{n}^{\sigma} \times \partial_y \mathbf{w}_{2})+\mathbf{n}^{\sigma}\cdot(\partial_x\mathbf{w}_{2}\times \partial_y \mathbf{n}^{\sigma} )+\mathbf{w}_{2}\cdot(\partial_x\mathbf{n}^{\sigma}\times \partial_y \mathbf{n}^{\sigma} )] +(1\leftrightarrow 2) \Big]\\ 
     + \frac{1}{|\mathbf{n}^{\sigma}|^3}\Big(\chi_{1} \ [\mathbf{n}^{\sigma}\cdot(\partial_x\mathbf{n}^{\sigma} \times \partial_y \mathbf{w}_{2})+\mathbf{n}^{\sigma}\cdot(\partial_x\mathbf{w}_{2}\times \partial_y \mathbf{n}^{\sigma} )+\mathbf{w}_{2}\cdot(\partial_x\mathbf{n}^{\sigma}\times \partial_y \mathbf{n}^{\sigma}) ] +(1\leftrightarrow 2) \Big)\Bigg\}
 \end{multline}
\end{widetext}
which has been used to calculate $\Lambda^{(2)}$ in Fig.~\ref{fig:Mz-gwave}.


\begin{thebibliography}{81}%
\makeatletter
\providecommand \@ifxundefined [1]{%
 \@ifx{#1\undefined}
}%
\providecommand \@ifnum [1]{%
 \ifnum #1\expandafter \@firstoftwo
 \else \expandafter \@secondoftwo
 \fi
}%
\providecommand \@ifx [1]{%
 \ifx #1\expandafter \@firstoftwo
 \else \expandafter \@secondoftwo
 \fi
}%
\providecommand \natexlab [1]{#1}%
\providecommand \enquote  [1]{``#1''}%
\providecommand \bibnamefont  [1]{#1}%
\providecommand \bibfnamefont [1]{#1}%
\providecommand \citenamefont [1]{#1}%
\providecommand \href@noop [0]{\@secondoftwo}%
\providecommand \href [0]{\begingroup \@sanitize@url \@href}%
\providecommand \@href[1]{\@@startlink{#1}\@@href}%
\providecommand \@@href[1]{\endgroup#1\@@endlink}%
\providecommand \@sanitize@url [0]{\catcode `\\12\catcode `\$12\catcode
  `\&12\catcode `\#12\catcode `\^12\catcode `\_12\catcode `\%12\relax}%
\providecommand \@@startlink[1]{}%
\providecommand \@@endlink[0]{}%
\providecommand \url  [0]{\begingroup\@sanitize@url \@url }%
\providecommand \@url [1]{\endgroup\@href {#1}{\urlprefix }}%
\providecommand \urlprefix  [0]{URL }%
\providecommand \Eprint [0]{\href }%
\providecommand \doibase [0]{http://dx.doi.org/}%
\providecommand \selectlanguage [0]{\@gobble}%
\providecommand \bibinfo  [0]{\@secondoftwo}%
\providecommand \bibfield  [0]{\@secondoftwo}%
\providecommand \translation [1]{[#1]}%
\providecommand \BibitemOpen [0]{}%
\providecommand \bibitemStop [0]{}%
\providecommand \bibitemNoStop [0]{.\EOS\space}%
\providecommand \EOS [0]{\spacefactor3000\relax}%
\providecommand \BibitemShut  [1]{\csname bibitem#1\endcsname}%
\let\auto@bib@innerbib\@empty
\bibitem [{\citenamefont {Šmejkal}\ \emph
  {et~al.}(2022{\natexlab{a}})\citenamefont {Šmejkal}, \citenamefont
  {Sinova},\ and\ \citenamefont {Jungwirth}}]{Smejkal:2022p031042}%
  \BibitemOpen
  \bibfield  {author} {\bibinfo {author} {\bibfnamefont {L.}~\bibnamefont
  {Šmejkal}}, \bibinfo {author} {\bibfnamefont {J.}~\bibnamefont {Sinova}}, \
  and\ \bibinfo {author} {\bibfnamefont {T.}~\bibnamefont {Jungwirth}},\ }\href
  {\doibase 10.1103/physrevx.12.031042} {\bibfield  {journal} {\bibinfo
  {journal} {Phys. Rev. X}\ }\textbf {\bibinfo {volume} {12}},\ \bibinfo
  {pages} {031042} (\bibinfo {year} {2022}{\natexlab{a}})}\BibitemShut
  {NoStop}%
\bibitem [{\citenamefont {Šmejkal}\ \emph
  {et~al.}(2022{\natexlab{b}})\citenamefont {Šmejkal}, \citenamefont
  {Sinova},\ and\ \citenamefont {Jungwirth}}]{Smejkal:2022p040501}%
  \BibitemOpen
  \bibfield  {author} {\bibinfo {author} {\bibfnamefont {L.}~\bibnamefont
  {Šmejkal}}, \bibinfo {author} {\bibfnamefont {J.}~\bibnamefont {Sinova}}, \
  and\ \bibinfo {author} {\bibfnamefont {T.}~\bibnamefont {Jungwirth}},\ }\href
  {\doibase 10.1103/physrevx.12.040501} {\bibfield  {journal} {\bibinfo
  {journal} {Phys. Rev. X}\ }\textbf {\bibinfo {volume} {12}},\ \bibinfo
  {pages} {040501} (\bibinfo {year} {2022}{\natexlab{b}})}\BibitemShut
  {NoStop}%
\bibitem [{\citenamefont {Mazin}\ and\ \citenamefont
  {Editors}(2022)}]{Mazin:2022p040002}%
  \BibitemOpen
  \bibfield  {author} {\bibinfo {author} {\bibfnamefont {I.}~\bibnamefont
  {Mazin}}\ and\ \bibinfo {author} {\bibfnamefont {T.~P.}\ \bibnamefont
  {Editors}},\ }\href {\doibase 10.1103/physrevx.12.040002} {\bibfield
  {journal} {\bibinfo  {journal} {Phys. Rev. X}\ }\textbf {\bibinfo {volume}
  {12}},\ \bibinfo {pages} {040002} (\bibinfo {year} {2022})}\BibitemShut
  {NoStop}%
\bibitem [{\citenamefont {Turek}(2022)}]{Turek:2022p094432}%
  \BibitemOpen
  \bibfield  {author} {\bibinfo {author} {\bibfnamefont {I.}~\bibnamefont
  {Turek}},\ }\href {\doibase 10.1103/physrevb.106.094432} {\bibfield
  {journal} {\bibinfo  {journal} {Phys. Rev. B}\ }\textbf {\bibinfo {volume}
  {106}},\ \bibinfo {pages} {094432} (\bibinfo {year} {2022})}\BibitemShut
  {NoStop}%
\bibitem [{\citenamefont {Mazin}\ \emph {et~al.}(2021)\citenamefont {Mazin},
  \citenamefont {Koepernik}, \citenamefont {Johannes}, \citenamefont
  {González-Hernández},\ and\ \citenamefont
  {Šmejkal}}]{Mazin:2021pe2108924118}%
  \BibitemOpen
  \bibfield  {author} {\bibinfo {author} {\bibfnamefont {I.~I.}\ \bibnamefont
  {Mazin}}, \bibinfo {author} {\bibfnamefont {K.}~\bibnamefont {Koepernik}},
  \bibinfo {author} {\bibfnamefont {M.~D.}\ \bibnamefont {Johannes}}, \bibinfo
  {author} {\bibfnamefont {R.}~\bibnamefont {González-Hernández}}, \ and\
  \bibinfo {author} {\bibfnamefont {L.}~\bibnamefont {Šmejkal}},\ }\href
  {\doibase 10.1073/pnas.2108924118} {\bibfield  {journal} {\bibinfo  {journal}
  {Proceedings of the National Academy of Sciences}\ }\textbf {\bibinfo
  {volume} {118}},\ \bibinfo {pages} {e2108924118} (\bibinfo {year}
  {2021})}\BibitemShut {NoStop}%
\bibitem [{\citenamefont {Song}\ \emph {et~al.}(2025)\citenamefont {Song},
  \citenamefont {Bai}, \citenamefont {Zhou}, \citenamefont {Han}, \citenamefont
  {Reichlova}, \citenamefont {Dil}, \citenamefont {Liu}, \citenamefont {Chen},\
  and\ \citenamefont {Pan}}]{Song:2025p473}%
  \BibitemOpen
  \bibfield  {author} {\bibinfo {author} {\bibfnamefont {C.}~\bibnamefont
  {Song}}, \bibinfo {author} {\bibfnamefont {H.}~\bibnamefont {Bai}}, \bibinfo
  {author} {\bibfnamefont {Z.}~\bibnamefont {Zhou}}, \bibinfo {author}
  {\bibfnamefont {L.}~\bibnamefont {Han}}, \bibinfo {author} {\bibfnamefont
  {H.}~\bibnamefont {Reichlova}}, \bibinfo {author} {\bibfnamefont {J.~H.}\
  \bibnamefont {Dil}}, \bibinfo {author} {\bibfnamefont {J.}~\bibnamefont
  {Liu}}, \bibinfo {author} {\bibfnamefont {X.}~\bibnamefont {Chen}}, \ and\
  \bibinfo {author} {\bibfnamefont {F.}~\bibnamefont {Pan}},\ }\href {\doibase
  10.1038/s41578-025-00779-1} {\bibfield  {journal} {\bibinfo  {journal} {Nat.
  Rev. Mater.}\ }\textbf {\bibinfo {volume} {10}},\ \bibinfo {pages} {473}
  (\bibinfo {year} {2025})}\BibitemShut {NoStop}%
\bibitem [{\citenamefont {Jungwirth}\ \emph {et~al.}(2025)\citenamefont
  {Jungwirth}, \citenamefont {Fernandes}, \citenamefont {Fradkin},
  \citenamefont {MacDonald}, \citenamefont {Sinova},\ and\ \citenamefont
  {Šmejkal}}]{Jungwirth:2025p100162}%
  \BibitemOpen
  \bibfield  {author} {\bibinfo {author} {\bibfnamefont {T.}~\bibnamefont
  {Jungwirth}}, \bibinfo {author} {\bibfnamefont {R.~M.}\ \bibnamefont
  {Fernandes}}, \bibinfo {author} {\bibfnamefont {E.}~\bibnamefont {Fradkin}},
  \bibinfo {author} {\bibfnamefont {A.~H.}\ \bibnamefont {MacDonald}}, \bibinfo
  {author} {\bibfnamefont {J.}~\bibnamefont {Sinova}}, \ and\ \bibinfo {author}
  {\bibfnamefont {L.}~\bibnamefont {Šmejkal}},\ }\href {\doibase
  10.1016/j.newton.2025.100162} {\bibfield  {journal} {\bibinfo  {journal}
  {Newton}\ }\textbf {\bibinfo {volume} {1}},\ \bibinfo {pages} {100162}
  (\bibinfo {year} {2025})}\BibitemShut {NoStop}%
\bibitem [{\citenamefont {Jungwirth}\ \emph {et~al.}(2026)\citenamefont
  {Jungwirth}, \citenamefont {Sinova}, \citenamefont {Fernandes}, \citenamefont
  {Liu}, \citenamefont {Watanabe}, \citenamefont {Murakami}, \citenamefont
  {Nakatsuji},\ and\ \citenamefont {Šmejkal}}]{Jungwirth:2026p837}%
  \BibitemOpen
  \bibfield  {author} {\bibinfo {author} {\bibfnamefont {T.}~\bibnamefont
  {Jungwirth}}, \bibinfo {author} {\bibfnamefont {J.}~\bibnamefont {Sinova}},
  \bibinfo {author} {\bibfnamefont {R.~M.}\ \bibnamefont {Fernandes}}, \bibinfo
  {author} {\bibfnamefont {Q.}~\bibnamefont {Liu}}, \bibinfo {author}
  {\bibfnamefont {H.}~\bibnamefont {Watanabe}}, \bibinfo {author}
  {\bibfnamefont {S.}~\bibnamefont {Murakami}}, \bibinfo {author}
  {\bibfnamefont {S.}~\bibnamefont {Nakatsuji}}, \ and\ \bibinfo {author}
  {\bibfnamefont {L.}~\bibnamefont {Šmejkal}},\ }\href {\doibase
  10.1038/s41586-025-09883-2} {\bibfield  {journal} {\bibinfo  {journal}
  {Nature}\ }\textbf {\bibinfo {volume} {649}},\ \bibinfo {pages} {837}
  (\bibinfo {year} {2026})}\BibitemShut {NoStop}%
\bibitem [{\citenamefont {Hayami}\ \emph {et~al.}(2019)\citenamefont {Hayami},
  \citenamefont {Yanagi},\ and\ \citenamefont {Kusunose}}]{Hayami:2019p123702}%
  \BibitemOpen
  \bibfield  {author} {\bibinfo {author} {\bibfnamefont {S.}~\bibnamefont
  {Hayami}}, \bibinfo {author} {\bibfnamefont {Y.}~\bibnamefont {Yanagi}}, \
  and\ \bibinfo {author} {\bibfnamefont {H.}~\bibnamefont {Kusunose}},\ }\href
  {\doibase 10.7566/jpsj.88.123702} {\bibfield  {journal} {\bibinfo  {journal}
  {J. Phys. Soc. Jpn.}\ }\textbf {\bibinfo {volume} {88}},\ \bibinfo {pages}
  {123702} (\bibinfo {year} {2019})}\BibitemShut {NoStop}%
\bibitem [{\citenamefont {Bai}\ \emph {et~al.}(2023)\citenamefont {Bai},
  \citenamefont {Zhang}, \citenamefont {Zhou}, \citenamefont {Chen},
  \citenamefont {Wan}, \citenamefont {Han}, \citenamefont {Zhu}, \citenamefont
  {Liang}, \citenamefont {Su}, \citenamefont {Han}, \citenamefont {Pan},\ and\
  \citenamefont {Song}}]{Bai:2023p216701}%
  \BibitemOpen
  \bibfield  {author} {\bibinfo {author} {\bibfnamefont {H.}~\bibnamefont
  {Bai}}, \bibinfo {author} {\bibfnamefont {Y.~C.}\ \bibnamefont {Zhang}},
  \bibinfo {author} {\bibfnamefont {Y.~J.}\ \bibnamefont {Zhou}}, \bibinfo
  {author} {\bibfnamefont {P.}~\bibnamefont {Chen}}, \bibinfo {author}
  {\bibfnamefont {C.~H.}\ \bibnamefont {Wan}}, \bibinfo {author} {\bibfnamefont
  {L.}~\bibnamefont {Han}}, \bibinfo {author} {\bibfnamefont {W.~X.}\
  \bibnamefont {Zhu}}, \bibinfo {author} {\bibfnamefont {S.~X.}\ \bibnamefont
  {Liang}}, \bibinfo {author} {\bibfnamefont {Y.~C.}\ \bibnamefont {Su}},
  \bibinfo {author} {\bibfnamefont {X.~F.}\ \bibnamefont {Han}}, \bibinfo
  {author} {\bibfnamefont {F.}~\bibnamefont {Pan}}, \ and\ \bibinfo {author}
  {\bibfnamefont {C.}~\bibnamefont {Song}},\ }\href {\doibase
  10.1103/physrevlett.130.216701} {\bibfield  {journal} {\bibinfo  {journal}
  {Phys. Rev. Lett.}\ }\textbf {\bibinfo {volume} {130}},\ \bibinfo {pages}
  {216701} (\bibinfo {year} {2023})}\BibitemShut {NoStop}%
\bibitem [{\citenamefont {Yuan}\ \emph {et~al.}(2021)\citenamefont {Yuan},
  \citenamefont {Wang}, \citenamefont {Luo},\ and\ \citenamefont
  {Zunger}}]{Yuan:2021p014409}%
  \BibitemOpen
  \bibfield  {author} {\bibinfo {author} {\bibfnamefont {L.-D.}\ \bibnamefont
  {Yuan}}, \bibinfo {author} {\bibfnamefont {Z.}~\bibnamefont {Wang}}, \bibinfo
  {author} {\bibfnamefont {J.-W.}\ \bibnamefont {Luo}}, \ and\ \bibinfo
  {author} {\bibfnamefont {A.}~\bibnamefont {Zunger}},\ }\href {\doibase
  10.1103/physrevmaterials.5.014409} {\bibfield  {journal} {\bibinfo  {journal}
  {Phys. Rev. Materials}\ }\textbf {\bibinfo {volume} {5}},\ \bibinfo {pages}
  {014409} (\bibinfo {year} {2021})}\BibitemShut {NoStop}%
\bibitem [{\citenamefont {Karube}\ \emph {et~al.}(2022)\citenamefont {Karube},
  \citenamefont {Tanaka}, \citenamefont {Sugawara}, \citenamefont {Kadoguchi},
  \citenamefont {Kohda},\ and\ \citenamefont {Nitta}}]{Karube:2022p137201}%
  \BibitemOpen
  \bibfield  {author} {\bibinfo {author} {\bibfnamefont {S.}~\bibnamefont
  {Karube}}, \bibinfo {author} {\bibfnamefont {T.}~\bibnamefont {Tanaka}},
  \bibinfo {author} {\bibfnamefont {D.}~\bibnamefont {Sugawara}}, \bibinfo
  {author} {\bibfnamefont {N.}~\bibnamefont {Kadoguchi}}, \bibinfo {author}
  {\bibfnamefont {M.}~\bibnamefont {Kohda}}, \ and\ \bibinfo {author}
  {\bibfnamefont {J.}~\bibnamefont {Nitta}},\ }\href {\doibase
  10.1103/physrevlett.129.137201} {\bibfield  {journal} {\bibinfo  {journal}
  {Phys. Rev. Lett.}\ }\textbf {\bibinfo {volume} {129}},\ \bibinfo {pages}
  {137201} (\bibinfo {year} {2022})}\BibitemShut {NoStop}%
\bibitem [{\citenamefont {Egorov}\ and\ \citenamefont
  {Evarestov}(2021)}]{Egorov:2021p2363}%
  \BibitemOpen
  \bibfield  {author} {\bibinfo {author} {\bibfnamefont {S.~A.}\ \bibnamefont
  {Egorov}}\ and\ \bibinfo {author} {\bibfnamefont {R.~A.}\ \bibnamefont
  {Evarestov}},\ }\href {\doibase 10.1021/acs.jpclett.1c00282} {\bibfield
  {journal} {\bibinfo  {journal} {The J. Phys. Chem. Letters}\ }\textbf
  {\bibinfo {volume} {12}},\ \bibinfo {pages} {2363} (\bibinfo {year}
  {2021})}\BibitemShut {NoStop}%
\bibitem [{\citenamefont {Guo}\ \emph {et~al.}(2023)\citenamefont {Guo},
  \citenamefont {Liu}, \citenamefont {Janson}, \citenamefont {Fulga},
  \citenamefont {Brink},\ and\ \citenamefont {Facio}}]{Guo:2023p100991}%
  \BibitemOpen
  \bibfield  {author} {\bibinfo {author} {\bibfnamefont {Y.}~\bibnamefont
  {Guo}}, \bibinfo {author} {\bibfnamefont {H.}~\bibnamefont {Liu}}, \bibinfo
  {author} {\bibfnamefont {O.}~\bibnamefont {Janson}}, \bibinfo {author}
  {\bibfnamefont {I.~C.}\ \bibnamefont {Fulga}}, \bibinfo {author}
  {\bibfnamefont {J.~v.~d.}\ \bibnamefont {Brink}}, \ and\ \bibinfo {author}
  {\bibfnamefont {J.~I.}\ \bibnamefont {Facio}},\ }\href {\doibase
  10.1016/j.mtphys.2023.100991} {\bibfield  {journal} {\bibinfo  {journal}
  {Materials Today Physics}\ }\textbf {\bibinfo {volume} {32}},\ \bibinfo
  {pages} {100991} (\bibinfo {year} {2023})}\BibitemShut {NoStop}%
\bibitem [{\citenamefont {Naka}\ \emph {et~al.}(2019)\citenamefont {Naka},
  \citenamefont {Hayami}, \citenamefont {Kusunose}, \citenamefont {Yanagi},
  \citenamefont {Motome},\ and\ \citenamefont {Seo}}]{Naka:2019p4305}%
  \BibitemOpen
  \bibfield  {author} {\bibinfo {author} {\bibfnamefont {M.}~\bibnamefont
  {Naka}}, \bibinfo {author} {\bibfnamefont {S.}~\bibnamefont {Hayami}},
  \bibinfo {author} {\bibfnamefont {H.}~\bibnamefont {Kusunose}}, \bibinfo
  {author} {\bibfnamefont {Y.}~\bibnamefont {Yanagi}}, \bibinfo {author}
  {\bibfnamefont {Y.}~\bibnamefont {Motome}}, \ and\ \bibinfo {author}
  {\bibfnamefont {H.}~\bibnamefont {Seo}},\ }\href {\doibase
  10.1038/s41467-019-12229-y} {\bibfield  {journal} {\bibinfo  {journal} {Nat.
  Commun.}\ }\textbf {\bibinfo {volume} {10}},\ \bibinfo {pages} {4305}
  (\bibinfo {year} {2019})}\BibitemShut {NoStop}%
\bibitem [{\citenamefont {Yuan}\ and\ \citenamefont
  {Zunger}(2023)}]{Yuan:2023pe2211966}%
  \BibitemOpen
  \bibfield  {author} {\bibinfo {author} {\bibfnamefont {L.}~\bibnamefont
  {Yuan}}\ and\ \bibinfo {author} {\bibfnamefont {A.}~\bibnamefont {Zunger}},\
  }\href {\doibase 10.1002/adma.202211966} {\bibfield  {journal} {\bibinfo
  {journal} {Adv. Mater.}\ }\textbf {\bibinfo {volume} {35}},\ \bibinfo {pages}
  {e2211966} (\bibinfo {year} {2023})}\BibitemShut {NoStop}%
\bibitem [{\citenamefont {Hayami}\ \emph {et~al.}(2020)\citenamefont {Hayami},
  \citenamefont {Yanagi},\ and\ \citenamefont {Kusunose}}]{Hayami:2020p144441}%
  \BibitemOpen
  \bibfield  {author} {\bibinfo {author} {\bibfnamefont {S.}~\bibnamefont
  {Hayami}}, \bibinfo {author} {\bibfnamefont {Y.}~\bibnamefont {Yanagi}}, \
  and\ \bibinfo {author} {\bibfnamefont {H.}~\bibnamefont {Kusunose}},\ }\href
  {\doibase 10.1103/physrevb.102.144441} {\bibfield  {journal} {\bibinfo
  {journal} {Phys. Rev. B}\ }\textbf {\bibinfo {volume} {102}},\ \bibinfo
  {pages} {144441} (\bibinfo {year} {2020})}\BibitemShut {NoStop}%
\bibitem [{\citenamefont {González-Hernández}\ \emph
  {et~al.}(2021)\citenamefont {González-Hernández}, \citenamefont {Šmejkal},
  \citenamefont {Výborný}, \citenamefont {Yahagi}, \citenamefont {Sinova},
  \citenamefont {Jungwirth},\ and\ \citenamefont
  {Železný}}]{Gonzalez-Hernandez:2021p127701}%
  \BibitemOpen
  \bibfield  {author} {\bibinfo {author} {\bibfnamefont {R.}~\bibnamefont
  {González-Hernández}}, \bibinfo {author} {\bibfnamefont {L.}~\bibnamefont
  {Šmejkal}}, \bibinfo {author} {\bibfnamefont {K.}~\bibnamefont {Výborný}},
  \bibinfo {author} {\bibfnamefont {Y.}~\bibnamefont {Yahagi}}, \bibinfo
  {author} {\bibfnamefont {J.}~\bibnamefont {Sinova}}, \bibinfo {author}
  {\bibfnamefont {T.}~\bibnamefont {Jungwirth}}, \ and\ \bibinfo {author}
  {\bibfnamefont {J.}~\bibnamefont {Železný}},\ }\href {\doibase
  10.1103/physrevlett.126.127701} {\bibfield  {journal} {\bibinfo  {journal}
  {Phys. Rev. Lett.}\ }\textbf {\bibinfo {volume} {126}},\ \bibinfo {pages}
  {127701} (\bibinfo {year} {2021})}\BibitemShut {NoStop}%
\bibitem [{\citenamefont {Naka}\ \emph {et~al.}(2021)\citenamefont {Naka},
  \citenamefont {Motome},\ and\ \citenamefont {Seo}}]{Naka:2021p125114}%
  \BibitemOpen
  \bibfield  {author} {\bibinfo {author} {\bibfnamefont {M.}~\bibnamefont
  {Naka}}, \bibinfo {author} {\bibfnamefont {Y.}~\bibnamefont {Motome}}, \ and\
  \bibinfo {author} {\bibfnamefont {H.}~\bibnamefont {Seo}},\ }\href {\doibase
  10.1103/physrevb.103.125114} {\bibfield  {journal} {\bibinfo  {journal}
  {Phys. Rev. B}\ }\textbf {\bibinfo {volume} {103}},\ \bibinfo {pages}
  {125114} (\bibinfo {year} {2021})}\BibitemShut {NoStop}%
\bibitem [{\citenamefont {Shao}\ \emph {et~al.}(2021)\citenamefont {Shao},
  \citenamefont {Zhang}, \citenamefont {Li}, \citenamefont {Eom},\ and\
  \citenamefont {Tsymbal}}]{Shao:2021p7061}%
  \BibitemOpen
  \bibfield  {author} {\bibinfo {author} {\bibfnamefont {D.-F.}\ \bibnamefont
  {Shao}}, \bibinfo {author} {\bibfnamefont {S.-H.}\ \bibnamefont {Zhang}},
  \bibinfo {author} {\bibfnamefont {M.}~\bibnamefont {Li}}, \bibinfo {author}
  {\bibfnamefont {C.-B.}\ \bibnamefont {Eom}}, \ and\ \bibinfo {author}
  {\bibfnamefont {E.~Y.}\ \bibnamefont {Tsymbal}},\ }\href {\doibase
  10.1038/s41467-021-26915-3} {\bibfield  {journal} {\bibinfo  {journal} {Nat.
  Commun.}\ }\textbf {\bibinfo {volume} {12}},\ \bibinfo {pages} {7061}
  (\bibinfo {year} {2021})}\BibitemShut {NoStop}%
\bibitem [{\citenamefont {Ma}\ \emph {et~al.}(2021)\citenamefont {Ma},
  \citenamefont {Hu}, \citenamefont {Li}, \citenamefont {Liu}, \citenamefont
  {Yao}, \citenamefont {Jia},\ and\ \citenamefont {Liu}}]{Ma:2021p2846}%
  \BibitemOpen
  \bibfield  {author} {\bibinfo {author} {\bibfnamefont {H.-Y.}\ \bibnamefont
  {Ma}}, \bibinfo {author} {\bibfnamefont {M.}~\bibnamefont {Hu}}, \bibinfo
  {author} {\bibfnamefont {N.}~\bibnamefont {Li}}, \bibinfo {author}
  {\bibfnamefont {J.}~\bibnamefont {Liu}}, \bibinfo {author} {\bibfnamefont
  {W.}~\bibnamefont {Yao}}, \bibinfo {author} {\bibfnamefont {J.-F.}\
  \bibnamefont {Jia}}, \ and\ \bibinfo {author} {\bibfnamefont
  {J.}~\bibnamefont {Liu}},\ }\href {\doibase 10.1038/s41467-021-23127-7}
  {\bibfield  {journal} {\bibinfo  {journal} {Nat. Commun.}\ }\textbf {\bibinfo
  {volume} {12}},\ \bibinfo {pages} {2846} (\bibinfo {year}
  {2021})}\BibitemShut {NoStop}%
\bibitem [{\citenamefont {Bose}\ \emph {et~al.}(2022)\citenamefont {Bose},
  \citenamefont {Schreiber}, \citenamefont {Jain}, \citenamefont {Shao},
  \citenamefont {Nair}, \citenamefont {Sun}, \citenamefont {Zhang},
  \citenamefont {Muller}, \citenamefont {Tsymbal}, \citenamefont {Schlom},\
  and\ \citenamefont {Ralph}}]{Bose:2022p267}%
  \BibitemOpen
  \bibfield  {author} {\bibinfo {author} {\bibfnamefont {A.}~\bibnamefont
  {Bose}}, \bibinfo {author} {\bibfnamefont {N.~J.}\ \bibnamefont {Schreiber}},
  \bibinfo {author} {\bibfnamefont {R.}~\bibnamefont {Jain}}, \bibinfo {author}
  {\bibfnamefont {D.-F.}\ \bibnamefont {Shao}}, \bibinfo {author}
  {\bibfnamefont {H.~P.}\ \bibnamefont {Nair}}, \bibinfo {author}
  {\bibfnamefont {J.}~\bibnamefont {Sun}}, \bibinfo {author} {\bibfnamefont
  {X.~S.}\ \bibnamefont {Zhang}}, \bibinfo {author} {\bibfnamefont {D.~A.}\
  \bibnamefont {Muller}}, \bibinfo {author} {\bibfnamefont {E.~Y.}\
  \bibnamefont {Tsymbal}}, \bibinfo {author} {\bibfnamefont {D.~G.}\
  \bibnamefont {Schlom}}, \ and\ \bibinfo {author} {\bibfnamefont {D.~C.}\
  \bibnamefont {Ralph}},\ }\href {\doibase 10.1038/s41928-022-00744-8}
  {\bibfield  {journal} {\bibinfo  {journal} {Nat. Electron.}\ }\textbf
  {\bibinfo {volume} {5}},\ \bibinfo {pages} {267} (\bibinfo {year}
  {2022})}\BibitemShut {NoStop}%
\bibitem [{\citenamefont {Liu}\ \emph {et~al.}(2022)\citenamefont {Liu},
  \citenamefont {Li}, \citenamefont {Han}, \citenamefont {Wan},\ and\
  \citenamefont {Liu}}]{Liu:2022p021016}%
  \BibitemOpen
  \bibfield  {author} {\bibinfo {author} {\bibfnamefont {P.}~\bibnamefont
  {Liu}}, \bibinfo {author} {\bibfnamefont {J.}~\bibnamefont {Li}}, \bibinfo
  {author} {\bibfnamefont {J.}~\bibnamefont {Han}}, \bibinfo {author}
  {\bibfnamefont {X.}~\bibnamefont {Wan}}, \ and\ \bibinfo {author}
  {\bibfnamefont {Q.}~\bibnamefont {Liu}},\ }\href {\doibase
  10.1103/physrevx.12.021016} {\bibfield  {journal} {\bibinfo  {journal} {Phys.
  Rev. X}\ }\textbf {\bibinfo {volume} {12}},\ \bibinfo {pages} {021016}
  (\bibinfo {year} {2022})}\BibitemShut {NoStop}%
\bibitem [{\citenamefont {Jiang}\ \emph {et~al.}(2024)\citenamefont {Jiang},
  \citenamefont {Song}, \citenamefont {Zhu}, \citenamefont {Fang},
  \citenamefont {Weng}, \citenamefont {Liu}, \citenamefont {Yang},\ and\
  \citenamefont {Fang}}]{Jiang:2024p031039}%
  \BibitemOpen
  \bibfield  {author} {\bibinfo {author} {\bibfnamefont {Y.}~\bibnamefont
  {Jiang}}, \bibinfo {author} {\bibfnamefont {Z.}~\bibnamefont {Song}},
  \bibinfo {author} {\bibfnamefont {T.}~\bibnamefont {Zhu}}, \bibinfo {author}
  {\bibfnamefont {Z.}~\bibnamefont {Fang}}, \bibinfo {author} {\bibfnamefont
  {H.}~\bibnamefont {Weng}}, \bibinfo {author} {\bibfnamefont {Z.-X.}\
  \bibnamefont {Liu}}, \bibinfo {author} {\bibfnamefont {J.}~\bibnamefont
  {Yang}}, \ and\ \bibinfo {author} {\bibfnamefont {C.}~\bibnamefont {Fang}},\
  }\href {\doibase 10.1103/physrevx.14.031039} {\bibfield  {journal} {\bibinfo
  {journal} {Phys. Rev. X}\ }\textbf {\bibinfo {volume} {14}},\ \bibinfo
  {pages} {031039} (\bibinfo {year} {2024})}\BibitemShut {NoStop}%
\bibitem [{\citenamefont {Xiao}\ \emph {et~al.}(2024)\citenamefont {Xiao},
  \citenamefont {Zhao}, \citenamefont {Li}, \citenamefont {Shindou},\ and\
  \citenamefont {Song}}]{Xiao:2024p031037}%
  \BibitemOpen
  \bibfield  {author} {\bibinfo {author} {\bibfnamefont {Z.}~\bibnamefont
  {Xiao}}, \bibinfo {author} {\bibfnamefont {J.}~\bibnamefont {Zhao}}, \bibinfo
  {author} {\bibfnamefont {Y.}~\bibnamefont {Li}}, \bibinfo {author}
  {\bibfnamefont {R.}~\bibnamefont {Shindou}}, \ and\ \bibinfo {author}
  {\bibfnamefont {Z.-D.}\ \bibnamefont {Song}},\ }\href {\doibase
  10.1103/physrevx.14.031037} {\bibfield  {journal} {\bibinfo  {journal} {Phys.
  Rev. X}\ }\textbf {\bibinfo {volume} {14}},\ \bibinfo {pages} {031037}
  (\bibinfo {year} {2024})}\BibitemShut {NoStop}%
\bibitem [{\citenamefont {Chen}\ \emph {et~al.}(2024)\citenamefont {Chen},
  \citenamefont {Ren}, \citenamefont {Zhu}, \citenamefont {Yu}, \citenamefont
  {Zhang}, \citenamefont {Liu}, \citenamefont {Li}, \citenamefont {Liu},
  \citenamefont {Li},\ and\ \citenamefont {Liu}}]{Chen:2024p031038}%
  \BibitemOpen
  \bibfield  {author} {\bibinfo {author} {\bibfnamefont {X.}~\bibnamefont
  {Chen}}, \bibinfo {author} {\bibfnamefont {J.}~\bibnamefont {Ren}}, \bibinfo
  {author} {\bibfnamefont {Y.}~\bibnamefont {Zhu}}, \bibinfo {author}
  {\bibfnamefont {Y.}~\bibnamefont {Yu}}, \bibinfo {author} {\bibfnamefont
  {A.}~\bibnamefont {Zhang}}, \bibinfo {author} {\bibfnamefont
  {P.}~\bibnamefont {Liu}}, \bibinfo {author} {\bibfnamefont {J.}~\bibnamefont
  {Li}}, \bibinfo {author} {\bibfnamefont {Y.}~\bibnamefont {Liu}}, \bibinfo
  {author} {\bibfnamefont {C.}~\bibnamefont {Li}}, \ and\ \bibinfo {author}
  {\bibfnamefont {Q.}~\bibnamefont {Liu}},\ }\href {\doibase
  10.1103/physrevx.14.031038} {\bibfield  {journal} {\bibinfo  {journal} {Phys.
  Rev. X}\ }\textbf {\bibinfo {volume} {14}},\ \bibinfo {pages} {031038}
  (\bibinfo {year} {2024})}\BibitemShut {NoStop}%
\bibitem [{\citenamefont {Schiff}\ \emph {et~al.}(2025)\citenamefont {Schiff},
  \citenamefont {Corticelli}, \citenamefont {Guerreiro}, \citenamefont
  {Romhányi},\ and\ \citenamefont {McClarty}}]{Schiff:2025p109}%
  \BibitemOpen
  \bibfield  {author} {\bibinfo {author} {\bibfnamefont {H.}~\bibnamefont
  {Schiff}}, \bibinfo {author} {\bibfnamefont {A.}~\bibnamefont {Corticelli}},
  \bibinfo {author} {\bibfnamefont {A.}~\bibnamefont {Guerreiro}}, \bibinfo
  {author} {\bibfnamefont {J.}~\bibnamefont {Romhányi}}, \ and\ \bibinfo
  {author} {\bibfnamefont {P.~A.}\ \bibnamefont {McClarty}},\ }\href {\doibase
  10.21468/scipostphys.18.3.109} {\bibfield  {journal} {\bibinfo  {journal}
  {SciPost Physics}\ }\textbf {\bibinfo {volume} {18}},\ \bibinfo {pages} {109}
  (\bibinfo {year} {2025})}\BibitemShut {NoStop}%
\bibitem [{\citenamefont {Fernandes}\ \emph {et~al.}(2024)\citenamefont
  {Fernandes}, \citenamefont {Carvalho}, \citenamefont {Birol},\ and\
  \citenamefont {Pereira}}]{Fernandes:2024p024404}%
  \BibitemOpen
  \bibfield  {author} {\bibinfo {author} {\bibfnamefont {R.~M.}\ \bibnamefont
  {Fernandes}}, \bibinfo {author} {\bibfnamefont {V.~S.~d.}\ \bibnamefont
  {Carvalho}}, \bibinfo {author} {\bibfnamefont {T.}~\bibnamefont {Birol}}, \
  and\ \bibinfo {author} {\bibfnamefont {R.~G.}\ \bibnamefont {Pereira}},\
  }\href {\doibase 10.1103/physrevb.109.024404} {\bibfield  {journal} {\bibinfo
   {journal} {Phys. Rev. B}\ }\textbf {\bibinfo {volume} {109}},\ \bibinfo
  {pages} {024404} (\bibinfo {year} {2024})}\BibitemShut {NoStop}%
\bibitem [{\citenamefont {Šmejkal}\ \emph
  {et~al.}(2022{\natexlab{c}})\citenamefont {Šmejkal}, \citenamefont
  {MacDonald}, \citenamefont {Sinova}, \citenamefont {Nakatsuji},\ and\
  \citenamefont {Jungwirth}}]{Smejkal:2022p482}%
  \BibitemOpen
  \bibfield  {author} {\bibinfo {author} {\bibfnamefont {L.}~\bibnamefont
  {Šmejkal}}, \bibinfo {author} {\bibfnamefont {A.~H.}\ \bibnamefont
  {MacDonald}}, \bibinfo {author} {\bibfnamefont {J.}~\bibnamefont {Sinova}},
  \bibinfo {author} {\bibfnamefont {S.}~\bibnamefont {Nakatsuji}}, \ and\
  \bibinfo {author} {\bibfnamefont {T.}~\bibnamefont {Jungwirth}},\ }\href
  {\doibase 10.1038/s41578-022-00430-3} {\bibfield  {journal} {\bibinfo
  {journal} {Nat. Rev. Mater.}\ }\textbf {\bibinfo {volume} {7}},\ \bibinfo
  {pages} {482} (\bibinfo {year} {2022}{\natexlab{c}})}\BibitemShut {NoStop}%
\bibitem [{\citenamefont {Antonenko}\ \emph {et~al.}(2025)\citenamefont
  {Antonenko}, \citenamefont {Fernandes},\ and\ \citenamefont
  {Venderbos}}]{Antonenko:2025p096703}%
  \BibitemOpen
  \bibfield  {author} {\bibinfo {author} {\bibfnamefont {D.~S.}\ \bibnamefont
  {Antonenko}}, \bibinfo {author} {\bibfnamefont {R.~M.}\ \bibnamefont
  {Fernandes}}, \ and\ \bibinfo {author} {\bibfnamefont {J.~W.~F.}\
  \bibnamefont {Venderbos}},\ }\href {\doibase 10.1103/physrevlett.134.096703}
  {\bibfield  {journal} {\bibinfo  {journal} {Phys. Rev. Lett.}\ }\textbf
  {\bibinfo {volume} {134}},\ \bibinfo {pages} {096703} (\bibinfo {year}
  {2025})}\BibitemShut {NoStop}%
\bibitem [{\citenamefont {McClarty}\ and\ \citenamefont
  {Rau}(2024)}]{McClarty:2024p176702}%
  \BibitemOpen
  \bibfield  {author} {\bibinfo {author} {\bibfnamefont {P.~A.}\ \bibnamefont
  {McClarty}}\ and\ \bibinfo {author} {\bibfnamefont {J.~G.}\ \bibnamefont
  {Rau}},\ }\href {\doibase 10.1103/physrevlett.132.176702} {\bibfield
  {journal} {\bibinfo  {journal} {Phys. Rev. Lett.}\ }\textbf {\bibinfo
  {volume} {132}},\ \bibinfo {pages} {176702} (\bibinfo {year}
  {2024})}\BibitemShut {NoStop}%
\bibitem [{\citenamefont {Bhowal}\ and\ \citenamefont
  {Spaldin}(2024)}]{Bhowal:2024p011019}%
  \BibitemOpen
  \bibfield  {author} {\bibinfo {author} {\bibfnamefont {S.}~\bibnamefont
  {Bhowal}}\ and\ \bibinfo {author} {\bibfnamefont {N.~A.}\ \bibnamefont
  {Spaldin}},\ }\href {\doibase 10.1103/physrevx.14.011019} {\bibfield
  {journal} {\bibinfo  {journal} {Phys. Rev. X}\ }\textbf {\bibinfo {volume}
  {14}},\ \bibinfo {pages} {011019} (\bibinfo {year} {2024})}\BibitemShut
  {NoStop}%
\bibitem [{\citenamefont {Aoyama}\ and\ \citenamefont
  {Ohgushi}(2024)}]{Aoyama:2024pL041402}%
  \BibitemOpen
  \bibfield  {author} {\bibinfo {author} {\bibfnamefont {T.}~\bibnamefont
  {Aoyama}}\ and\ \bibinfo {author} {\bibfnamefont {K.}~\bibnamefont
  {Ohgushi}},\ }\href {\doibase 10.1103/physrevmaterials.8.l041402} {\bibfield
  {journal} {\bibinfo  {journal} {Phys. Rev. Materials}\ }\textbf {\bibinfo
  {volume} {8}},\ \bibinfo {pages} {L041402} (\bibinfo {year}
  {2024})}\BibitemShut {NoStop}%
\bibitem [{\citenamefont {Steward}\ \emph {et~al.}(2023)\citenamefont
  {Steward}, \citenamefont {Fernandes},\ and\ \citenamefont
  {Schmalian}}]{Steward:2023p144418}%
  \BibitemOpen
  \bibfield  {author} {\bibinfo {author} {\bibfnamefont {C.~R.~W.}\
  \bibnamefont {Steward}}, \bibinfo {author} {\bibfnamefont {R.~M.}\
  \bibnamefont {Fernandes}}, \ and\ \bibinfo {author} {\bibfnamefont
  {J.}~\bibnamefont {Schmalian}},\ }\href {\doibase
  10.1103/physrevb.108.144418} {\bibfield  {journal} {\bibinfo  {journal}
  {Phys. Rev. B}\ }\textbf {\bibinfo {volume} {108}},\ \bibinfo {pages}
  {144418} (\bibinfo {year} {2023})}\BibitemShut {NoStop}%
\bibitem [{\citenamefont {Takahashi}\ \emph {et~al.}(2025)\citenamefont
  {Takahashi}, \citenamefont {Steward}, \citenamefont {Ogata}, \citenamefont
  {Fernandes},\ and\ \citenamefont {Schmalian}}]{Takahashi:2025p184408}%
  \BibitemOpen
  \bibfield  {author} {\bibinfo {author} {\bibfnamefont {K.}~\bibnamefont
  {Takahashi}}, \bibinfo {author} {\bibfnamefont {C.~R.~W.}\ \bibnamefont
  {Steward}}, \bibinfo {author} {\bibfnamefont {M.}~\bibnamefont {Ogata}},
  \bibinfo {author} {\bibfnamefont {R.~M.}\ \bibnamefont {Fernandes}}, \ and\
  \bibinfo {author} {\bibfnamefont {J.}~\bibnamefont {Schmalian}},\ }\href
  {\doibase 10.1103/physrevb.111.184408} {\bibfield  {journal} {\bibinfo
  {journal} {Phys. Rev. B}\ }\textbf {\bibinfo {volume} {111}},\ \bibinfo
  {pages} {184408} (\bibinfo {year} {2025})}\BibitemShut {NoStop}%
\bibitem [{\citenamefont {Li}\ \emph {et~al.}(2024)\citenamefont {Li},
  \citenamefont {Fan}, \citenamefont {Wang}, \citenamefont {Zhang},\ and\
  \citenamefont {Li}}]{Li:2024p222404}%
  \BibitemOpen
  \bibfield  {author} {\bibinfo {author} {\bibfnamefont {J.-Y.}\ \bibnamefont
  {Li}}, \bibinfo {author} {\bibfnamefont {A.-D.}\ \bibnamefont {Fan}},
  \bibinfo {author} {\bibfnamefont {Y.-K.}\ \bibnamefont {Wang}}, \bibinfo
  {author} {\bibfnamefont {Y.}~\bibnamefont {Zhang}}, \ and\ \bibinfo {author}
  {\bibfnamefont {S.}~\bibnamefont {Li}},\ }\href {\doibase 10.1063/5.0242426}
  {\bibfield  {journal} {\bibinfo  {journal} {Applied Physics Letters}\
  }\textbf {\bibinfo {volume} {125}},\ \bibinfo {pages} {222404} (\bibinfo
  {year} {2024})}\BibitemShut {NoStop}%
\bibitem [{\citenamefont {Chakraborty}\ \emph {et~al.}(2024)\citenamefont
  {Chakraborty}, \citenamefont {Hernández}, \citenamefont {Šmejkal},\ and\
  \citenamefont {Sinova}}]{Chakraborty:2024p144421}%
  \BibitemOpen
  \bibfield  {author} {\bibinfo {author} {\bibfnamefont {A.}~\bibnamefont
  {Chakraborty}}, \bibinfo {author} {\bibfnamefont {R.~G.}\ \bibnamefont
  {Hernández}}, \bibinfo {author} {\bibfnamefont {L.}~\bibnamefont
  {Šmejkal}}, \ and\ \bibinfo {author} {\bibfnamefont {J.}~\bibnamefont
  {Sinova}},\ }\href {\doibase 10.1103/physrevb.109.144421} {\bibfield
  {journal} {\bibinfo  {journal} {Phys. Rev. B}\ }\textbf {\bibinfo {volume}
  {109}},\ \bibinfo {pages} {144421} (\bibinfo {year} {2024})}\BibitemShut
  {NoStop}%
\bibitem [{\citenamefont {Karetta}\ \emph {et~al.}(2025)\citenamefont
  {Karetta}, \citenamefont {Verbeek}, \citenamefont {Jaeschke-Ubiergo},
  \citenamefont {Šmejkal},\ and\ \citenamefont
  {Sinova}}]{Karetta:2025p094454}%
  \BibitemOpen
  \bibfield  {author} {\bibinfo {author} {\bibfnamefont {B.}~\bibnamefont
  {Karetta}}, \bibinfo {author} {\bibfnamefont {X.~H.}\ \bibnamefont
  {Verbeek}}, \bibinfo {author} {\bibfnamefont {R.}~\bibnamefont
  {Jaeschke-Ubiergo}}, \bibinfo {author} {\bibfnamefont {L.}~\bibnamefont
  {Šmejkal}}, \ and\ \bibinfo {author} {\bibfnamefont {J.}~\bibnamefont
  {Sinova}},\ }\href {\doibase 10.1103/pbbr-hwz4} {\bibfield  {journal}
  {\bibinfo  {journal} {Phys. Rev. B}\ }\textbf {\bibinfo {volume} {112}},\
  \bibinfo {pages} {094454} (\bibinfo {year} {2025})}\BibitemShut {NoStop}%
\bibitem [{\citenamefont {Belashchenko}(2025)}]{Belashchenko:2025p086701}%
  \BibitemOpen
  \bibfield  {author} {\bibinfo {author} {\bibfnamefont {K.~D.}\ \bibnamefont
  {Belashchenko}},\ }\href {\doibase 10.1103/physrevlett.134.086701} {\bibfield
   {journal} {\bibinfo  {journal} {Phys. Rev. Lett.}\ }\textbf {\bibinfo
  {volume} {134}},\ \bibinfo {pages} {086701} (\bibinfo {year}
  {2025})}\BibitemShut {NoStop}%
\bibitem [{\citenamefont {Naka}\ \emph {et~al.}(2025)\citenamefont {Naka},
  \citenamefont {Motome}, \citenamefont {Miyazaki},\ and\ \citenamefont
  {Seo}}]{Naka:2025p083702}%
  \BibitemOpen
  \bibfield  {author} {\bibinfo {author} {\bibfnamefont {M.}~\bibnamefont
  {Naka}}, \bibinfo {author} {\bibfnamefont {Y.}~\bibnamefont {Motome}},
  \bibinfo {author} {\bibfnamefont {T.}~\bibnamefont {Miyazaki}}, \ and\
  \bibinfo {author} {\bibfnamefont {H.}~\bibnamefont {Seo}},\ }\href {\doibase
  10.7566/jpsj.94.083702} {\bibfield  {journal} {\bibinfo  {journal} {J. Phys.
  Soc. Jpn.}\ }\textbf {\bibinfo {volume} {94}},\ \bibinfo {pages} {083702}
  (\bibinfo {year} {2025})}\BibitemShut {NoStop}%
\bibitem [{\citenamefont {Zhang}\ \emph
  {et~al.}(2025{\natexlab{a}})\citenamefont {Zhang}, \citenamefont {Zheng},
  \citenamefont {Liu}, \citenamefont {Zhang}, \citenamefont {Xiong},\ and\
  \citenamefont {Lu}}]{Zhang:2025p024415}%
  \BibitemOpen
  \bibfield  {author} {\bibinfo {author} {\bibfnamefont {W.}~\bibnamefont
  {Zhang}}, \bibinfo {author} {\bibfnamefont {M.}~\bibnamefont {Zheng}},
  \bibinfo {author} {\bibfnamefont {Y.}~\bibnamefont {Liu}}, \bibinfo {author}
  {\bibfnamefont {Z.}~\bibnamefont {Zhang}}, \bibinfo {author} {\bibfnamefont
  {R.}~\bibnamefont {Xiong}}, \ and\ \bibinfo {author} {\bibfnamefont
  {Z.}~\bibnamefont {Lu}},\ }\href {\doibase 10.1103/8zlt-mlms} {\bibfield
  {journal} {\bibinfo  {journal} {Phys. Rev. B}\ }\textbf {\bibinfo {volume}
  {112}},\ \bibinfo {pages} {024415} (\bibinfo {year}
  {2025}{\natexlab{a}})}\BibitemShut {NoStop}%
\bibitem [{\citenamefont {Chen}\ \emph {et~al.}(2025)\citenamefont {Chen},
  \citenamefont {He}, \citenamefont {Xiong}, \citenamefont {Quan},
  \citenamefont {Hou}, \citenamefont {Ji}, \citenamefont {Yang},\ and\
  \citenamefont {Li}}]{Chen:2025p102409}%
  \BibitemOpen
  \bibfield  {author} {\bibinfo {author} {\bibfnamefont {C.}~\bibnamefont
  {Chen}}, \bibinfo {author} {\bibfnamefont {X.}~\bibnamefont {He}}, \bibinfo
  {author} {\bibfnamefont {Q.}~\bibnamefont {Xiong}}, \bibinfo {author}
  {\bibfnamefont {C.}~\bibnamefont {Quan}}, \bibinfo {author} {\bibfnamefont
  {H.}~\bibnamefont {Hou}}, \bibinfo {author} {\bibfnamefont {S.}~\bibnamefont
  {Ji}}, \bibinfo {author} {\bibfnamefont {J.}~\bibnamefont {Yang}}, \ and\
  \bibinfo {author} {\bibfnamefont {X.}~\bibnamefont {Li}},\ }\href {\doibase
  10.1063/5.0277631} {\bibfield  {journal} {\bibinfo  {journal} {Applied
  Physics Letters}\ }\textbf {\bibinfo {volume} {127}},\ \bibinfo {pages}
  {102409} (\bibinfo {year} {2025})}\BibitemShut {NoStop}%
\bibitem [{\citenamefont {Xun}\ \emph {et~al.}(2025)\citenamefont {Xun},
  \citenamefont {Liu}, \citenamefont {Zhang}, \citenamefont {Wu},\ and\
  \citenamefont {Li}}]{Xun:2025p161903}%
  \BibitemOpen
  \bibfield  {author} {\bibinfo {author} {\bibfnamefont {W.}~\bibnamefont
  {Xun}}, \bibinfo {author} {\bibfnamefont {X.}~\bibnamefont {Liu}}, \bibinfo
  {author} {\bibfnamefont {Y.}~\bibnamefont {Zhang}}, \bibinfo {author}
  {\bibfnamefont {Y.-Z.}\ \bibnamefont {Wu}}, \ and\ \bibinfo {author}
  {\bibfnamefont {P.}~\bibnamefont {Li}},\ }\href {\doibase 10.1063/5.0267525}
  {\bibfield  {journal} {\bibinfo  {journal} {Applied Physics Letters}\
  }\textbf {\bibinfo {volume} {126}},\ \bibinfo {pages} {161903} (\bibinfo
  {year} {2025})}\BibitemShut {NoStop}%
\bibitem [{\citenamefont {Jiang}\ \emph
  {et~al.}(2025{\natexlab{a}})\citenamefont {Jiang}, \citenamefont {Zhang},
  \citenamefont {Bai}, \citenamefont {Tian}, \citenamefont {Zhang},
  \citenamefont {Gong},\ and\ \citenamefont {Kong}}]{Jiang:2025p053102}%
  \BibitemOpen
  \bibfield  {author} {\bibinfo {author} {\bibfnamefont {Y.}~\bibnamefont
  {Jiang}}, \bibinfo {author} {\bibfnamefont {X.}~\bibnamefont {Zhang}},
  \bibinfo {author} {\bibfnamefont {H.}~\bibnamefont {Bai}}, \bibinfo {author}
  {\bibfnamefont {Y.}~\bibnamefont {Tian}}, \bibinfo {author} {\bibfnamefont
  {B.}~\bibnamefont {Zhang}}, \bibinfo {author} {\bibfnamefont {W.-J.}\
  \bibnamefont {Gong}}, \ and\ \bibinfo {author} {\bibfnamefont
  {X.}~\bibnamefont {Kong}},\ }\href {\doibase 10.1063/5.0252374} {\bibfield
  {journal} {\bibinfo  {journal} {Applied Physics Letters}\ }\textbf {\bibinfo
  {volume} {126}},\ \bibinfo {pages} {053102} (\bibinfo {year}
  {2025}{\natexlab{a}})}\BibitemShut {NoStop}%
\bibitem [{\citenamefont {Khodas}\ \emph {et~al.}(2025)\citenamefont {Khodas},
  \citenamefont {Mu}, \citenamefont {Mazin},\ and\ \citenamefont
  {Belashchenko}}]{Khodas:2025arXiv06}%
  \BibitemOpen
  \bibfield  {author} {\bibinfo {author} {\bibfnamefont {M.}~\bibnamefont
  {Khodas}}, \bibinfo {author} {\bibfnamefont {S.}~\bibnamefont {Mu}}, \bibinfo
  {author} {\bibfnamefont {I.~I.}\ \bibnamefont {Mazin}}, \ and\ \bibinfo
  {author} {\bibfnamefont {K.~D.}\ \bibnamefont {Belashchenko}},\ }\href
  {\doibase 10.48550/arxiv.2506.06257} {\bibfield  {journal} {\bibinfo
  {journal} {arXiv}\ } (\bibinfo {year} {2025}),\
  10.48550/arxiv.2506.06257}\BibitemShut {NoStop}%
\bibitem [{\citenamefont {Zhai}\ \emph {et~al.}(2025)\citenamefont {Zhai},
  \citenamefont {Yu}, \citenamefont {Lv}, \citenamefont {Zhang},\ and\
  \citenamefont {Zhao}}]{Zhai:arXiv2025}%
  \BibitemOpen
  \bibfield  {author} {\bibinfo {author} {\bibfnamefont {Y.}~\bibnamefont
  {Zhai}}, \bibinfo {author} {\bibfnamefont {L.}~\bibnamefont {Yu}}, \bibinfo
  {author} {\bibfnamefont {J.}~\bibnamefont {Lv}}, \bibinfo {author}
  {\bibfnamefont {W.}~\bibnamefont {Zhang}}, \ and\ \bibinfo {author}
  {\bibfnamefont {H.~J.}\ \bibnamefont {Zhao}},\ }\href {\doibase
  10.48550/arxiv.2506.07447} {\bibfield  {journal} {\bibinfo  {journal}
  {arXiv}\ } (\bibinfo {year} {2025}),\ 10.48550/arxiv.2506.07447}\BibitemShut
  {NoStop}%
\bibitem [{\citenamefont {Smolenski}\ \emph {et~al.}(2025)\citenamefont
  {Smolenski}, \citenamefont {Mao}, \citenamefont {Zhang}, \citenamefont {Guo},
  \citenamefont {Shawon}, \citenamefont {Xu}, \citenamefont {Downey},
  \citenamefont {Musall}, \citenamefont {Yi}, \citenamefont {Xie},
  \citenamefont {Jozwiak}, \citenamefont {Bostwick}, \citenamefont {Tamura},
  \citenamefont {Rotenberg}, \citenamefont {Li}, \citenamefont {Sun},
  \citenamefont {Zhang},\ and\ \citenamefont {Jo}}]{Smolenski:arXiv2025}%
  \BibitemOpen
  \bibfield  {author} {\bibinfo {author} {\bibfnamefont {S.}~\bibnamefont
  {Smolenski}}, \bibinfo {author} {\bibfnamefont {N.}~\bibnamefont {Mao}},
  \bibinfo {author} {\bibfnamefont {D.}~\bibnamefont {Zhang}}, \bibinfo
  {author} {\bibfnamefont {Y.}~\bibnamefont {Guo}}, \bibinfo {author}
  {\bibfnamefont {A.~K. M.~A.}\ \bibnamefont {Shawon}}, \bibinfo {author}
  {\bibfnamefont {M.}~\bibnamefont {Xu}}, \bibinfo {author} {\bibfnamefont
  {E.}~\bibnamefont {Downey}}, \bibinfo {author} {\bibfnamefont
  {T.}~\bibnamefont {Musall}}, \bibinfo {author} {\bibfnamefont
  {M.}~\bibnamefont {Yi}}, \bibinfo {author} {\bibfnamefont {W.}~\bibnamefont
  {Xie}}, \bibinfo {author} {\bibfnamefont {C.}~\bibnamefont {Jozwiak}},
  \bibinfo {author} {\bibfnamefont {A.}~\bibnamefont {Bostwick}}, \bibinfo
  {author} {\bibfnamefont {N.}~\bibnamefont {Tamura}}, \bibinfo {author}
  {\bibfnamefont {E.}~\bibnamefont {Rotenberg}}, \bibinfo {author}
  {\bibfnamefont {L.}~\bibnamefont {Li}}, \bibinfo {author} {\bibfnamefont
  {K.}~\bibnamefont {Sun}}, \bibinfo {author} {\bibfnamefont {Y.}~\bibnamefont
  {Zhang}}, \ and\ \bibinfo {author} {\bibfnamefont {N.~H.}\ \bibnamefont
  {Jo}},\ }\href {\doibase 10.48550/arxiv.2509.21481} {\bibfield  {journal}
  {\bibinfo  {journal} {arXiv}\ } (\bibinfo {year} {2025}),\
  10.48550/arxiv.2509.21481}\BibitemShut {NoStop}%
\bibitem [{Sor(2025)}]{Sorn:2025p245115}%
  \BibitemOpen
  \href {\doibase 10.1103/vzmh-mxlz} {\bibfield  {journal} {\bibinfo  {journal}
  {Phys. Rev. B}\ }\textbf {\bibinfo {volume} {112}},\ \bibinfo {pages}
  {245115} (\bibinfo {year} {2025})}\BibitemShut {NoStop}%
\bibitem [{\citenamefont {Ye}\ \emph {et~al.}(2026)\citenamefont {Ye},
  \citenamefont {Tenzin}, \citenamefont {Sławińska},\ and\ \citenamefont
  {Autieri}}]{Ye:2026p014413}%
  \BibitemOpen
  \bibfield  {author} {\bibinfo {author} {\bibfnamefont {C.~C.}\ \bibnamefont
  {Ye}}, \bibinfo {author} {\bibfnamefont {K.}~\bibnamefont {Tenzin}}, \bibinfo
  {author} {\bibfnamefont {J.}~\bibnamefont {Sławińska}}, \ and\ \bibinfo
  {author} {\bibfnamefont {C.}~\bibnamefont {Autieri}},\ }\href {\doibase
  10.1103/g32j-hnvz} {\bibfield  {journal} {\bibinfo  {journal} {Phys. Rev. B}\
  }\textbf {\bibinfo {volume} {113}},\ \bibinfo {pages} {014413} (\bibinfo
  {year} {2026})}\BibitemShut {NoStop}%
\bibitem [{\citenamefont {Brekke}\ \emph {et~al.}(2023)\citenamefont {Brekke},
  \citenamefont {Brataas},\ and\ \citenamefont {Sudbø}}]{Brekke:2023p224421}%
  \BibitemOpen
  \bibfield  {author} {\bibinfo {author} {\bibfnamefont {B.}~\bibnamefont
  {Brekke}}, \bibinfo {author} {\bibfnamefont {A.}~\bibnamefont {Brataas}}, \
  and\ \bibinfo {author} {\bibfnamefont {A.}~\bibnamefont {Sudbø}},\ }\href
  {\doibase 10.1103/physrevb.108.224421} {\bibfield  {journal} {\bibinfo
  {journal} {Phys. Rev. B}\ }\textbf {\bibinfo {volume} {108}},\ \bibinfo
  {pages} {224421} (\bibinfo {year} {2023})}\BibitemShut {NoStop}%
\bibitem [{\citenamefont {Roig}\ \emph {et~al.}(2024)\citenamefont {Roig},
  \citenamefont {Kreisel}, \citenamefont {Yu}, \citenamefont {Andersen},\ and\
  \citenamefont {Agterberg}}]{Roig:2024p144412}%
  \BibitemOpen
  \bibfield  {author} {\bibinfo {author} {\bibfnamefont {M.}~\bibnamefont
  {Roig}}, \bibinfo {author} {\bibfnamefont {A.}~\bibnamefont {Kreisel}},
  \bibinfo {author} {\bibfnamefont {Y.}~\bibnamefont {Yu}}, \bibinfo {author}
  {\bibfnamefont {B.~M.}\ \bibnamefont {Andersen}}, \ and\ \bibinfo {author}
  {\bibfnamefont {D.~F.}\ \bibnamefont {Agterberg}},\ }\href {\doibase
  10.1103/physrevb.110.144412} {\bibfield  {journal} {\bibinfo  {journal}
  {Phys. Rev. B}\ }\textbf {\bibinfo {volume} {110}},\ \bibinfo {pages}
  {144412} (\bibinfo {year} {2024})}\BibitemShut {NoStop}%
\bibitem [{\citenamefont {Maier}\ and\ \citenamefont
  {Okamoto}(2023)}]{Maier:2023pL100402}%
  \BibitemOpen
  \bibfield  {author} {\bibinfo {author} {\bibfnamefont {T.~A.}\ \bibnamefont
  {Maier}}\ and\ \bibinfo {author} {\bibfnamefont {S.}~\bibnamefont
  {Okamoto}},\ }\href {\doibase 10.1103/physrevb.108.l100402} {\bibfield
  {journal} {\bibinfo  {journal} {Phys. Rev. B}\ }\textbf {\bibinfo {volume}
  {108}},\ \bibinfo {pages} {L100402} (\bibinfo {year} {2023})}\BibitemShut
  {NoStop}%
\bibitem [{\citenamefont {Xiao}\ \emph {et~al.}(2005)\citenamefont {Xiao},
  \citenamefont {Shi},\ and\ \citenamefont {Niu}}]{Xiao:2005p137204}%
  \BibitemOpen
  \bibfield  {author} {\bibinfo {author} {\bibfnamefont {D.}~\bibnamefont
  {Xiao}}, \bibinfo {author} {\bibfnamefont {J.}~\bibnamefont {Shi}}, \ and\
  \bibinfo {author} {\bibfnamefont {Q.}~\bibnamefont {Niu}},\ }\href {\doibase
  10.1103/physrevlett.95.137204} {\bibfield  {journal} {\bibinfo  {journal}
  {Phys. Rev. Lett.}\ }\textbf {\bibinfo {volume} {95}},\ \bibinfo {pages}
  {137204} (\bibinfo {year} {2005})}\BibitemShut {NoStop}%
\bibitem [{\citenamefont {Thonhauser}\ \emph {et~al.}(2005)\citenamefont
  {Thonhauser}, \citenamefont {Ceresoli}, \citenamefont {Vanderbilt},\ and\
  \citenamefont {Resta}}]{Thonhauser:2005p137205}%
  \BibitemOpen
  \bibfield  {author} {\bibinfo {author} {\bibfnamefont {T.}~\bibnamefont
  {Thonhauser}}, \bibinfo {author} {\bibfnamefont {D.}~\bibnamefont
  {Ceresoli}}, \bibinfo {author} {\bibfnamefont {D.}~\bibnamefont
  {Vanderbilt}}, \ and\ \bibinfo {author} {\bibfnamefont {R.}~\bibnamefont
  {Resta}},\ }\href {\doibase 10.1103/physrevlett.95.137205} {\bibfield
  {journal} {\bibinfo  {journal} {Phys. Rev. Lett.}\ }\textbf {\bibinfo
  {volume} {95}},\ \bibinfo {pages} {137205} (\bibinfo {year}
  {2005})}\BibitemShut {NoStop}%
\bibitem [{\citenamefont {Ceresoli}\ \emph {et~al.}(2006)\citenamefont
  {Ceresoli}, \citenamefont {Thonhauser}, \citenamefont {Vanderbilt},\ and\
  \citenamefont {Resta}}]{Ceresoli:2006p024408}%
  \BibitemOpen
  \bibfield  {author} {\bibinfo {author} {\bibfnamefont {D.}~\bibnamefont
  {Ceresoli}}, \bibinfo {author} {\bibfnamefont {T.}~\bibnamefont
  {Thonhauser}}, \bibinfo {author} {\bibfnamefont {D.}~\bibnamefont
  {Vanderbilt}}, \ and\ \bibinfo {author} {\bibfnamefont {R.}~\bibnamefont
  {Resta}},\ }\href {\doibase 10.1103/physrevb.74.024408} {\bibfield  {journal}
  {\bibinfo  {journal} {Phys. Rev. B}\ }\textbf {\bibinfo {volume} {74}},\
  \bibinfo {pages} {024408} (\bibinfo {year} {2006})}\BibitemShut {NoStop}%
\bibitem [{\citenamefont {Mitscherling}\ \emph {et~al.}(2025)\citenamefont
  {Mitscherling}, \citenamefont {Priessnitz},\ and\ \citenamefont
  {Šmejkal}}]{Mitscherling:2025arXiv12}%
  \BibitemOpen
  \bibfield  {author} {\bibinfo {author} {\bibfnamefont {J.}~\bibnamefont
  {Mitscherling}}, \bibinfo {author} {\bibfnamefont {J.}~\bibnamefont
  {Priessnitz}}, \ and\ \bibinfo {author} {\bibfnamefont {L.}~\bibnamefont
  {Šmejkal}},\ }\href {\doibase 10.48550/arxiv.2512.09051} {\bibfield
  {journal} {\bibinfo  {journal} {arXiv}\ } (\bibinfo {year} {2025}),\
  10.48550/arxiv.2512.09051}\BibitemShut {NoStop}%
\bibitem [{Note1()}]{Note1}%
  \BibitemOpen
  \bibinfo {note} {Technically, the Hamiltonian decomposes into two diagonal
  mirror sectors. The spin label $\sigma $ may be used here to label these
  mirror sectors.}\BibitemShut {Stop}%
\bibitem [{\citenamefont {Yershov}\ \emph {et~al.}(2024)\citenamefont
  {Yershov}, \citenamefont {Kravchuk}, \citenamefont {Daghofer},\ and\
  \citenamefont {Brink}}]{Yershov:2024p144421}%
  \BibitemOpen
  \bibfield  {author} {\bibinfo {author} {\bibfnamefont {K.~V.}\ \bibnamefont
  {Yershov}}, \bibinfo {author} {\bibfnamefont {V.~P.}\ \bibnamefont
  {Kravchuk}}, \bibinfo {author} {\bibfnamefont {M.}~\bibnamefont {Daghofer}},
  \ and\ \bibinfo {author} {\bibfnamefont {J.~v.~d.}\ \bibnamefont {Brink}},\
  }\href {\doibase 10.1103/physrevb.110.144421} {\bibfield  {journal} {\bibinfo
   {journal} {Phys. Rev. B}\ }\textbf {\bibinfo {volume} {110}},\ \bibinfo
  {pages} {144421} (\bibinfo {year} {2024})}\BibitemShut {NoStop}%
\bibitem [{\citenamefont {Kaushal}\ and\ \citenamefont
  {Franz}(2025)}]{Kaushal:2025p156502}%
  \BibitemOpen
  \bibfield  {author} {\bibinfo {author} {\bibfnamefont {N.}~\bibnamefont
  {Kaushal}}\ and\ \bibinfo {author} {\bibfnamefont {M.}~\bibnamefont
  {Franz}},\ }\href {\doibase 10.1103/s31h-hk2v} {\bibfield  {journal}
  {\bibinfo  {journal} {Phys. Rev. Lett.}\ }\textbf {\bibinfo {volume} {135}},\
  \bibinfo {pages} {156502} (\bibinfo {year} {2025})}\BibitemShut {NoStop}%
\bibitem [{\citenamefont {Dürrnagel}\ \emph {et~al.}(2025)\citenamefont
  {Dürrnagel}, \citenamefont {Hohmann}, \citenamefont {Maity}, \citenamefont
  {Seufert}, \citenamefont {Klett}, \citenamefont {Klebl},\ and\ \citenamefont
  {Thomale}}]{Durrnagel:2025p036502}%
  \BibitemOpen
  \bibfield  {author} {\bibinfo {author} {\bibfnamefont {M.}~\bibnamefont
  {Dürrnagel}}, \bibinfo {author} {\bibfnamefont {H.}~\bibnamefont {Hohmann}},
  \bibinfo {author} {\bibfnamefont {A.}~\bibnamefont {Maity}}, \bibinfo
  {author} {\bibfnamefont {J.}~\bibnamefont {Seufert}}, \bibinfo {author}
  {\bibfnamefont {M.}~\bibnamefont {Klett}}, \bibinfo {author} {\bibfnamefont
  {L.}~\bibnamefont {Klebl}}, \ and\ \bibinfo {author} {\bibfnamefont
  {R.}~\bibnamefont {Thomale}},\ }\href {\doibase 10.1103/2g3v-z76q} {\bibfield
   {journal} {\bibinfo  {journal} {Phys. Rev. Lett.}\ }\textbf {\bibinfo
  {volume} {135}},\ \bibinfo {pages} {036502} (\bibinfo {year}
  {2025})}\BibitemShut {NoStop}%
\bibitem [{\citenamefont {Fu}\ \emph {et~al.}(2025)\citenamefont {Fu},
  \citenamefont {Hu}, \citenamefont {Li}, \citenamefont {Duan}, \citenamefont
  {Liu},\ and\ \citenamefont {Ouyang}}]{Fu:2025arXiv07}%
  \BibitemOpen
  \bibfield  {author} {\bibinfo {author} {\bibfnamefont {Z.}~\bibnamefont
  {Fu}}, \bibinfo {author} {\bibfnamefont {M.}~\bibnamefont {Hu}}, \bibinfo
  {author} {\bibfnamefont {A.}~\bibnamefont {Li}}, \bibinfo {author}
  {\bibfnamefont {H.}~\bibnamefont {Duan}}, \bibinfo {author} {\bibfnamefont
  {J.}~\bibnamefont {Liu}}, \ and\ \bibinfo {author} {\bibfnamefont
  {F.}~\bibnamefont {Ouyang}},\ }\href {\doibase 10.48550/arxiv.2507.22474}
  {\bibfield  {journal} {\bibinfo  {journal} {arXiv}\ } (\bibinfo {year}
  {2025}),\ 10.48550/arxiv.2507.22474}\BibitemShut {NoStop}%
\bibitem [{\citenamefont {Sun}\ \emph {et~al.}(2009)\citenamefont {Sun},
  \citenamefont {Yao}, \citenamefont {Fradkin},\ and\ \citenamefont
  {Kivelson}}]{Sun:2009p046811}%
  \BibitemOpen
  \bibfield  {author} {\bibinfo {author} {\bibfnamefont {K.}~\bibnamefont
  {Sun}}, \bibinfo {author} {\bibfnamefont {H.}~\bibnamefont {Yao}}, \bibinfo
  {author} {\bibfnamefont {E.}~\bibnamefont {Fradkin}}, \ and\ \bibinfo
  {author} {\bibfnamefont {S.~A.}\ \bibnamefont {Kivelson}},\ }\href {\doibase
  10.1103/physrevlett.103.046811} {\bibfield  {journal} {\bibinfo  {journal}
  {Phys. Rev. Lett.}\ }\textbf {\bibinfo {volume} {103}},\ \bibinfo {pages}
  {046811} (\bibinfo {year} {2009})}\BibitemShut {NoStop}%
\bibitem [{\citenamefont {Weeks}\ and\ \citenamefont
  {Franz}(2010)}]{Weeks:2010p085310}%
  \BibitemOpen
  \bibfield  {author} {\bibinfo {author} {\bibfnamefont {C.}~\bibnamefont
  {Weeks}}\ and\ \bibinfo {author} {\bibfnamefont {M.}~\bibnamefont {Franz}},\
  }\href {\doibase 10.1103/physrevb.82.085310} {\bibfield  {journal} {\bibinfo
  {journal} {Phys. Rev. B}\ }\textbf {\bibinfo {volume} {82}},\ \bibinfo
  {pages} {085310} (\bibinfo {year} {2010})}\BibitemShut {NoStop}%
\bibitem [{\citenamefont {Sun}\ \emph {et~al.}(2017)\citenamefont {Sun},
  \citenamefont {Zhang}, \citenamefont {Liu}, \citenamefont {Felser},\ and\
  \citenamefont {Yan}}]{Sun:2017p235104}%
  \BibitemOpen
  \bibfield  {author} {\bibinfo {author} {\bibfnamefont {Y.}~\bibnamefont
  {Sun}}, \bibinfo {author} {\bibfnamefont {Y.}~\bibnamefont {Zhang}}, \bibinfo
  {author} {\bibfnamefont {C.-X.}\ \bibnamefont {Liu}}, \bibinfo {author}
  {\bibfnamefont {C.}~\bibnamefont {Felser}}, \ and\ \bibinfo {author}
  {\bibfnamefont {B.}~\bibnamefont {Yan}},\ }\href {\doibase
  10.1103/physrevb.95.235104} {\bibfield  {journal} {\bibinfo  {journal} {Phys.
  Rev. B}\ }\textbf {\bibinfo {volume} {95}},\ \bibinfo {pages} {235104}
  (\bibinfo {year} {2017})}\BibitemShut {NoStop}%
\bibitem [{\citenamefont {Berlijn}\ \emph {et~al.}(2016)\citenamefont
  {Berlijn}, \citenamefont {Snijders}, \citenamefont {Delaire}, \citenamefont
  {Zhou}, \citenamefont {Maier}, \citenamefont {Cao}, \citenamefont {Chi},
  \citenamefont {Matsuda}, \citenamefont {Wang}, \citenamefont {Koehler},
  \citenamefont {Kent},\ and\ \citenamefont {Weitering}}]{Berlijn:2017p077201}%
  \BibitemOpen
  \bibfield  {author} {\bibinfo {author} {\bibfnamefont {T.}~\bibnamefont
  {Berlijn}}, \bibinfo {author} {\bibfnamefont {P.~C.}\ \bibnamefont
  {Snijders}}, \bibinfo {author} {\bibfnamefont {O.}~\bibnamefont {Delaire}},
  \bibinfo {author} {\bibfnamefont {H.-D.}\ \bibnamefont {Zhou}}, \bibinfo
  {author} {\bibfnamefont {T.~A.}\ \bibnamefont {Maier}}, \bibinfo {author}
  {\bibfnamefont {H.-B.}\ \bibnamefont {Cao}}, \bibinfo {author} {\bibfnamefont
  {S.-X.}\ \bibnamefont {Chi}}, \bibinfo {author} {\bibfnamefont
  {M.}~\bibnamefont {Matsuda}}, \bibinfo {author} {\bibfnamefont
  {Y.}~\bibnamefont {Wang}}, \bibinfo {author} {\bibfnamefont {M.~R.}\
  \bibnamefont {Koehler}}, \bibinfo {author} {\bibfnamefont {P.~R.~C.}\
  \bibnamefont {Kent}}, \ and\ \bibinfo {author} {\bibfnamefont {H.~H.}\
  \bibnamefont {Weitering}},\ }\href {\doibase 10.1103/physrevlett.118.077201}
  {\bibfield  {journal} {\bibinfo  {journal} {Phys. Rev. Lett.}\ }\textbf
  {\bibinfo {volume} {118}},\ \bibinfo {pages} {077201} (\bibinfo {year}
  {2016})}\BibitemShut {NoStop}%
\bibitem [{\citenamefont {Ahn}\ \emph {et~al.}(2019)\citenamefont {Ahn},
  \citenamefont {Hariki}, \citenamefont {Lee},\ and\ \citenamefont
  {Kuneš}}]{Ahn:2019p184432}%
  \BibitemOpen
  \bibfield  {author} {\bibinfo {author} {\bibfnamefont {K.-H.}\ \bibnamefont
  {Ahn}}, \bibinfo {author} {\bibfnamefont {A.}~\bibnamefont {Hariki}},
  \bibinfo {author} {\bibfnamefont {K.-W.}\ \bibnamefont {Lee}}, \ and\
  \bibinfo {author} {\bibfnamefont {J.}~\bibnamefont {Kuneš}},\ }\href
  {\doibase 10.1103/physrevb.99.184432} {\bibfield  {journal} {\bibinfo
  {journal} {Phys. Rev. B}\ }\textbf {\bibinfo {volume} {99}},\ \bibinfo
  {pages} {184432} (\bibinfo {year} {2019})}\BibitemShut {NoStop}%
\bibitem [{\citenamefont {Šmejkal}\ \emph {et~al.}(2020)\citenamefont
  {Šmejkal}, \citenamefont {González-Hernández}, \citenamefont {Jungwirth},\
  and\ \citenamefont {Sinova}}]{Smejkal:2020peaaz8809}%
  \BibitemOpen
  \bibfield  {author} {\bibinfo {author} {\bibfnamefont {L.}~\bibnamefont
  {Šmejkal}}, \bibinfo {author} {\bibfnamefont {R.}~\bibnamefont
  {González-Hernández}}, \bibinfo {author} {\bibfnamefont {T.}~\bibnamefont
  {Jungwirth}}, \ and\ \bibinfo {author} {\bibfnamefont {J.}~\bibnamefont
  {Sinova}},\ }\href {\doibase 10.1126/sciadv.aaz8809} {\bibfield  {journal}
  {\bibinfo  {journal} {Sci. Adv.}\ }\textbf {\bibinfo {volume} {6}},\ \bibinfo
  {pages} {eaaz8809} (\bibinfo {year} {2020})}\BibitemShut {NoStop}%
\bibitem [{\citenamefont {Feng}\ \emph {et~al.}(2022)\citenamefont {Feng},
  \citenamefont {Zhou}, \citenamefont {Šmejkal}, \citenamefont {Wu},
  \citenamefont {Zhu}, \citenamefont {Guo}, \citenamefont
  {González-Hernández}, \citenamefont {Wang}, \citenamefont {Yan},
  \citenamefont {Qin}, \citenamefont {Zhang}, \citenamefont {Wu}, \citenamefont
  {Chen}, \citenamefont {Meng}, \citenamefont {Liu}, \citenamefont {Xia},
  \citenamefont {Sinova}, \citenamefont {Jungwirth},\ and\ \citenamefont
  {Liu}}]{Feng:2022p735}%
  \BibitemOpen
  \bibfield  {author} {\bibinfo {author} {\bibfnamefont {Z.}~\bibnamefont
  {Feng}}, \bibinfo {author} {\bibfnamefont {X.}~\bibnamefont {Zhou}}, \bibinfo
  {author} {\bibfnamefont {L.}~\bibnamefont {Šmejkal}}, \bibinfo {author}
  {\bibfnamefont {L.}~\bibnamefont {Wu}}, \bibinfo {author} {\bibfnamefont
  {Z.}~\bibnamefont {Zhu}}, \bibinfo {author} {\bibfnamefont {H.}~\bibnamefont
  {Guo}}, \bibinfo {author} {\bibfnamefont {R.}~\bibnamefont
  {González-Hernández}}, \bibinfo {author} {\bibfnamefont {X.}~\bibnamefont
  {Wang}}, \bibinfo {author} {\bibfnamefont {H.}~\bibnamefont {Yan}}, \bibinfo
  {author} {\bibfnamefont {P.}~\bibnamefont {Qin}}, \bibinfo {author}
  {\bibfnamefont {X.}~\bibnamefont {Zhang}}, \bibinfo {author} {\bibfnamefont
  {H.}~\bibnamefont {Wu}}, \bibinfo {author} {\bibfnamefont {H.}~\bibnamefont
  {Chen}}, \bibinfo {author} {\bibfnamefont {Z.}~\bibnamefont {Meng}}, \bibinfo
  {author} {\bibfnamefont {L.}~\bibnamefont {Liu}}, \bibinfo {author}
  {\bibfnamefont {Z.}~\bibnamefont {Xia}}, \bibinfo {author} {\bibfnamefont
  {J.}~\bibnamefont {Sinova}}, \bibinfo {author} {\bibfnamefont
  {T.}~\bibnamefont {Jungwirth}}, \ and\ \bibinfo {author} {\bibfnamefont
  {Z.}~\bibnamefont {Liu}},\ }\href {\doibase 10.1038/s41928-022-00866-z}
  {\bibfield  {journal} {\bibinfo  {journal} {Nat. Electron.}\ }\textbf
  {\bibinfo {volume} {5}},\ \bibinfo {pages} {735} (\bibinfo {year}
  {2022})}\BibitemShut {NoStop}%
\bibitem [{\citenamefont {Yuan}\ \emph {et~al.}(2020)\citenamefont {Yuan},
  \citenamefont {Wang}, \citenamefont {Luo}, \citenamefont {Rashba},\ and\
  \citenamefont {Zunger}}]{Yuan:2020p014422}%
  \BibitemOpen
  \bibfield  {author} {\bibinfo {author} {\bibfnamefont {L.-D.}\ \bibnamefont
  {Yuan}}, \bibinfo {author} {\bibfnamefont {Z.}~\bibnamefont {Wang}}, \bibinfo
  {author} {\bibfnamefont {J.-W.}\ \bibnamefont {Luo}}, \bibinfo {author}
  {\bibfnamefont {E.~I.}\ \bibnamefont {Rashba}}, \ and\ \bibinfo {author}
  {\bibfnamefont {A.}~\bibnamefont {Zunger}},\ }\href {\doibase
  10.1103/physrevb.102.014422} {\bibfield  {journal} {\bibinfo  {journal}
  {Phys. Rev. B}\ }\textbf {\bibinfo {volume} {102}},\ \bibinfo {pages}
  {014422} (\bibinfo {year} {2020})}\BibitemShut {NoStop}%
\bibitem [{\citenamefont {Glazer}(1972)}]{Glazer:1972p3384}%
  \BibitemOpen
  \bibfield  {author} {\bibinfo {author} {\bibfnamefont {A.~M.}\ \bibnamefont
  {Glazer}},\ }\href {\doibase 10.1107/s0567740872007976} {\bibfield  {journal}
  {\bibinfo  {journal} {Acta Crystallographica Section B}\ }\textbf {\bibinfo
  {volume} {28}},\ \bibinfo {pages} {3384} (\bibinfo {year}
  {1972})}\BibitemShut {NoStop}%
\bibitem [{\citenamefont {Schmitz}\ \emph {et~al.}(2005)\citenamefont
  {Schmitz}, \citenamefont {Entin-Wohlman}, \citenamefont {Aharony},
  \citenamefont {Harris},\ and\ \citenamefont
  {Müller-Hartmann}}]{Schmitz:2005p144412}%
  \BibitemOpen
  \bibfield  {author} {\bibinfo {author} {\bibfnamefont {R.}~\bibnamefont
  {Schmitz}}, \bibinfo {author} {\bibfnamefont {O.}~\bibnamefont
  {Entin-Wohlman}}, \bibinfo {author} {\bibfnamefont {A.}~\bibnamefont
  {Aharony}}, \bibinfo {author} {\bibfnamefont {A.~B.}\ \bibnamefont {Harris}},
  \ and\ \bibinfo {author} {\bibfnamefont {E.}~\bibnamefont
  {Müller-Hartmann}},\ }\href {\doibase 10.1103/physrevb.71.144412} {\bibfield
   {journal} {\bibinfo  {journal} {Phys. Rev. B}\ }\textbf {\bibinfo {volume}
  {71}},\ \bibinfo {pages} {144412} (\bibinfo {year} {2005})}\BibitemShut
  {NoStop}%
\bibitem [{\citenamefont {Bousquet}\ and\ \citenamefont
  {Spaldin}(2011)}]{Bousquet:2011p197603}%
  \BibitemOpen
  \bibfield  {author} {\bibinfo {author} {\bibfnamefont {E.}~\bibnamefont
  {Bousquet}}\ and\ \bibinfo {author} {\bibfnamefont {N.}~\bibnamefont
  {Spaldin}},\ }\href {\doibase 10.1103/physrevlett.107.197603} {\bibfield
  {journal} {\bibinfo  {journal} {Phys. Rev. Lett.}\ }\textbf {\bibinfo
  {volume} {107}},\ \bibinfo {pages} {197603} (\bibinfo {year}
  {2011})}\BibitemShut {NoStop}%
\bibitem [{\citenamefont {Wang}\ \emph {et~al.}(2022)\citenamefont {Wang},
  \citenamefont {Gautreau}, \citenamefont {Birol},\ and\ \citenamefont
  {Fernandes}}]{Wang:2022p144404}%
  \BibitemOpen
  \bibfield  {author} {\bibinfo {author} {\bibfnamefont {Z.}~\bibnamefont
  {Wang}}, \bibinfo {author} {\bibfnamefont {D.}~\bibnamefont {Gautreau}},
  \bibinfo {author} {\bibfnamefont {T.}~\bibnamefont {Birol}}, \ and\ \bibinfo
  {author} {\bibfnamefont {R.~M.}\ \bibnamefont {Fernandes}},\ }\href {\doibase
  10.1103/physrevb.105.144404} {\bibfield  {journal} {\bibinfo  {journal}
  {Phys. Rev. B}\ }\textbf {\bibinfo {volume} {105}},\ \bibinfo {pages}
  {144404} (\bibinfo {year} {2022})}\BibitemShut {NoStop}%
\bibitem [{\citenamefont {Rooj}\ \emph {et~al.}(2025)\citenamefont {Rooj},
  \citenamefont {Saxena},\ and\ \citenamefont {Ganguli}}]{Rooj:2025p014434}%
  \BibitemOpen
  \bibfield  {author} {\bibinfo {author} {\bibfnamefont {S.}~\bibnamefont
  {Rooj}}, \bibinfo {author} {\bibfnamefont {S.}~\bibnamefont {Saxena}}, \ and\
  \bibinfo {author} {\bibfnamefont {N.}~\bibnamefont {Ganguli}},\ }\href
  {\doibase 10.1103/physrevb.111.014434} {\bibfield  {journal} {\bibinfo
  {journal} {Phys. Rev. B}\ }\textbf {\bibinfo {volume} {111}},\ \bibinfo
  {pages} {014434} (\bibinfo {year} {2025})}\BibitemShut {NoStop}%
\bibitem [{\citenamefont {Venderbos}(2025)}]{Venderbos:arXiv2025-2}%
  \BibitemOpen
  \bibfield  {author} {\bibinfo {author} {\bibfnamefont {J.~W.~F.}\
  \bibnamefont {Venderbos}},\ }\href {\doibase 10.48550/arxiv.2512.19380}
  {\bibfield  {journal} {\bibinfo  {journal} {arXiv}\ } (\bibinfo {year}
  {2025}),\ 10.48550/arxiv.2512.19380}\BibitemShut {NoStop}%
\bibitem [{\citenamefont {Radhakrishnan}\ \emph {et~al.}(2026)\citenamefont
  {Radhakrishnan}, \citenamefont {Bell}, \citenamefont {Ortix},\ and\
  \citenamefont {Venderbos}}]{Radhakrishnan:2016arXiv02}%
  \BibitemOpen
  \bibfield  {author} {\bibinfo {author} {\bibfnamefont {H.}~\bibnamefont
  {Radhakrishnan}}, \bibinfo {author} {\bibfnamefont {B.}~\bibnamefont {Bell}},
  \bibinfo {author} {\bibfnamefont {C.}~\bibnamefont {Ortix}}, \ and\ \bibinfo
  {author} {\bibfnamefont {J.~W.~F.}\ \bibnamefont {Venderbos}},\ }\href
  {\doibase 10.48550/arxiv.2602.05894} {\bibfield  {journal} {\bibinfo
  {journal} {arXiv}\ } (\bibinfo {year} {2026}),\ 10.48550/arxiv.2602.05894},\
  \Eprint {http://arxiv.org/abs/2602.05894} {2602.05894} \BibitemShut {NoStop}%
\bibitem [{\citenamefont {Lin}\ \emph {et~al.}(2018)\citenamefont {Lin},
  \citenamefont {Si}, \citenamefont {Zhu}, \citenamefont {Cai}, \citenamefont
  {Li}, \citenamefont {Kong}, \citenamefont {Yu},\ and\ \citenamefont
  {Wen}}]{Lin:2018p075132}%
  \BibitemOpen
  \bibfield  {author} {\bibinfo {author} {\bibfnamefont {H.}~\bibnamefont
  {Lin}}, \bibinfo {author} {\bibfnamefont {J.}~\bibnamefont {Si}}, \bibinfo
  {author} {\bibfnamefont {X.}~\bibnamefont {Zhu}}, \bibinfo {author}
  {\bibfnamefont {K.}~\bibnamefont {Cai}}, \bibinfo {author} {\bibfnamefont
  {H.}~\bibnamefont {Li}}, \bibinfo {author} {\bibfnamefont {L.}~\bibnamefont
  {Kong}}, \bibinfo {author} {\bibfnamefont {X.}~\bibnamefont {Yu}}, \ and\
  \bibinfo {author} {\bibfnamefont {H.-H.}\ \bibnamefont {Wen}},\ }\href
  {\doibase 10.1103/physrevb.98.075132} {\bibfield  {journal} {\bibinfo
  {journal} {Phys. Rev. B}\ }\textbf {\bibinfo {volume} {98}},\ \bibinfo
  {pages} {075132} (\bibinfo {year} {2018})}\BibitemShut {NoStop}%
\bibitem [{\citenamefont {Wei}\ \emph {et~al.}(2025)\citenamefont {Wei},
  \citenamefont {Li}, \citenamefont {Hatt}, \citenamefont {Huai}, \citenamefont
  {Liu}, \citenamefont {Singh}, \citenamefont {Kim}, \citenamefont {Fernandes},
  \citenamefont {Cardon}, \citenamefont {Zhao}, \citenamefont {Tran},
  \citenamefont {Frandsen}, \citenamefont {Burch}, \citenamefont {Liu},\ and\
  \citenamefont {Ji}}]{Wei:2025p024402}%
  \BibitemOpen
  \bibfield  {author} {\bibinfo {author} {\bibfnamefont {C.-C.}\ \bibnamefont
  {Wei}}, \bibinfo {author} {\bibfnamefont {X.}~\bibnamefont {Li}}, \bibinfo
  {author} {\bibfnamefont {S.}~\bibnamefont {Hatt}}, \bibinfo {author}
  {\bibfnamefont {X.}~\bibnamefont {Huai}}, \bibinfo {author} {\bibfnamefont
  {J.}~\bibnamefont {Liu}}, \bibinfo {author} {\bibfnamefont {B.}~\bibnamefont
  {Singh}}, \bibinfo {author} {\bibfnamefont {K.-M.}\ \bibnamefont {Kim}},
  \bibinfo {author} {\bibfnamefont {R.~M.}\ \bibnamefont {Fernandes}}, \bibinfo
  {author} {\bibfnamefont {P.}~\bibnamefont {Cardon}}, \bibinfo {author}
  {\bibfnamefont {L.}~\bibnamefont {Zhao}}, \bibinfo {author} {\bibfnamefont
  {T.~T.}\ \bibnamefont {Tran}}, \bibinfo {author} {\bibfnamefont {B.~A.}\
  \bibnamefont {Frandsen}}, \bibinfo {author} {\bibfnamefont {K.~S.}\
  \bibnamefont {Burch}}, \bibinfo {author} {\bibfnamefont {F.}~\bibnamefont
  {Liu}}, \ and\ \bibinfo {author} {\bibfnamefont {H.}~\bibnamefont {Ji}},\
  }\href {\doibase 10.1103/physrevmaterials.9.024402} {\bibfield  {journal}
  {\bibinfo  {journal} {Phys. Rev. Materials}\ }\textbf {\bibinfo {volume}
  {9}},\ \bibinfo {pages} {024402} (\bibinfo {year} {2025})}\BibitemShut
  {NoStop}%
\bibitem [{\citenamefont {Jiang}\ \emph
  {et~al.}(2025{\natexlab{b}})\citenamefont {Jiang}, \citenamefont {Hu},
  \citenamefont {Bai}, \citenamefont {Song}, \citenamefont {Mu}, \citenamefont
  {Qu}, \citenamefont {Li}, \citenamefont {Zhu}, \citenamefont {Pi},
  \citenamefont {Wei}, \citenamefont {Sun}, \citenamefont {Huang},
  \citenamefont {Zheng}, \citenamefont {Peng}, \citenamefont {He},
  \citenamefont {Li}, \citenamefont {Luo}, \citenamefont {Li}, \citenamefont
  {Chen}, \citenamefont {Li}, \citenamefont {Weng},\ and\ \citenamefont
  {Qian}}]{Jiang:2025p754}%
  \BibitemOpen
  \bibfield  {author} {\bibinfo {author} {\bibfnamefont {B.}~\bibnamefont
  {Jiang}}, \bibinfo {author} {\bibfnamefont {M.}~\bibnamefont {Hu}}, \bibinfo
  {author} {\bibfnamefont {J.}~\bibnamefont {Bai}}, \bibinfo {author}
  {\bibfnamefont {Z.}~\bibnamefont {Song}}, \bibinfo {author} {\bibfnamefont
  {C.}~\bibnamefont {Mu}}, \bibinfo {author} {\bibfnamefont {G.}~\bibnamefont
  {Qu}}, \bibinfo {author} {\bibfnamefont {W.}~\bibnamefont {Li}}, \bibinfo
  {author} {\bibfnamefont {W.}~\bibnamefont {Zhu}}, \bibinfo {author}
  {\bibfnamefont {H.}~\bibnamefont {Pi}}, \bibinfo {author} {\bibfnamefont
  {Z.}~\bibnamefont {Wei}}, \bibinfo {author} {\bibfnamefont {Y.-J.}\
  \bibnamefont {Sun}}, \bibinfo {author} {\bibfnamefont {Y.}~\bibnamefont
  {Huang}}, \bibinfo {author} {\bibfnamefont {X.}~\bibnamefont {Zheng}},
  \bibinfo {author} {\bibfnamefont {Y.}~\bibnamefont {Peng}}, \bibinfo {author}
  {\bibfnamefont {L.}~\bibnamefont {He}}, \bibinfo {author} {\bibfnamefont
  {S.}~\bibnamefont {Li}}, \bibinfo {author} {\bibfnamefont {J.}~\bibnamefont
  {Luo}}, \bibinfo {author} {\bibfnamefont {Z.}~\bibnamefont {Li}}, \bibinfo
  {author} {\bibfnamefont {G.}~\bibnamefont {Chen}}, \bibinfo {author}
  {\bibfnamefont {H.}~\bibnamefont {Li}}, \bibinfo {author} {\bibfnamefont
  {H.}~\bibnamefont {Weng}}, \ and\ \bibinfo {author} {\bibfnamefont
  {T.}~\bibnamefont {Qian}},\ }\href {\doibase 10.1038/s41567-025-02822-y}
  {\bibfield  {journal} {\bibinfo  {journal} {Nat. Phys.}\ }\textbf {\bibinfo
  {volume} {21}},\ \bibinfo {pages} {754} (\bibinfo {year}
  {2025}{\natexlab{b}})}\BibitemShut {NoStop}%
\bibitem [{\citenamefont {Zhang}\ \emph
  {et~al.}(2025{\natexlab{b}})\citenamefont {Zhang}, \citenamefont {Cheng},
  \citenamefont {Yin}, \citenamefont {Liu}, \citenamefont {Deng}, \citenamefont
  {Qiao}, \citenamefont {Shi}, \citenamefont {Zhang}, \citenamefont {Lin},
  \citenamefont {Liu}, \citenamefont {Ye}, \citenamefont {Huang}, \citenamefont
  {Meng}, \citenamefont {Zhang}, \citenamefont {Okuda}, \citenamefont
  {Shimada}, \citenamefont {Cui}, \citenamefont {Zhao}, \citenamefont {Cao},
  \citenamefont {Qiao}, \citenamefont {Liu},\ and\ \citenamefont
  {Chen}}]{Zhang:2025p760}%
  \BibitemOpen
  \bibfield  {author} {\bibinfo {author} {\bibfnamefont {F.}~\bibnamefont
  {Zhang}}, \bibinfo {author} {\bibfnamefont {X.}~\bibnamefont {Cheng}},
  \bibinfo {author} {\bibfnamefont {Z.}~\bibnamefont {Yin}}, \bibinfo {author}
  {\bibfnamefont {C.}~\bibnamefont {Liu}}, \bibinfo {author} {\bibfnamefont
  {L.}~\bibnamefont {Deng}}, \bibinfo {author} {\bibfnamefont {Y.}~\bibnamefont
  {Qiao}}, \bibinfo {author} {\bibfnamefont {Z.}~\bibnamefont {Shi}}, \bibinfo
  {author} {\bibfnamefont {S.}~\bibnamefont {Zhang}}, \bibinfo {author}
  {\bibfnamefont {J.}~\bibnamefont {Lin}}, \bibinfo {author} {\bibfnamefont
  {Z.}~\bibnamefont {Liu}}, \bibinfo {author} {\bibfnamefont {M.}~\bibnamefont
  {Ye}}, \bibinfo {author} {\bibfnamefont {Y.}~\bibnamefont {Huang}}, \bibinfo
  {author} {\bibfnamefont {X.}~\bibnamefont {Meng}}, \bibinfo {author}
  {\bibfnamefont {C.}~\bibnamefont {Zhang}}, \bibinfo {author} {\bibfnamefont
  {T.}~\bibnamefont {Okuda}}, \bibinfo {author} {\bibfnamefont
  {K.}~\bibnamefont {Shimada}}, \bibinfo {author} {\bibfnamefont
  {S.}~\bibnamefont {Cui}}, \bibinfo {author} {\bibfnamefont {Y.}~\bibnamefont
  {Zhao}}, \bibinfo {author} {\bibfnamefont {G.-H.}\ \bibnamefont {Cao}},
  \bibinfo {author} {\bibfnamefont {S.}~\bibnamefont {Qiao}}, \bibinfo {author}
  {\bibfnamefont {J.}~\bibnamefont {Liu}}, \ and\ \bibinfo {author}
  {\bibfnamefont {C.}~\bibnamefont {Chen}},\ }\href {\doibase
  10.1038/s41567-025-02864-2} {\bibfield  {journal} {\bibinfo  {journal} {Nat.
  Phys.}\ }\textbf {\bibinfo {volume} {21}},\ \bibinfo {pages} {760} (\bibinfo
  {year} {2025}{\natexlab{b}})}\BibitemShut {NoStop}%
\bibitem [{\citenamefont {Chang}\ \emph {et~al.}(2025)\citenamefont {Chang},
  \citenamefont {Mazin},\ and\ \citenamefont
  {Belashchenko}}]{Chang:2025arXiv08}%
  \BibitemOpen
  \bibfield  {author} {\bibinfo {author} {\bibfnamefont {P.-H.}\ \bibnamefont
  {Chang}}, \bibinfo {author} {\bibfnamefont {I.~I.}\ \bibnamefont {Mazin}}, \
  and\ \bibinfo {author} {\bibfnamefont {K.~D.}\ \bibnamefont {Belashchenko}},\
  }\href {\doibase 10.48550/arxiv.2508.04839} {\bibfield  {journal} {\bibinfo
  {journal} {arXiv}\ } (\bibinfo {year} {2025}),\
  10.48550/arxiv.2508.04839}\BibitemShut {NoStop}%
\end{thebibliography}
\end{document}